\documentclass[a4paper,onecolumn,11pt,accepted=2025-12-17]{quantumarticle}
\pdfoutput=1
\usepackage[utf8]{inputenc}
\usepackage[english]{babel}
\usepackage[T1]{fontenc}
\usepackage{tikz}
\usepackage[numbers,sort&compress]{natbib}

\usepackage{mathtools}
\usepackage{siunitx}
\usepackage{amsfonts}
\usepackage{amssymb}
\usepackage[hidelinks]{hyperref}
\usepackage{svg}
\usepackage{float}
\usepackage[braket, qm]{qcircuit}
\usepackage{bbm}
\usepackage{bm}
\usepackage{lipsum} 
\usepackage{tocloft}
\usepackage{titletoc}
\usepackage{comment}
\usepackage[normalem]{ulem}
\usepackage{stackrel}
\usepackage{algorithm2e}
\usepackage[dvipsnames]{xcolor}
\usepackage{subfiles}
\graphicspath{ {Figures/} }

\newcommand{\NS}{S} 
\newcommand{\rhou}{\rho} 
\newcommand{\upr}{U_{\rm enc}}
\newcommand{\udec}{U_{\rm dec}}
\newcommand{\usense}{U_{\rm sense}} 
\newcommand{\usensegen}{U_{\rm sense}^{\rm general}}
\newcommand{\boltzmann}{k_{\rm B}}
\DeclareMathOperator{\sinc}{sinc}

\begin{document}

\title{Exponential advantage in quantum sensing of correlated parameters}

\author{Sridhar~Prabhu$^{*}$}
\affiliation{School of Applied and Engineering Physics, Cornell University, Ithaca, NY 14853, USA}
\affiliation{Department of Physics, Cornell University, Ithaca, NY 14853, USA}
\orcid{0009-0005-4575-3916}
\author{Vladimir~Kremenetski$^{*}$}
\affiliation{School of Applied and Engineering Physics, Cornell University, Ithaca, NY 14853, USA}
\orcid{0000-0002-6260-4171}
\author{Saeed~A.~Khan}
\affiliation{School of Applied and Engineering Physics, Cornell University, Ithaca, NY 14853, USA}
\orcid{0000-0002-8047-4657}
\author{Ryotatsu~Yanagimoto}
\affiliation{School of Applied and Engineering Physics, Cornell University, Ithaca, NY 14853, USA}
\affiliation{Physics \& Informatics Laboratories, NTT Research, Inc., Sunnyvale, CA 94085, USA}
\orcid{0000-0002-9609-0076}
\author{Peter~L.~McMahon$^{\dagger}$}
\affiliation{School of Applied and Engineering Physics, Cornell University, Ithaca, NY 14853, USA}
\affiliation{Kavli Institute at Cornell for Nanoscale Science, Cornell University, Ithaca, NY 14853, USA}
\orcid{0000-0002-1177-9887}

\def\thefootnote{*}\footnotetext{These authors contributed equally.}
\def\thefootnote{$\dagger$}\footnotetext{To whom correspondence should be addressed: svp36@cornell.edu, vk387@cornell.edu, pmcmahon@cornell.edu}

\maketitle
\begin{abstract}
  Conventionally in quantum sensing, the goal is to estimate one or more unknown parameters that are assumed to be deterministic---that is, they do not change between shots of the quantum-sensing protocol. We instead consider the setting where the parameters are stochastic: each shot of the quantum-sensing protocol senses parameter values that come from independent random draws. In this work, we explore three examples where the stochastic parameters are \textit{correlated} and show how using entanglement provides a benefit in classification or estimation tasks: (1) a two-parameter classification task, for which an entangled sensor can achieve a large improvement in classification accuracy over an unentangled sensor given a low shot budget (e.g., 97\% versus 80\% given a budget of 50 shots); (2) an $N$-parameter estimation task and a classification variant of it, for which an entangled sensor requires just a constant number (independent of $N$) shots to achieve the same error as an unentangled sensor using exponentially many (${\sim}2^N$) shots---with numerical experiments showing that an entangled sensor needs several orders of magnitude fewer shots than an unentangled sensor even in cases when $N$ is small; (3) classifying the magnetization of a spin chain in thermal equilibrium, where the individual spins fluctuate but the total spin in one direction is conserved---this gives a practical setting in which stochastic parameters are correlated in a way that an entangled sensor can be designed to exploit, and numerical experiments showing that even with small $N$, an entangled sensor can achieve the same accuracy as an unentangled sensor but use ${\sim}100\times$ fewer shots. We also present a theoretical framework for assessing, for a given choice of entangled sensing protocol and distributions to discriminate between, how much advantage the entangled sensor would have over an unentangled sensor. Our work motivates the further study of sensing correlated stochastic parameters using entangled quantum sensors---and since classical sensors by definition cannot be entangled, our work shows the possibility for entangled quantum sensors to achieve an exponential advantage in sample complexity over classical sensors, in contrast to the typical quadratic advantage.
\end{abstract}

\section{Introduction}
\label{sec:intro}

Quantum sensing involves the use of quantum resources (such as entanglement) to improve the sensitivity of a sensor in measuring parameters beyond the classical limit, also known as the standard quantum limit (SQL)~\cite{degen_quantum_2017, giovannetti_quantum_2006}. In some cases, this allows for an improvement in sensitivity scaling with resources beyond that set by the SQL. In the prototypical example of phase-sensing with qubits, where the phase is a rotation around the Pauli-Z axis, interferometry protocols using the Greenberger-Horne-Zeilinger (GHZ) state~\cite{liu_2021, greenberger1989going} are known to be optimal in sensitivity. In this case, the GHZ state with $N$ qubits  achieves an error in estimating the phase that scales as $1/N$, known as the Heisenberg limit. In contrast, the SQL, which is achievable with $N$-qubit unentangled qubits, can only achieve an error that scales as $1/\sqrt{N}$. In other words, the unentangled sensing-protocol requires a number of samples that scales quadratically in $N$ to achieve the same error as the entangled GHZ sensing-protocol. There are many other well-established quantum-sensing protocols for other sensing tasks and systems, which use quantum states such as spin-squeezed states~\cite{Colombo_spinsqueeze_2022} and $N00N$ states~\cite{Jones_N00N-2009,Dowling_N00N_2008} to name a few. In these sensing tasks, the parameter to be sensed remains fixed during the shots of the quantum-sensing protocol, as shown in Fig.~\ref{Fig:Main1} a).

An alternative setting is when the parameters are \textit{stochastic}, that is, varying between shots of the quantum-sensing protocol~\cite{tsang_quantum_2016, mouradian_quantum_2021, gefen2019overcoming, shi2023ultimate, gardner2025stochastic, tsang2023quantum, gardner2025lindblad, tsang2011fundamental, tsang2012fundamental, ng2016spectrum} (see Fig.~\ref{Fig:Main1} b)). Such a regime includes many physical scenarios, such as fluctuations in thermodynamic quantities of classically-interacting systems, or background spectral noise in electromagnetic signals. In this setting, the parameters to be sensed can be described as samples from an underlying probability distribution~\footnote{Throughout this manuscript, the terms shots and samples are interchangeable because we assume that each shot of a quantum-sensing protocol senses a single new sample of the stochastic parameters.}. Over multiple shots, the measurement outcomes of the quantum-sensing protocol can be used to infer the properties of the probability distribution. The quantum-sensing advantage and the role of entanglement in this setting remain largely open questions. 

The aim of the sensing tasks we consider in this work are the binary-classification of probability distributions and estimation of the parameters of probability distributions that govern the stochastic signals received by the quantum sensor. In this work, we focus on phase-sensing with qubits and show that even simple sensing tasks involving the discrimination of two distributions of multiple, correlated, stochastic parameters can yield an exponential sensing advantage, where an entangled quantum sensor can achieve the same classification accuracy as an unentangled quantum sensor with exponentially fewer samples. We illustrate this with a pedagogical example of two qubits to sense two phases, before generalizing to an $N$-qubit $N$-phases task. The phase sensed by each qubit varies between shots of the sensing protocol. However, the sum of the phases is constrained to take a fixed value. We consider the task of estimating this sum. We show how an entangled quantum-sensing protocol requires exponentially fewer samples than the optimal unentangled quantum-sensing protocol to achieve the same error of the estimate, as a function of $N$. We also consider a binary-classification task~\cite{chin2024quantum, zhuang_physical-layer_2019, Sinanan_Singh_2024, rossi_quantum_2022}, where the two classes have different values for the sum of the phases. This exponential advantage in sample complexity arises from the fine-tuned distribution of the parameters we choose. While such a scaling is difficult to obtain more generally, we numerically show that the advantage can still be large. To motivate this, we consider a stochastic sensing task which can arise in physical systems. We consider the task of sensing the conserved total spin Z-component of the classical XXZ spin-chain model. We show that this advantage stems from the ability of an entangled quantum sensor to directly probe the low-noise, information-carrying combinations of otherwise stochastic parameters.

The rest of this paper is organized as follows. In Sec.~\ref{sec:stoch}, we introduce the paradigm of quantum sensing of stochastic parameters, in contrast to conventional quantum sensing. In Sec.~\ref{sec:corrgauss}, we describe a simple example of stochastic sensing for the task of discriminating between two two-dimensional Gaussian distributions. In particular, we show how a two-qubit Bell state can provide an improvement in classification accuracy over the optimal unentangled-sensing protocol. In Sec.~\ref{sec:exp}, we discuss two related $N$-parameter stochastic sensing tasks for which an entangled quantum sensor can achieve an exponential sensing advantage in sample complexity, discussing the reasons for such an advantage and illustrating our theoretical results using simulations. The precise nature of the probability distribution of the $N$-parameter phases allows us to realize this advantage, but limits the scope of realistic scenarios where such distributions naturally arise. In Sec.~\ref{sec:xxz}, we motivate how entanglement can improve sensitivity for correlated distributions that can arise in physical systems, studying a sensing task involving the classical XXZ spin-system. Sec.~\ref{sec:theory} introduces a general framework to assess whether an entangled sensor has a substantial scaling advantage over an unentangled sensor, and if it does, to estimate how much advantage it has (i.e. polynomial, exponential, etc.). We also present conditions that guarantee or forbid exponential an entanglement-enabled exponential advantage for binary-classification between parameter distributions. Sec.~\ref{sec:conc} summarizes our results, followed by an outlook on avenues for future exploration.

\section{Our Setting: Quantum Sensing of Stochastic Parameters}
\label{sec:stoch}

\begin{figure}[h!]
    \centering
    \includegraphics[width=0.8\linewidth]{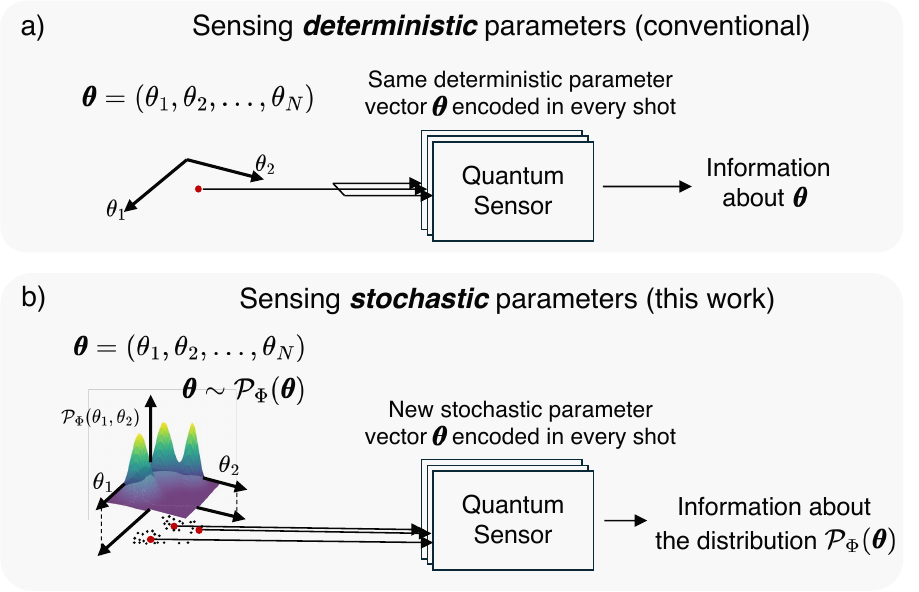}
    \caption{\textbf{Quantum sensing of deterministic versus stochastic parameters} \textbf{a)} Conventional quantum sensing tasks estimate or classify an (most generally) $N$-dimensional parameter vector $\bm{\theta}$ using samples obtained from a quantum sensor. For each sample, the same \textit{deterministic} $\bm{\theta}$ is received by the quantum sensor. \textbf{b)} In quantum sensing of \textit{stochastic} parameters, which we consider in this work, the $N$-dimensional parameter vector $\bm{\theta}$ received by the quantum sensor is sampled from a probability distribution $\mathcal{P}_{\Phi}(\bm{\theta})$. Each sample obtained from the quantum sensor experiences a different, \textit{stochastic} $\bm{\theta}$. Estimates made using quantum sensor measurement outcomes in this setting consequently depend on the properties of the underlying distribution $\mathcal{P}_{\Phi}(\bm{\theta})$. In this work, we show the advantages of entanglement in the quantum sensor in this latter setting.}
    \label{Fig:Main1}
\end{figure}

Conventionally, it is usually assumed that under repeated executions of a quantum-sensing protocol, the unknown parameter being sensed remains fixed~(see Fig.~\ref{Fig:Main1} a)). In this work, we explore the paradigm of sensing an (most generally) $N$-dimensional vector of parameters $\bm{\theta}$ that is itself described by a probability distribution, as illustrated in Fig.~\ref{Fig:Main1} b). In this setting, every repeated measurement of the sensing protocol experiences a potentially different parameter. Such a framework can describe the sensing of parameters generated by a stochastic physical process, where $\bm{\theta}$ is described by a random parameter sampled from a parameterized probability distribution $\mathcal{P}_{\Phi}(\bm{\theta})$. Here $\Phi$ describes all the features of the probability distribution, such as its mean, variance, covariance, etc. We discuss this framework in the context of binary classification, with the goal of predicting which of two distributions $\mathcal{P}_{\rm A}(\bm{\theta})$ or $\mathcal{P}_{\rm B}(\bm{\theta})$ generates the stochastic parameters received by the quantum sensor. Our framework can be naturally extended to multi-class classification tasks, and estimation tasks (with the goal of estimating continuous parameters of the distribution). We explore the latter in Sec.~\ref{sec:exp}.

The general protocol we consider consists of a quantum sensor comprising $N$ qubits  initialized to the all-zero state $\ket{\psi_0}$ prior to the sensing protocol. An encoding unitary $\upr$ is designed to create the optimal sensing probe state $\ket{\psi_{\rm probe}} = \upr \ket{\psi_0}$, followed by the sensing unitary $\usense(\bm{\theta})$, which depends on the instance of stochastic parameters $\bm{\theta}$. This sensing unitary is determined by the physical interaction between the signal and the sensor. In this work, we consider the prototypical scenario of phase sensing with qubits, namely:
\begin{align}
    \usense(\bm{\theta}) = e^{-\frac{i}{2} \sum_i \theta_i \hat{\sigma}_i^z},
    \label{eq:usense}
\end{align}
where $\hat{\sigma}_i^z$ is the Pauli-Z operator acting on the $i^{\rm th}$ qubit. We consider more general sensing unitaries in Appendices~\ref{s: general framework proofs} and~\ref{s:nonlinear constraint noncommuting example}, both for the cases where the terms in the exponent of the unitary commute with each other and where they do not. Following the sensing unitary, the quantum state is projected in the optimal measurement basis. We represent this action by a $\udec$ unitary, which produces the state $\ket{\psi_{\rm f}(\bm{\theta})} \equiv \udec\usense(\bm{\theta})\upr\ket{\psi_0}$. Finally, a measurement in the computational bit-string basis, defined by the positive operator-valued measure (POVM) elements $\{\hat{M_k}\}$, results in a single sample from the quantum sensor. The expectation of the measurement outcomes is formally obtained by averaging over the (unknown) probability distribution functions of a class
\begin{align}
    p_k(\Phi) = \int d\bm{\theta}~\mathcal{P}_{\Phi}(\bm{\theta}){\rm Tr}(\hat{M_k} \ket{\psi_{\rm f}(\bm{\theta})}\bra{\psi_{\rm f}(\bm{\theta})})
    \label{eq:pkInputAv}.
\end{align}
The distribution of measurement outcomes, therefore, depends not on a specific instance of random values $\bm{\theta}$, but instead on the parameters $\Phi$ governing the probability distribution describing these random parameters.

In this work, we consider tasks where the underlying probability distributions give rise to correlations between multiple stochastic parameters. We identify and discuss the potential for a quantum entanglement sensing advantage by comparing the optimal sensing protocol with entanglement (where $\upr$ and $\udec$ contain gates acting on multiple qubits), against one without entanglement (where $\upr$ and $\udec$ are restricted to single-qubit gates). While our analyses and results are formally about identifying advantages between quantum-sensing protocols using entanglement and protocols not using any entanglement, we can also interpret our results as showing an advantage of quantum sensing with entanglement versus classical sensing, since classical sensors by definition cannot be entangled. We first present a simple two-parameter example in the next section.

\section{Two-qubit example: discriminating bivariate distributions} 
\label{sec:corrgauss}

\subsection{Introduction}

In this section, we present an example that illustrates the advantage that can be gained by using an entangled quantum-sensing protocol for sensing stochastic parameters. We consider two probability distributions over a pair of phases such that the average difference between the phases is distinct, while the average sum of the phases is the same. For both probability distributions, the difference between the phases has lower noise than the sum of the phases. An entangled quantum-sensing protocol with the Bell probe state $(\ket{01}+\ket{10})/\sqrt{2}$ and measurement in the Bell state basis can directly extract information about the low-noise difference directly, allowing it to accurately discriminate between the two probability distributions. In contrast, an unentangled probe state and measurement basis can only access this information through individual phases on each qubit, which are noisier than the difference. This ability of an entangled quantum-sensing protocol to access less-noisy functions of the phases leads to an entanglement-enabled quantum-sensing advantage.

\begin{figure}[h!]
    \centering
    \includegraphics[width=0.95\linewidth]{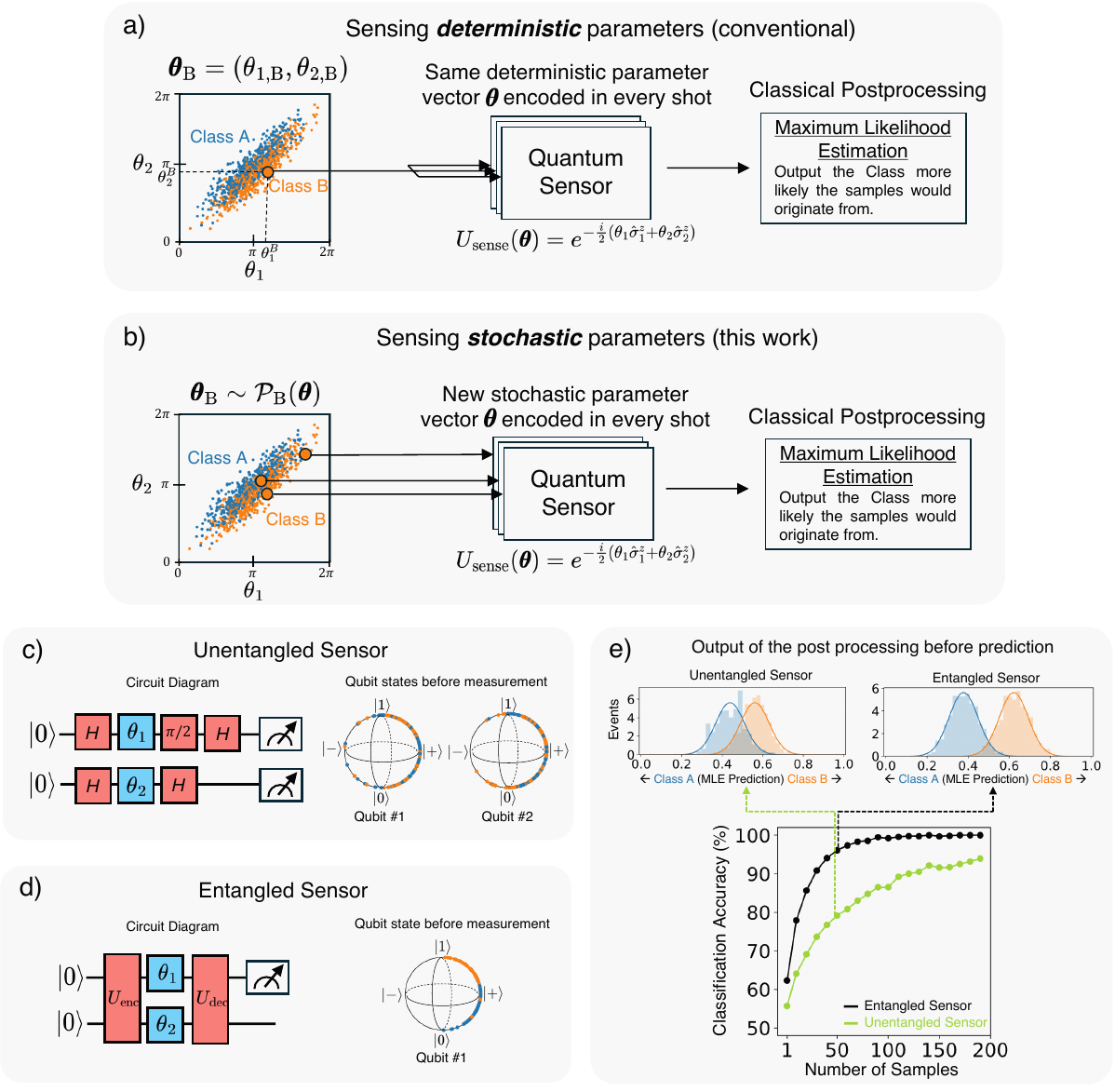}
    \caption{\textbf{Example of two-qubit stochastic sensing task, discriminating two Gaussian distributions of correlated parameters $\bm{\theta}=(\theta_1,\theta_2)$}. \textbf{a)} Conventional sensing for binary distribution discrimination. A fixed sample $\bm{\theta}_{L}$ sampled from one of two Gaussian distributions $L=\rm A, \rm B$ is received by the quantum sensor $S$ times. The sensing interaction described by the unitary operator $\usense(\bm{\theta}) = e^{-\frac{i}{2}(\theta_1 \hat{\sigma}^z_1 + \theta_2 \hat{\sigma}^z_2)}$. Using measurement outcomes averaged over $S$ samples, a maximum likelihood estimator (MLE) is used to determine which class the samples $\bm{\theta}_{L}$ originated from. \textbf{b)} Same task as \textbf{a)}, but where the parameters are stochastic. Now each new shot of the quantum-sensing protocol receives a new stochastic vector $\bm{\theta}_L$. \textbf{c)} Circuit diagram for the unentangled sensor, which leads to a qubit distribution that is spread out over the Bloch sphere due to the stochasticity of the parameters. \textbf{d)} Circuit diagram for the entangled sensor (see Appendix~\ref{s: corrgauss} for more details). By preparing an appropriate entangled state (here the Bell state), the entangled sensor is able to measure directly along the low-noise axis $\theta_1-\theta_2$. The decoding unitary disentangles the two qubits and maps the information of the class to the state of the first qubit, which is measured. The second qubit is always in the ground state $\ket{0}$, and hence does not need to be measured. The delocalization of the first qubit state is then suppressed by the ability of the entangled state to avoid noise along the large noise axis $\theta_1+\theta_2$. \textbf{e)} Classification accuracy performance for the schemes in c) and d). The entangled sensor outperforms the unentangled sensor, achieving a higher classification accuracy for a given number of shots. Insets show the histogram of the MLE prediction, for 50 samples of the quantum sensor. Values less than $0.5$ are predicted to be in Class A. Values greater than $0.5$ are predicted to be in Class B.}
    \label{Fig:Main2}
\end{figure}

\subsection{Theory}

We consider the example of discriminating between two bivariate-Gaussian distributions $\mathcal{P}_{L}(\bm{\theta})$ of random parameters $\bm{\theta} = (\theta_1,\theta_2)$ for $L= \rm A, \rm B$, characterized by their mean values $\bm{\bar{\theta}}_{L}=(\bar{\theta}_{1, L},\bar{\theta}_{2, L})$ and covariance matrices $\mathbf{V}_{L}$. The distributions we consider possess identical covariance matrices, $\mathbf{V}_{\rm A} = \mathbf{V}_{\rm B}$, and are only distinguished by their mean values, which we set to have an identical sum $\bar{\theta}_{1, L} + \bar{\theta}_{2, L}=0~\forall~L$, but a distinct difference $\bar{\theta}_{1, L} - \bar{\theta}_{2, L} = \pm C$ for $L=\rm A, \rm B$ respectively. The identical covariance matrices of the distributions are non-diagonal, $\mathbf{V}_{\rm A} = \mathbf{V}_{\rm B} = \begin{psmallmatrix} \sigma^2 & \sigma_{\rm corr}^2 \\ \sigma_{\rm corr}^2 & \sigma^2 \end{psmallmatrix}$, indicative of non-zero correlations between the random parameters $\theta_j$. Here $\sigma^2$ defines the local (marginal) variance of parameters $\theta_1$ and $\theta_2$, which we set to be equal. $\sigma_{\rm corr}^2$ denotes the magnitude of cross-correlations between the random parameters $\theta_1,\theta_2$: for $\sigma_{\rm corr}^2 = 0$ the parameters are completely uncorrelated, while for $\sigma_{\rm corr}^2 = \sigma^2$ they are perfectly correlated. Due to these correlations, the combination $\theta_1 - \theta_2$ has a reduced variance $\sigma_-^2 = \sigma^2 - \sigma_{\rm corr}^2$, while $\theta_1 + \theta_2$ has an increased variance $\sigma_+^2 = \sigma^2 + \sigma_{\rm corr}^2$.

The task is to determine which of the two underlying Gaussian distributions the quantum sensor is sampling from, given $\NS$ stochastic samples of $\bm{\theta}$. As the only distinction between the distributions lies in the random parameter $\theta_1-\theta_2$, we are equivalently tasked with constructing an estimator for the mean of this quantity. These two parameters $\theta_1$ and $\theta_2$ are encoded in the sensing unitary as rotations along the Pauli-Z axis of two different qubits, $\usense({\bm{\theta}}) = e^{-\frac{i}{2}(\theta_1 \hat{\sigma}^z_1 + \theta_2 \hat{\sigma}^z_2)}$. Using $\NS$ samples extracted using computational-basis measurements, outcomes $x_k \in \{x_{11},x_{10},x_{01},x_{00}\}$ are unbiased estimators of the probabilities of measuring bit-strings corresponding to computational-basis states $\ket{11},\ket{10},\ket{01},\ket{00}$ respectively. The probabilities of these bit-strings will be different for the two classes. From this, we can estimate the performance of a quantum-sensing protocol as a function of the number of samples $\NS$ (see Fig.~\ref{Fig:Main2} b)). 

The standard unentangled sensing strategy would be to consider two independent single-qubit sensors, to estimate $\bar{\theta}_j$, as depicted in Fig~\ref{Fig:Main2} c). In this case, $\upr = \otimes_{j=1,2}~{\rm H_j}$ where ${\rm H_j}$ is the Hadamard gate on the $j^{\rm th}$ qubit, and $\udec = R^z_1(\pi/2)\otimes_{j=1,2}~{\rm H_j}$, where $R^z_1(\pi/2)$ performs a rotation of $\pi/2$ around the Pauli-Z axis on the first qubit (see Appendix~\ref{s: corrgauss} for more details). The simplest post-processing step is to only use local measurements on the $j^{\rm th}$ qubit, neglecting any correlations. The estimators for these single qubit excitation probabilities simply become $x_1 \equiv x_{11} + x_{10}$ and $x_2 \equiv x_{11} + x_{01}$. It will now prove useful to define a linear estimator for this unentangled protocol as $y^{u, \rm 1q} = x_1-x_2$, which, in the limit of $|\bar{\theta}_j| \ll 1$, can be shown to have the expected value $\mathbb{E}[y^{u, \rm 1q}] = \frac{1}{2}e^{-\frac{1}{2}\sigma^2}(\bar{\theta}_1-\bar{\theta}_2)$; as a result, $y^{u, \rm 1q}$ estimates the difference of means we intend to sense. For classification, we are ultimately interested in the discriminator between classes, $y^{u,\rm 1q}_{\rm A}-y^{u,\rm 1q}_{\rm B}$. In the limit of large $S$, the qubit measurement results approximate Gaussian distributions, whose distinguishability can be quantified using Fisher's discriminant~\cite{ripley_statistical_1996}, which is just the signal-to-noise ratio of the discriminator between classes:
\begin{align}
    D({\rm n}) = \frac{(\mathbb{E}[y_{\rm A}^{\rm n}]-\mathbb{E}[y_{\rm B}^{\rm n}])^2}{ \frac{1}{2}({\rm Var}[y_{\rm A}^{\rm n}] + {\rm Var}[y_{\rm B}^{\rm n}]) },
    \label{eq:snr}
\end{align}
where `${\rm n}$' is a label that defines different quantum-sensing protocols. We will use Fisher's discriminant as a useful analytic expression to compare the various protocols. However, we emphasize that in actual numerical calculations of classification accuracy, we are not restricted to large $S$ or to the use of Fisher's discriminant. For the unentangled protocol with only single-qubit measurements, $D({u\rm,1q})$ takes the form $D({u\rm,1q})\simeq 4\NS~e^{-\sigma^2} C^2$. We therefore see that the separability is strongly limited by the marginal variance $\sigma^2$.

However, this analysis overlooks any information present in the correlations between qubit measurements in each sample. Such correlations manifest even in the absence of entanglement, purely from the correlations of the stochastic parameters themselves, dictated by $\mathcal{P}_{L}(\bm{\theta})$. A better estimator can be constructed by considering all four bit-string outcomes, which can account for all qubit measurement correlations. We show that in the limit of large variance $\sigma^2 \gtrsim 1$ and $|C|\ll 1$, the estimator $y^{u\rm,2q} = x_{11}+x_{00}$ has the expected value $\mathbb{E}[y^{u\rm,2q}] \simeq \frac{1}{2} + \frac{1}{4}e^{-\sigma_-^2} (\bar{\theta}_1-\bar{\theta}_2)$, estimating the difference of means (see Appendix~\ref{s: corrgauss}). We consider the case of $|C|\ll 1$ and equal covariance matrices of the two distributions ($\mathbf{V}_{\rm A} = \mathbf{V}_{\rm B}$) only to provide simple analytic expressions for the discussion here. Importantly, the performance improvement enabled by the entangled protocol due to its ability to directly measure the useful combination $\theta_1-\theta_2$ is also valid for larger $C$ (see Appendix~\ref{s: corrgauss}). The corresponding value of Fisher's discriminant becomes $D({u\rm,2q}) \simeq \NS~e^{-2\sigma_-^2} C^2$. We note that the sensitivity is now determined by the reduced variance $\sigma_-^2$. For strong enough correlations where $\sigma_-^2 \ll \sigma^2$, this presents a significant improvement over the prior case ignoring two-qubit correlations. To visualize discrimination using two-qubit measurements, we consider an instance of this task for $\sigma = 1.5, \sigma_{\rm corr}^2 = 0.99\sigma^2$, and $C = -0.25$; samples from the two distributions defined by these parameters are shown in the scatter plot in Fig~\ref{Fig:Main2} b). Then, we perform simulations of the unentangled protocol for a finite $\NS=50$ shots for both classes, and plot histograms of the estimator $y^{u\rm,2q}$ in Fig.~\ref{Fig:Main2} e). Also shown are the Gaussian profiles determined by analytic expressions for $\mathbb{E}[y^{u\rm,2q}]$ and ${\rm Var}[y^{u\rm,2q}]$.

We now show how entanglement can be used to take advantage of classical correlations among the stochastic parameters. The key idea is to use an encoding protocol $\upr$ that can generate entanglement to prepare a state that is sensitive only to the low-noise combination $\theta_1 - \theta_2$. For this task the entangling protocol requires preparing the Bell state $\upr\ket{\psi_0} = \ket{\Psi^+}$, where $\ket{\Psi^+} = \frac{1}{\sqrt{2}}(\ket{01} + \ket{10})$, and also measuring in this basis, as depicted in Fig~\ref{Fig:Main2} d). The decoding unitary disentangles the two qubits, with the predicted class label information encoded in the first qubit, while the second qubit is always in the ground state. In this case, again assuming $|C| \ll 1$, but now without any constraints on $\sigma^2$ (to which the entangled protocol is completely insensitive), we find that the single-qubit excitation probability $y_e = x_1$ is an estimator of the desired difference of means, $\mathbb{E}[y^e] = \frac{1}{2} + \frac{1}{2}e^{-\sigma_-^2} (\bar{\theta}_1-\bar{\theta}_2)$. More importantly, Fisher's discriminant now becomes $D({e}) \simeq 4\NS~e^{-2\sigma_-^2} C^2$. The entangled protocol is therefore affected only by the small variance $\sigma_-^2$, and not the larger marginal variance $\sigma^2$ of the individual phases $\theta_1, \theta_2$, or the variance $\sigma_+^2$ of the high-noise combination $\theta_1 + \theta_2$. Furthermore, by requiring only measurements of a single qubit, the entangled protocol provides an enhancement in sensitivity to the difference in mean values of the two distributions by a factor of $4$ in comparison to the unentangled protocol that accounts for two-qubit correlations in measurement. For comparison with the unentangled case, in Fig.~\ref{Fig:Main2} e) we now plot histograms of the estimator $y_e$ sampled using $\NS=50$ shots for the two classes now from the entangled sensor, together with fits to Gaussian profiles determined by the aforementioned analytic expressions. We find an enhanced separation at the same number of shots $\NS$.

\subsection{Numerical Results}

We simulate both the unentangled and the entangled protocols for this discrimination task and compute the classification accuracy as a function of $\NS$ samples obtained from the quantum sensor. The predicted class label is determined using a maximum likelihood estimator constructed from all normalized bit-string frequencies $\{x_{11},x_{10},x_{01},x_{00}\}$, which account for all correlations in qubit measurements. We observe an increase in classification using the entangled protocol over the unentangled protocol, finding an approximately $4 \times$ reduction in the number of samples $\NS$ required by the entangled scheme to reach a desired accuracy, consistent with the analytic findings~(for further details of the various protocols, simulations, and post-processing, see Appendix~\ref{s: corrgauss}). We also see a large gap in classification accuracy between the two protocols in the low shot regime. The classification accuracy at $\NS=50$ for the entangled quantum-sensing protocol is $97\%$, compared to the unentangled quantum-sensing protocol, which only achieves $80\%$.

This toy example illustrates how entanglement can provide an advantage for sensing of correlated stochastic parameters. Using only local measurements, the discrimination is limited by the local variance $\sigma^2$. An improvement is enabled by considering two-qubit correlations, leading to discrimination limited by the smaller variance $\sigma_-^2$, but which requires the estimation of two-qubit correlations that have a higher sample complexity. The entangled protocol circumvents both these issues, by directly sensing the low noise combination of correlated parameters. This intuition also naturally provides hints as to how the entanglement advantage could be maximized. First, for large enough $\sigma^2$, local measurements carry no information about the class label, requiring the computation of more sample-expensive correlations. Secondly, if the task demanded sensing $N$ correlated parameters (instead of 2), the required correlation function order could also increase, making the task harder for the unentangled protocol. We formalize this intuition in the next section to show that the entangled quantum-sensing protocol can exhibit an exponential quantum advantage in the number of sensing parameters $N$ over its unentangled quantum-sensing protocol counterpart for suitable tasks with stochastic parameters.

\section{Exponential stochastic sensing advantage} 
\label{sec:exp}

\subsection{Introduction}

The example in the previous section illustrates how using an entangled quantum-sensing protocol can yield an advantage over an unentangled one for the specific case of a two-qubit system sensing two correlated phases. In this section, we show that the sensing advantage can scale exponentially in the number of qubits in the protocol. As an illustration, we consider a setting in which the phases of $\usense$ are generated at random from a uniform distribution such that the sum of the phases $\theta_i$ equals some fixed unknown quantity $\sum_i \theta_i = C$. We consider binary-classification tasks in which we must distinguish between two values of $C$, or estimation tasks to estimate $C$ within some error (see Fig.~\ref{Fig:Main3} a)). We show how an entangled quantum-sensing protocol with the GHZ state can achieve an exponential sensing advantage over the optimal unentangled quantum-sensing protocol. We will consider the estimation task in detail (the details of the binary-classification task are similar and can be found in Appendix~\ref{s: exponential advantage}).

\subsection{Theory}

\begin{figure}[H]
    \centering
    \includegraphics[width=0.8\linewidth]{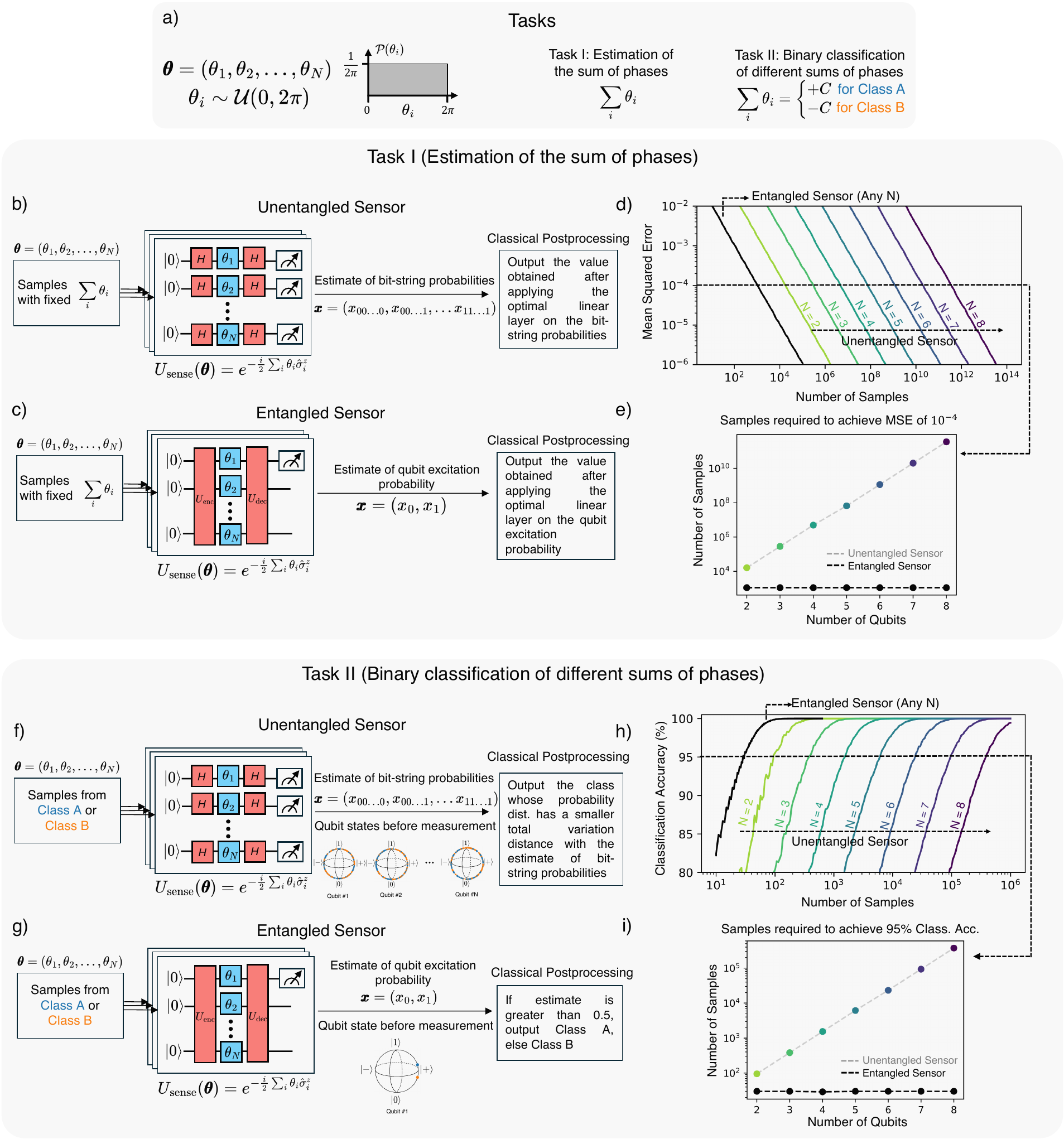}
    \caption{\textbf{Exponential advantage for estimation and classification tasks involving sensing $N$ correlated stochastic parameters with $N$ qubits.} \textbf{a)} Random phases are sampled an $N$-parameter probability distribution whose marginal distributions are uniform over a full period $\mathcal{U}(0,2\pi)$, but which has a fixed value of $\sum_i\theta_i$. We consider an estimation task where the goal is to estimate this sum (which is assumed to be close to $0$) and a binary classification task where the two classes to be distinguished have different values $\pm C$ of this constraint, where $C=0.3$ for the simulation results. \textbf{b)} Optimal sensing protocol using an unentangled $N$-qubit quantum sensor for the estimation task. Measurement outcomes $\{x_k\}$ computed using $\NS$ shots estimate bit-string probabilities of the qubits. The estimate of $C$ is obtained by applying a linear layer on these results $\{x_k\}$. \textbf{c)} Optimal sensing protocol using an entangled $N$-qubit GHZ-state. The measurement outcomes is the estimate of qubit excitation probability of a single qubit; which is then used to obtain the estimate of the sum of phases (see Appendix~\ref{s: exponential advantage}). \textbf{d)} Mean squared error (MSE) of the estimate to the true value as a function of the number of samples of the quantum-sensing protocol. The performance of the entangled-sensing protocol is independent of $N$. \textbf{e)} Samples required to achieve an MSE of $10^{-4}$, for the entangled and unentangled quantum-sensing protocols. The sample requirement scales exponentially in $N$ for the unentangled quantum-sensing protocol. \textbf{f)} Optimal sensing protocol using an unentangled $N$-qubit quantum sensor for the binary classification task. The qubit states are spread out entirely over the Bloch spheres prior to measurement. \textbf{g)} Optimal sensing protocol using an entangled $N$-qubit GHZ-state. Like before, all the information is encoded in a single qubit. Qubit states before measurement now show no spreading over the Bloch sphere for different random samples. \textbf{h)} Classification accuracy as a function of the number of samples of the quantum-sensing protocol for different number of qubits. The performance of the entangled-sensing protocol is independent of $N$. \textbf{i)} Samples required to achieve a classification accuracy of $95\%$, for the entangled and unentangled quantum-sensing protocols. The sample requirement scales exponentially in $N$ for the unentangled quantum-sensing protocol.}
    \label{Fig:Main3}
\end{figure}

Consider a protocol that uses entanglement to prepare the GHZ-state $\upr \ket{\psi_0} = \ket{\psi_{\rm probe}} = \frac{1}{\sqrt{2}}(\ket{00\dots0} + \ket{11\dots1})$. The GHZ-state performs a Ramsey measurement on the sum of phases experienced by all qubits, $\sum_i \theta_i = C$, and therefore is only sensitive to the value of the constraint. The state after sensing is $\usense\ket{\psi_{\rm probe}} = \frac{1}{\sqrt{2}}(e^{-iC/2}\ket{00\dots0} + e^{iC/2}\ket{11\dots1})$ for every sample of parameters from the underlying probability distribution. This state is measured in the GHZ-basis with a $\frac{\pi}{2}$ phase offset $\frac{1}{\sqrt{2}}(\ket{00\dots0} + i \ket{11\dots1})$ that maps the information of the state on to a single qubit. The phase offset increases the sensitivity of the protocol when $C$ is close to $0$. This is achieved with the decoding unitary $\udec$ that disentangles the $N$-qubits. $\udec$ is the Hermitian conjugate of the entangling unitary $\upr$, followed by a Hadamard operation on the first qubit. The probability of this qubit being in the excited state is $\frac{1 + \sin{C}}{2}$. All other $N-1$ qubits end up in the ground state, irrespective of the input, and therefore do not need to be measured (see Fig.~\ref{Fig:Main3} c)). Consequently, we can estimate the value of $C$ with any level of confidence within a number of shots that is independent of the number of qubits $N$, as shown in Fig.~\ref{Fig:Main3} d). The entangled protocol is completely insensitive to the stochastic nature of each individual phase $\theta_i$. We show the detailed circuit diagram in Appendix~\ref{s: exponential advantage}.

For the unentangled protocol, a Ramsey sequence is performed on each qubit, similar to the protocol described in the previous section (see Fig.~\ref{Fig:Main3} b)). To achieve the optimal sensitivity, the protocol includes a phase offset on any one qubit for when the number of qubits is odd (this subtlety is discussed in Appendix~\ref{s: exponential advantage}, but is not relevant for the analysis of the sample complexity with $N$). Since each phase $\theta_i$ is uniformly sampled from the entire domain of $0$ to $2\pi$, the average qubit excitation probability for any qubit is $0.5$, independent of the value of $C$. The only method to infer the total sum of phases is to consider correlations in the measurement outcomes across multiple qubits. However, any correlation among a subset of qubits will average out to be independent of the value of $\sum_i \theta_i = C$. For example, the $(N-1)^{\rm th}$-order correlation among the first $N-1$ qubits will only be sensitive to linear combinations of the first $N-1$ phases, such as $\sum_i^{N-1} \theta_i$. Due to the uniform nature of the individual qubit distributions, these linear combinations of phases also have uniform distributions over the entire domain of 0 to $2\pi$ (modulo $2\pi$). Therefore, the only information about $C$ is in the $N^{\rm th}$-order correlation of all phases, which requires a sample complexity that scales exponentially in $N$ to estimate within any small error. A similar analysis also applies for the binary classification task, where the goal is to distinguish among two probability distributions with different values for $C$. The details of this are presented in Appendix.~\ref{s: exponential advantage}. In Section \ref{s: general conditions for exponential advantage}, we further use Proposition 1 to prove that no protocol using an unentangled probe state and local measurements can discriminate between different values of $C$ in an amount of shots that is polynomial in the number of qubits. Even more strongly, for this example with completely uniformly-random sources of noise, we can use Theorem D.3.2 in Appendix~\ref{s: theorems and proofs} to show that an unentangled protocol cannot perform such a discrimination efficiently even if allowed to perform measurements in arbitrary bases. 

However, we note that our analysis is in the context of assuming a unitary model under fixed sensing-time assumptions, where the sensed parameters originate from a distribution. For scenarios where the protocol can optimize over the sensing-time, the exponential advantage in sample-complexity can reduce to a polynomial Heisenberg advantage. We discuss this further in Appendix~\ref{s: general framework proofs: continuous time}.

\subsection{Numerical Results}

For these numerical simulations, we apply a trained linear layer that takes in the estimate of probabilities of all possible bit-strings to obtain an estimate of the sum of phases (see Appendix~\ref{s: exponential advantage} for more details). By considering the statistics of all $2^N$ bit-string outcomes, this method can take into account all order of correlations in the measurement outcomes. We repeat this simulation multiple times as a function of the number of shots of the quantum-sensing protocol to obtain the mean squared error of the estimate from the true value. We train the linear layer with a sufficiently large dataset size so that any errors in the values of the linear layer are much smaller than the error of the estimate. The results of the simulations are shown in Fig.~\ref{Fig:Main3} d). In Fig.~\ref{Fig:Main3} e), we compare the number of samples required to achieve a mean squared error of $10^{-4}$, as a function of the number of qubits. While the GHZ-sensing protocol has a sample requirement independent of the number of qubits, the unentangled protocol has a scaling exponential in the number of qubits.

The unentangled and entangled quantum-sensing protocols for the task of binary classification are similar, and are illustrated in Fig.~\ref{Fig:Main3} f) and g). The two classes, A and B, have different values of the total sum of phases $\pm C$ (we set $C=0.3$ for the simulation results presented in Fig.~\ref{Fig:Main3}). For this task, we use the total variation distance to predict the class label for a given number of samples. We do this by first computing the probability distribution of all bit-strings for each class with a sufficiently large dataset (see Appendix~\ref{s: exponential advantage} for more details). We then generate a set of samples, from which we estimate the bit-string probabilities. The total variational distance is computed between this estimate and the probability distribution for each class. The predicted class is the one with the lower value. We repeat this simulation as a function of the number of qubits (see Fig.~\ref{Fig:Main3} h)). In Fig.~\ref{Fig:Main3} i), we plot the samples needed to achieve $95\%$ accuracy, which illustrates the exponentially increasing sample requirement for the unentangled protocol. This is in contrast with the GHZ-protocol, which requires the same number of samples for any number of qubits.

This task and quantum-sensing protocol can be extended to the case where the constraint is not determined by the total sum of the individual phases $\theta_i$. We can consider a linear combination $\sum_i \alpha_i \theta_i = \pm C$, where $\pm C$ is the constraint to be satisfied, and $\alpha_i \in [-1,+1]$. In this case, the task is, for example, the constraint $\sum_i \alpha_i \theta_i = +C$ for Class A and  $\sum_i \alpha_i \theta_i = -C$ for Class B. The optimal entangled and unentangled quantum-sensing protocol can be determined by considering the following argument that allows us to map this problem to the previous classification task. If we `invert' the labeling of the qubit states (that is, relabeling $\ket{1}$ as a $\ket{0}$, and vice versa), then the effect of a phase $-\theta_i$ in the original basis transforms to a phase $\theta_i$ in the new basis. This can be achieved in the sensing protocol by sandwiching the sensing unitary by $\pi$ pulse gates before and after the sensing unitary for the relevant qubits. If we perform this operation for the qubits for which $\alpha_i = -1$, then we recover the original task of the constraint being the sum of phases $\theta_i$.

The exponential sensing advantage can persist when the constraint is not strictly satisfied. For example, consider the task where each phase is uniformly distributed, $\theta_i \sim \mathcal{U}(0,2\pi)$, but the sum satisfies a fixed value up to some fluctuations. For example, consider this to be a normal distribution: $\sum_i \theta_i \sim \mathcal{N}(\mu, \sigma^2)$, and that this distribution is different for the two classes. As long as the mean and variance do not scale with the number of qubits $N$, then these two distributions can still be separated within a number of shots independent of $N$ with an entangled-sensing protocol. We generally hope to expect an exponential sensing advantage when the information distinguishing the two distributions depend on some global property of the sensed parameters. For instance, if the two distributions have different marginal variance of a single qubit, there would not be a scaling advantage in $N$. We consider these more general scenarios in Appendix~\ref{sec:theory}, where we discuss the properties of the probability distributions which enable a exponential advantage in sample complexity.

The same advantage can be observed when a single qubit is used to sequentially sense the $N$ stochastic parameters over time, instead of considering $N$ phases on $N$ qubits. In this case, we define the SQL protocol limited to an incoherent qubit. This protocol is confined to sensing each phase separately. It can be shown that the bit-string probability distributions, obtained from the $N$ measurements, has the same distribution as $N$ unentangled qubits, each experiencing a single phase $\theta_i$. On the other hand, the quantum-enhanced protocol involves a qubit that can coherently sense multiple phase. In this case, the optimal protocol is the standard Ramsey protocol that coherently integrates all the phases, and hence experiences the total sum of the phases $\sum_i \theta_i$. It can therefore be shown that the performance of this protocol directly maps onto the case of the GHZ-protocol, where the $i^{\rm th}$ qubit experiences $\theta_i$. In summary, this provides an exponential sensing advantage with a single qubit. This is further discussed in Appendix~\ref{s: exponential advantage}.

\section{Application: Sensing conserved quantities} 
\label{sec:xxz}

\subsection{Introduction}

\begin{figure}[h!]
    \centering
    \includegraphics[width=0.75\linewidth]{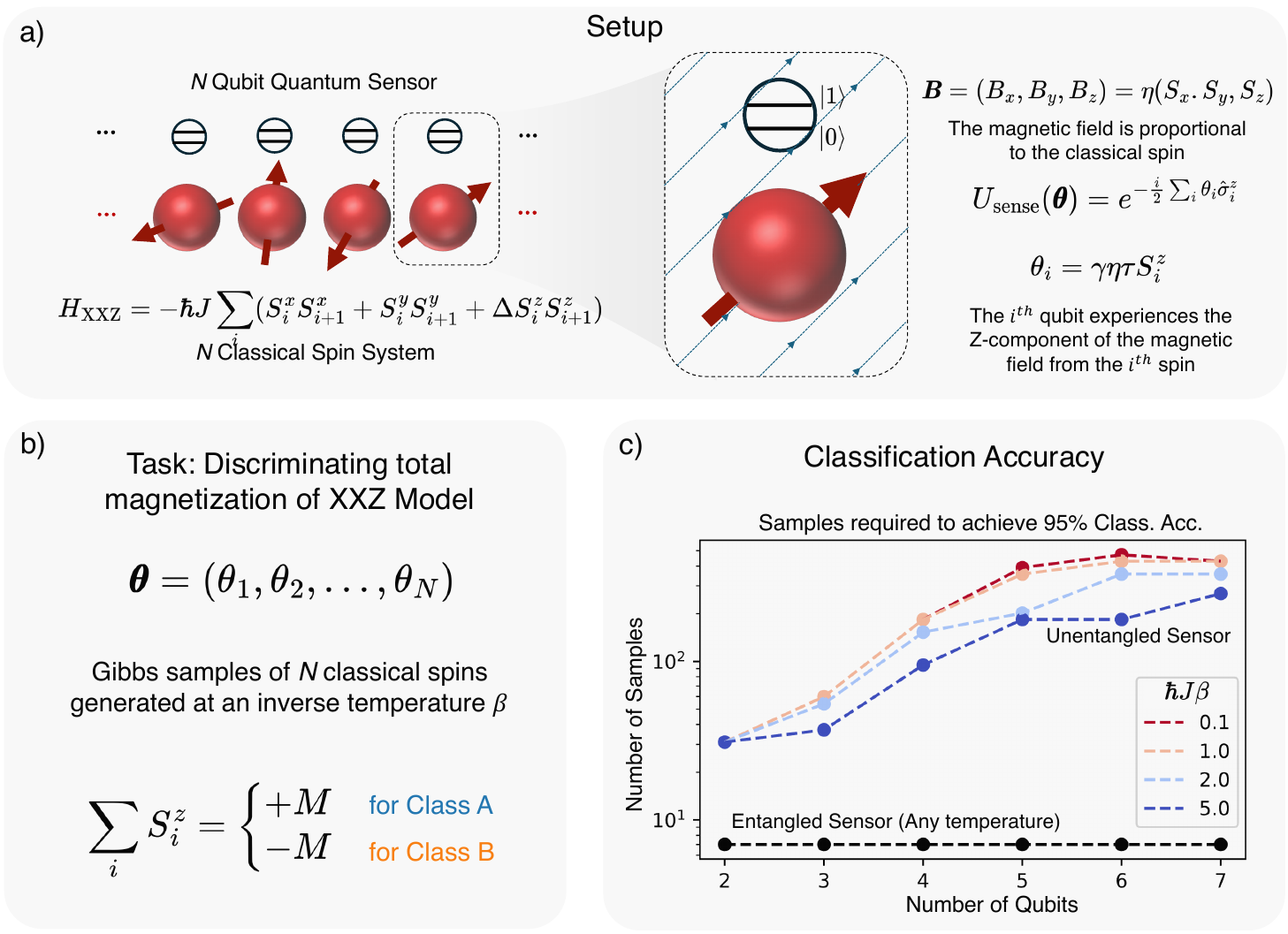}
    \caption{\textbf{Discriminating two distributions with different total magnetization of classically interacting spins} \textbf{a)} An $N$-qubit quantum sensor senses the magnetic field of a system of $N$ classical spins at temperature $T$, interacting via the XXZ Hamiltonian. Qubit $i$ of the quantum sensor is only sensitive to the local Z component of the magnetic field, which in turn is proportional to the $S_i^z$ of the $i^{\rm th}$ spin. The XXZ interaction conserves the total spin component in the Z direction. \textbf{b)} The classification task is then to distinguish between two values of the conserved quantity $\sum_i S_i^z = \pm M$. Classical spin system configurations follow the Boltzmann distribution at finite $T$, and are sampled using a Metropolis Monte Carlo algorithm. \textbf{c)} Number of samples required by the unentangled and entangled protocols to reach 95\% classification accuracy. Fluctuations in $S_i^z$ values among samples results in the unentangled sensor requiring a number of samples dependent on $N$ and the $T$. Increasing $T$ increases the amount of local fluctuations, making the task harder for the unentangled sensor. In contrast, the entangled sensor is insensitive to local fluctuations and requires a constant number of samples independent of $N$ and $T$. In these simulations, we set the value of anisotropy to be $\Delta = 0.75 $. The value of $\gamma \eta \tau $ sets the scaling between the $S^z_i$ values the phase $\theta_i$ experienced by the qubit. We set this value to be $\pi$, such that the total range of phase that can be experienced is between $-\pi$ and $\pi$ corresponding to the $S^z_i$ values ranging between $-1$ and $1$.}
    \label{Fig:Main4}
\end{figure}

The situation of stochastic parameters described in the previous subsection can arise in physical systems exhibiting global conservation laws, but whose local variables are free to fluctuate on a timescale much shorter than the quantum-sensing timescale. A specific example we consider is that of $N$ classical spins $\mathbf{S}_i = (S_i^x,S_i^y,S_i^z) \in \mathbb{R}^3$, with $i \in \{1, \dotsm, N\}$, and $|\mathbf{S}_i| = 1$, on a one-dimensional chain with periodic boundary conditions, interacting via the \textit{classical} XXZ Hamiltonian:
\begin{align}
    H_{\rm XXZ} = - \hbar J \sum_i (S_i^xS_{i+1}^x + S_i^y S_{i+1}^y +\Delta  S_i^z S_{i+1}^z),
    \label{eq:hxxz}
\end{align}
with coupling parameter $J$ and the dimensionless anisotropy parameter $\Delta$~(see Fig.~\ref{Fig:Main4} a))~\cite{girvin2019modern, prosen_macroscopic_2013, das2020nonlinear}. While the above Hamiltonian does not conserve the local spin component in the Z direction, the total spin component of all qubits in the Z direction, $\sum_i S_i^z$, is a conserved quantity~(see Appendix~\ref{s: xxz}). As a result, spin dynamics generated by the XXZ Hamiltonian allow for the local magnetization $S_i^z$ of classical spins to fluctuate, but the total Z direction magnetization remains fixed. If the classical spin system is thermalized at a finite temperature $\beta = 1/\boltzmann T$, possible spin configurations form a Gibbs ensemble defined by the Boltzmann distribution $\mathcal{P}(\bm{S}) \propto e^{-\beta H_{\rm XXZ}(\bm{S})}$, where we have defined $\bm{S} = ((S_1^x, S_1^y, S_1^z),\ldots, (S_N^x, S_N^y, S_N^z))$ to represent the spin configuration.

\subsection{Theory}

We then envision an $N$-qubit quantum sensor placed to sense the local magnetic fields from this system of $N$ interacting classical spins, as depicted in Fig.~\ref{Fig:Main4} a). Each qubit of the quantum sensor is designated to sense the Z component of the local field of a single classical spin $B^z_i$, which is in turn proportional to $S^z_i, B^z_i = \eta S^z_i$. This can arise when the sensor is physically close to the corresponding spin, so that the magnetic field generated by the corresponding spin is dominant compared to the contributions from the rest of spins. The Hamiltonian associated with this magnetic field for qubit $i$ is $H_i = \frac{\hbar \gamma}{2} B^z_i \hat{\sigma}_i^z$, where $\gamma$ is the gyromagnetic ratio. The qubit experiences the magnetic field for a sensing integration time of $\tau$, which results in the sensing unitary $\usense(\bm{\theta}) = e^{-i (\sum_i H_i) \tau / \hbar} = e^{-\frac{i}{2} \sum_i \theta_i \hat{\sigma}_i^z}$, where

\begin{align}
    \theta_i = \gamma \eta \tau S^z_i.
\end{align}
For the simulations, we choose $\gamma \eta \tau = \pi$, so that $\theta_i \in [-\pi,\pi]~\forall~i$. A physical implementation of the quantum sensor can be SQUID-based sensors~\cite{fagaly_superconducting_2006}, which can be designed to only sense the Z component of the magnetic field. The task is to determine which of two distinct values of the conserved quantity $\sum_i S_i^z = \pm M$ describes the classical spin system. To analyze this stochastic sensing paradigm, we simulate a Gibbs ensemble of the classical spins at different temperature values $\beta$.

\subsection{Numerical Results}

The simulation is performed using a Metropolis algorithm, a standard Monte Carlo sampling scheme for classical spin systems~\cite{prosen_macroscopic_2013}, with a constrained transition rule that ensures conservation of the total Z component of the magnetization, as illustrated in Fig.~\ref{Fig:Main4} b). Then, each sample available to the quantum sensor is provided by a single $N$-spin configuration from this Gibbs ensemble. In these simulations, we set the anisotropy to the value $\Delta = 0.75$. In Fig.~\ref{Fig:Main4} c), we show the number of samples required to achieve 95\% classification accuracy for different problem sizes (number of classical spins $N$) and thermodynamic temperature $T$, using both unentangled and entangled-sensing protocols. We clearly see that the number of shots required by the unentangled protocol varies with $N$ and $T$. On the other hand, the optimal entangled-sensing protocol prepares a GHZ-state that is sensitive to $\sum_i S_i^z$ only. As a result, its sample requirement is independent of the temperature and system size. Note further that the unentangled protocol fares worse with increasing temperature $T$ of the classical spin system. $T$ determines the strength of local spin fluctuations; thus with increasing $T$ the local $S_i^z$ distributions become less-localized, rendering the discrimination of global properties over all $N$ spins increasingly more difficult. In contrast, the entangled-sensing protocol directly senses the conserved combination $\pm M$ and is therefore insensitive to the temperature $T$ of the classical spin system. Unlike the task in the previous section, we do not observe an exponential sample complexity in $N$. This is because the distribution of local spins is not uniform, even at infinite temperature. The distribution of $\theta_i$ asymptotically becomes the uniform distribution in the limit $\gamma \eta \tau \to \infty$. However, the entangled protocol still outperforms the unentangled sensor by requiring orders-of-magnitude fewer samples. This observation indicates that true exponential advantages in stochastic sensing of correlated signals might be hard to achieve in tasks closer in describing physical scenarios, but perhaps more importantly that even in such cases entanglement can be a useful resource.

\section{Theoretical framework for evaluating entanglement sensing advantage for binary classification of stochastic parameters}
\label{sec:theory}

\subsection{Introduction}
\label{s: introduction section vi}

Our analysis has shown that entangled quantum-sensing protocols can provide a substantial, even exponential, advantage in the sensing of stochastic parameters for classification and estimation tasks, over unentangled quantum-sensing protocols. However, we have shown this advantage for stochastic parameters that satisfy very specific linear constraints. It is therefore natural to ask: can general conditions be identified for which such an advantage exists? We address this question in this section. More precisely, suppose we wish to distinguish two distributions $\mathcal{P}_{\rm A}(\bm{\theta})$ and $\mathcal{P}_{\rm B}(\bm{\theta})$ using stochastic parameters $\bm{\theta}$ sampled from the distributions. \textit{Can we identify properties of these distributions such that an unentangled quantum-sensing protocol requires exponentially many shots in the number of parameters $N$ to distinguish them, while an entangled quantum-sensing protocol requires only polynomially many?} We refer to cases where this is true as enabling an exponential entanglement advantage for sensing stochastic parameters. To address this question, we first build a theoretical framework that quantifies how a quantum system senses properties of a probability distribution when sensing stochastic parameters. This allows us to identify the key feature of such distributions that enables an entangled probe state to be much more sensitive than an unentangled one. While our current results do not completely determine where these advantages occur, they do provide some sufficient conditions for their presence or absence. An additional note is that our framework relies on evaluating or bounding the total variational distance between distributions of measurements, which only determines the shot scaling up to a quadratic factor. While this is sufficient to determine exponential advantage and strongly separated polynomial scalings, it cannot determine the superiority of two protocols with shot scalings of similar order.

\subsection{Framework}
\label{s: framework}

We now provide a framework to analyze more general sensing tasks for binary classification of stochastic parameters. In this case, an optimal POVM can be defined that has only two possible outcomes, each corresponding to one of the two classes. The possible outcomes can be defined by the expectation of a single operator $\hat{O}$ that can be written using Eq.~(\ref{eq:pkInputAv}), the expression $\ket{\psi_{\rm f}(\bm{\theta})} = \udec\usense(\bm{\theta})\upr\ket{\psi_0}$, and some rearrangement:
\begin{align}
    \mathbb{E}[\hat{O}] = {\rm Tr}\Big( (\udec^{\dagger}{M}_0 \udec) \int d\bm{\theta}~\mathcal{P}_{\Phi}(\bm{\theta})\usense(\bm{\theta})\rhou_{\rm probe}\usense^{\dagger}(\bm{\theta}) \Big).
    \label{eq:p0}
\end{align}
\begin{align}
    \usense = \exp\{-i\hat{G}(\theta)\} = \exp\left\{-i\sum_j \theta_j \hat{G}_j\right\},\ \ [\hat{G}_j,\hat{G}_k]=0 \ \forall j,k
    \label{eq:usense general}
\end{align}
is the fully general structure of $\usense$ that our framework applies to, with the additional constraint that the eigenstates of $\usense$ are product states. Here we have introduced the probe state $\rhou_{\rm probe} = \ket{\psi_{\rm probe}}\bra{\psi_{\rm probe}} = \upr \ket{\psi_0}\bra{\psi_0}\upr^{\dagger}$. $\hat{O} \equiv (\udec^{\dagger} M_0 \udec)$ is then an effective POVM; if $M_0$ describes a projective measurement ($M_0^{\dagger} M_0 = M_0$), then $\hat{O}$ is also projective. If the expected values of the projectors of the two distributions are exponentially close together, then distinguishing the two distributions requires exponentially many shots (for more details see Appendix~\ref{s: general framework proofs}). Consequently, we can concern ourselves only with changes in the expected value of the projector between the two distributions of $\bm{\theta}$. The distribution of measurement outcomes $\mathbb{E}[\hat{O}]$ is determined by the action of $\usense(\bm{\theta})$ on $\rhou_{\rm probe}$; this dependence is most easily expressed by writing $\rhou_{\rm probe}$ in the qubit bit-string representation, $\rhou_{\rm probe} = \sum_{\bm{a},\bm{b}} \rho_{\bm{a},\bm{b}} | \bm{a} \rangle \langle \bm{b} |$ using the eigenstates $\ket{\bm{a}}$ of $\hat{G}(\bm{\theta})$. This means $e^{-i\hat{G}(\bm{\theta})}\ket{\bm{a}} = e^{-i\bm{q}(\bm{a})\cdot\bm{\theta}}\ket{\bm{a}}$ where $q:\mathbb{R}^N\rightarrow \mathbb{R}^N$ is a function mapping the bit-string of $\bm{a}$ to a vector whose entries are dependent on the eigenvalues that $\ket{\bm{a}}$ has for each $\hat{G}_j$. $\bm{a}$ denotes the vectorized version of the bit-string in $\ket{a}$, where the $j$-th entry of  $\bm{a}$ is equal to the $j$-th bit of $\ket{a}$. Since $\usense$ has an eigenbasis of product states independent of $\bm{\theta}$, we can always write our expression in this basis. Consequently, Eq.~(\ref{eq:p0}) can equivalently be written as:

\begin{align}
    \mathbb{E}[\hat{O}] &= \sum_{\bm{a},\bm{b}} {\rm Tr}\Big( \hat{O}~\rho_{\bm{a},\bm{b}} \int d\bm{\theta}~\mathcal{P}_{\Phi}(\bm{\theta}) e^{-i(\bm{q}(\bm{a})-\bm{q}(\bm{b}))\cdot \bm{\theta}}| \bm{a} \rangle \langle \bm{b} | \Big) \nonumber \\
    &\equiv \sum_{\bm{a},\bm{b}} O_{\bm{a},\bm{b}}~\rho_{\bm{a},\bm{b}}~\chi(\bm{q}(\bm{a})-\bm{q}(\bm{b})), 
\end{align}
where $O_{\bm{a},\bm{b}} = {\rm Tr}\Big\{ \hat{O}| \bm{a} \rangle \langle \bm{b} | \Big\}$ and $\chi(\bm{k})$ is the \textit{characteristic function}~\cite{lukacs_characteristic_1970} of the classical probability distribution $\mathcal{P}_{\Phi}(\bm{\theta})$, which is defined, as usual, as the Fourier transform of $\mathcal{P}_{\Phi}(\bm{\theta})$,

\begin{equation}
    \chi(\bm{k}) = \int d\bm{\theta}~\mathcal{P}_{\Phi}(\bm{\theta}) e^{-i\bm{k}\cdot \bm{\theta}}.
    \label{eq:characteristic function}
\end{equation}

\begin{figure}[H]
    \centering
    \includegraphics[width=0.8\linewidth]{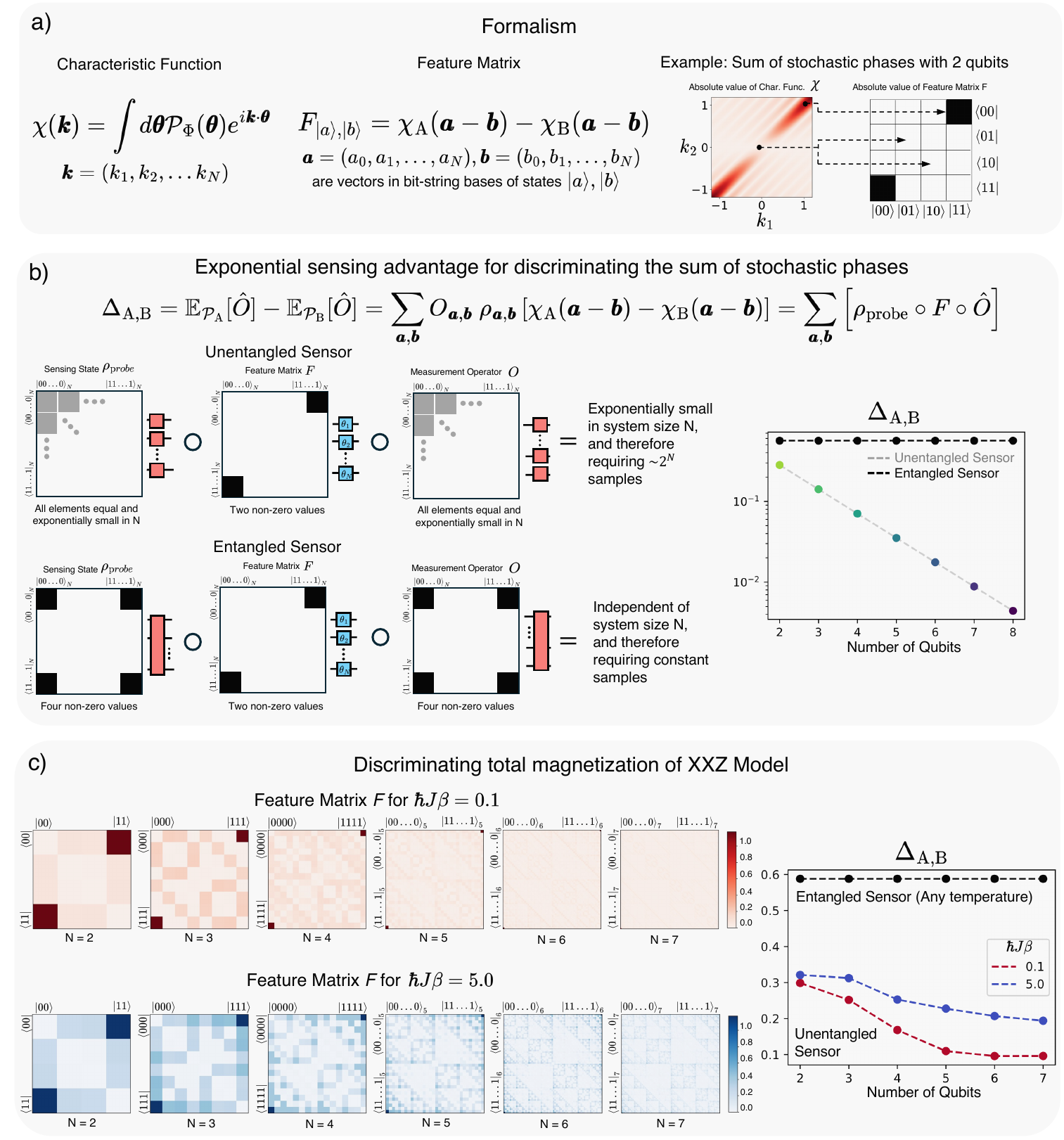}
    \caption{\textbf{Framework for sensing advantage of stochastic parameters} \textbf{a)} The feature matrix is constructed based on the difference of the characteristic functions and the choice of $\usense$. In particular, $\usense$ determines at what points we take the difference of the two characteristic functions. For our local choice of $\usense$ with $\hat{G}_j = 1/2 (\hat{\sigma}_j^z + \mathbb{\hat{I}})$ in Eq.~\ref{eq:usense general}, the point we evaluate at is simply given by the difference of the vector of bits from the pair of states $\ket{i},\ket{j}$ in the eigenbasis of $\usense$. Notably, many different pairs of bit-strings can produce the same $\bm{k}$ (for example, $2^N$ pairs of identical states produce an all-zero vector for $\bm{k}$), so $\usense$ can evaluate the same point in the characteristic functions many times. \textbf{b)} We can define the separation value as the difference in the expected value of a projector $\hat{O}$ between the two distributions of the parameters $\bm{\theta}$. For a given choice of $\hat{O}$ and $\rho_{\text{probe}}$, the number of shots required to distinguish the two distributions scales at least as $\sim 1/\Delta_{\rm A,\rm B}$. This separation value $\Delta_{\rm A,\rm B}$ is equivalent to the sum over the elements of the Schur product between the matrices of $\rho_{\text{probe}},F,\hat{O}$ written in the eigenbasis of $\usense$. Due to the limited structure of product state probes and unentangled measurement bases, sufficiently concentrated feature matrices cannot be efficiently distinguished by product states, as they can only weigh most entries exponentially little, and so produce an exponentially small separation value. Conversely, an entangled probe state and measurement basis can pick out specific entries to weigh substantially more than others, and this produces a non-vanishing separation value. For our example in the plot, we present the separation value vs. number of qubits and feature matrices for an entangled sensor (GHZ-state probe and measurement basis) and an unentangled sensor (Hadamard probe state - the optimal unentangled probe - and product basis). The task is the same as in Sec~\ref{sec:exp}. \textbf{c)} Plotting the feature matrices for XXZ tasks at different temperatures. The concentration of features occurs more strongly and quickly for the higher temperature case, though both eventually plateau due a bounded amount of noise possible with a fixed temperature. On the right, we display the best possible separation value achievable using entangled and unentangled probe states for different temperatures.}
    \label{Fig:Main5}
\end{figure}

Since we are primarily interested in exploring the advantages of entanglement from the probe state, we will focus on sensing interactions that are not entangling. In particular, we consider the case of single-qubit sensing, specified by the generator of the sensing unitary $\hat{G}(\bm{\theta}) = \sum_j \theta_j \hat{G}_j$ given by Eq.~(\ref{eq:usense}) with $\hat{G}_j = (1/2)(\hat{\sigma}_j^z + \mathbb{\hat{I}})$ (the additional factor $\propto \mathbb{\hat{I}}$ is introduced for convenience). For this choice of $\usense$, we have $\bm{q}(\bm{a}) = \bm{a}$.  We note that many of our results hold for more general sensing unitaries, as will be seen later.

To study the conditions that produce an exponential advantage, we introduce the separation value $\Delta_{\rm A,\rm B}$:

\begin{align}
    \Delta_{\rm A,\rm B} = \mathbb{E}_{\mathcal{P}_{\rm A}}[\hat{O}]-\mathbb{E}_{\mathcal{P}_{\rm B}}[\hat{O}] = \sum_{\bm{a},\bm{b}} O_{\bm{a},\bm{b}}~\rho_{\bm{a},\bm{b}}\left[ \chi_{\rm A}(\bm{a}-\bm{b}) -  \chi_{\rm B}(\bm{a}-\bm{b}) \right] = \sum_{\bm{a},\bm{b}}\left[\rho_{\text{probe}} \circ F \circ \hat{O} \right].
\label{e: separation value}
\end{align}

Here, ``$\circ$" denotes the Schur product (i.e. element-wise product) between the matrices, and $F$ denotes the ``characteristic feature" matrix, defined as: 
\begin{align}
    F_{\ket{a},\ket{b}} = \chi_{\rm A}(\bm{a}-\bm{b}) - \chi_{\rm B}(\bm{a}-\bm{b}).
\label{e: feature matrix}
\end{align}

Note that the elements of the feature matrix depend only on the characteristic functions of the underlying problem distributions, $\mathcal{P}_{\rm A}(\bm{\theta}),\mathcal{P}_{\rm B}(\bm{\theta})$, and on our choice of $\usense$. The latter determines which point in the Fourier transform a particular entry in the feature matrix corresponds to (for our local choice of $\usense = e^{-\frac{i}{2}\sum_i \theta_i \hat{\sigma}^z_i}$, it is simply the vectorization of the state bit-strings as mentioned earlier). We schematically represent the construction of the feature matrix in Fig.~\ref{Fig:Main5} a), for the task described in Sec.~\ref{sec:exp} for two qubits, and work out additional examples in Appendix~\ref{s: feature matrix examples}. 

\subsection{General conditions for exponential advantage}
\label{s: general conditions for exponential advantage}
The distribution of feature matrix element magnitudes determines whether there is an exponential advantage for an entangled probe state over an unentangled probe state. A feature matrix with a concentrated number of non-vanishing entries can only lead to a non-vanishing separation value if the corresponding probe state and observable also have concentrated weights at those entries. That is, the combined weights of the probe state and observable must pick out the ``informative" entries of the feature matrix. While this is possible for an entangled probe state, a product state necessarily weighs almost all possible entries in the matrix by an exponentially small amount. The latter can lead to an exponentially vanishing separation value, if the useful entries of the feature vector are too concentrated, as the product state lacks enough structure to capture them. The separation value can suffer even more if the measurement basis is chosen to be strictly local as well. 

The concentration of feature matrix elements is a function of both the underlying distributions ($\mathcal{P}_{\rm A}(\bm{\theta}) , \mathcal{P}_{\rm B}(\bm{\theta})$) and the points in the characteristic function difference that a given $\usense$ evaluates. A guaranteed case of exponential sensing advantage could arise because the difference of characteristic functions itself is exponentially concentrated over its domain and $\usense$ samples a small, nonzero number of points at these concentrations, as seen in Fig.~\ref{Fig:Main5} b). On the other hand, an exponentially concentrated difference of characteristic functions could fail to yield a case of exponential advantage because $\usense$ only evaluates the transform at vanishing points. In this case, both an entangled and unentangled probe state would require exponentially many shots. Alternatively, $\usense$ could evaluate the difference mostly at the non-vanishing points. In this case, both the entangled and unentangled probe state would require only polynomially many shots for the discrimination. Finally, a fairly unconcentrated and undulating difference of characteristic functions could produce a case of exponential advantage because $\usense$ happens to sample the transform almost exclusively at some vanishing points, with a precious few selected with significant magnitude. 

Consequently, much of the question of advantage must be decided in the application setting, for a particular distribution and $\usense$. The general propositions that we present for guaranteed exponential sensing advantage, or its absence, are therefore framed in terms of the distribution of values in the characteristic feature matrix $F$, rather than e.g. the probability distributions $\mathcal{P}_{\rm A,\rm B}(\bm{\theta})$ themselves. While many of the propositions assume a single-qubit, possibly scaled choice of $\hat{G}_j$, the formalism of Eqs.~(\ref{e: separation value}) and (\ref{e: feature matrix}) holds for any choice of commuting $\hat{G}_j$. The formal statements and proofs of our main propositions, as well as additional theorems and corollaries, are presented in Appendix~\ref{s: general framework proofs}.

The first situation we must consider is one where each term in the feature matrix shrinks so quickly with system size that no choice of probe state distinguishes the distributions with polynomially many shots. Consequently, there is no exponential sensing advantage because both entangled and unentangled probes require exponentially many shots to distinguish the distributions. As an illustrative example, consider a problem in the form of linear constraints. The examples we have considered in the paper thus far, of linear constraints on stochastic parameters, are an example of this situation. In particular, we have 
\begin{equation}
    f(\theta_1,...,\theta_N) = \sum_{j=1}^N\alpha_j\theta_j = C, \alpha_j\in \mathbb{R},\ \ \ \ \ \mathcal{P}_C(\bm{\theta}) = \delta\left(\theta_N - \frac{1}{\alpha_N}(C-\sum_{j=1}^{N-1}\alpha_j\theta_j)\right)\prod_{j=1}^{N-1}p_j(\theta_j),
\label{e: linear constraint example}
\end{equation}
where $p_j$ represent independent distributions on each parameter $\theta_j \in \mathbb{R}$, each a uniform distribution from 0 to $2\pi$. In essence, we generate the first $N-1$ parameters independently of each other and of $C$, and then choose the remaining parameter $\theta_N$ to satisfy the linear constraint. We try to distinguish two distributions, one where the $\theta_j$ are generated to satisfy a constraint $C_1$, and another we're they're generated to satisfy a constraint $C_2$. Consider the specific case where $\alpha_1 = 2$ and $\alpha_{j>1}=1$. The magnitude of the difference between the distributions' characteristic functions is given by $|\chi_{C_1}(\bm{k})-\chi_{C_2}(\bm{k})| = \sin(k_N(C_1-C_2)/2)\sinc(\pi(k_1-2k_N))\prod_{j=2}^{N-1}\sinc(\pi(k_j-k_N))$. For our local choice of $\usense$, we are restricted to choosing $k_j\in \{-1,0,1\}$. $k_N$ must equal $-1$ or $1$ to prevent the first term from vanishing, but regardless of what value we choose for $k_1$, the first $\sinc$ term is evaluated at a non-zero integer multiple of $\pi$, and so we cannot avoid a zero in the product. Since all elements of the feature matrix go to zero, distinguishing the two distributions becomes impossible. Note that the difference in characteristic functions is still exponentially concentrated, with  significant non-vanishing points (for example, $k_N=1=k_{j\neq 1}$ and $k_1=2$). Nevertheless, there is still no exponential sensing advantage because our choice of $\usense$ does not evaluate this difference at non-vanishing points. We explore this example further in Appendix~\ref{s: multiple copies}, where we get an exponential advantage with a modified $\usense$ - one that is equivalent to having multiple copies of $\bm{\theta}$. This example highlights the importance of a $\usense$ that matches the problem in some way. In contrast, if there is at least one entry in the feature matrix that is polynomially small, a general probe state can be tailored to predominantly weigh that entry and its conjugate.

In order to isolate the effects of entanglement on sensing advantage, the following proposition specifically considers the case where the sensing unitary consists entirely of single qubit rotation gates, with one parameter per gate as in Eq.~\ref{eq:usense}. Specifically, we want to bound how well an unentangled probe state and measurement basis can do, given the properties of certain average values of the feature matrix. The proposition statement is in terms of averages since an unentangled probe state lacks the structure to exploit more fine-grained properties of the feature matrix. \\

\textit{\textbf{Proposition 1 (Theorem D.3.3 in Appendix \ref{s: theorems and proofs})-- Sufficient condition for hard tasks for product probe states and unentangled measurement bases:} Let $F_{\text{\rm RMS}}(d)$ be the root mean square magnitude of $|F_{a,b}|$ for states differing by $d$ bits. Defining $\epsilon = d/N$, if  $F_{\text{\rm RMS}}(d)  \lessapprox r^{-N}$ with $r>\sqrt{\frac{(1/\epsilon -1)^\epsilon}{1-\epsilon}}$  for every $d$, then exponentially many shots are required for the discrimination for protocols using unentangled probe states and unentangled measurement bases.} \\

The proof of this statement, presented fully in Appendix \ref{s: theorems and proofs}, uses the fact that a feature matrix with the above property has such a concentrated distribution of non-vanishing elements that any product probe state and local measurement basis must necessarily average over mostly very-quickly-vanishing terms. This leads to an exponentially vanishing separation between measurement outcomes for the possible distributions. Consider the linear example in Eq.~(\ref{e: linear constraint example}), with $\alpha_j=1~\forall~j$ and local $\usense$,  $p_j(\theta_j)$ all uniform distributions on interval $[\phi_0, \phi_0 +\phi_{\delta}]$.  The difference in characteristic functions is given by $|\chi^{C_1}(\bm{k})-\chi^{C_2}(\bm{k})| = |\sin(k_n(C_1-C_2)/2)\prod_{j=1}^{N-1}\sinc(\phi_{\delta}(k_n-k_j)/2)|$. As before, $k_j\in \{\pm 1,0\}$. Note that $k_n=\pm 1$, otherwise the value goes to zero. Consequently, for strings with Hamming distance $d$, $N-d$ values of $k_j$ must be set to zero, introducing a factor of $\sinc^{N-d}(\phi_\delta/2)$. The remaining set of terms is an average of the $2^{d-1}$ ways one can choose $k_j=\pm 1$, which introduces another factor of either $1$ or $|\sinc(\phi_{\delta})|$. Therefore the mean square of the terms is $(1/2^d)\sinc^{2(N-d)}(\phi_{\delta}/2)(1+\binom{N}{1}\sinc^2(\phi_{\delta})+\ldots) = \sinc^{2(N-d)}(\phi_{\delta}/2)(\frac{1+\sinc^2(\phi_{\delta}/2)}{2})^{d}$. The root mean square of the magnitude is therefore $\sinc^{N-d}(\phi_{\delta}/2)$ $\left(\frac{1+\sinc^2(\phi_{\delta}/2)}{2}\right)^{d/2}$. Solving for the choice of $\phi_{\delta}$ that would satisfy the inequality for every $d$, gives us a sufficient condition of $ \phi_{\delta} \geq 0.9\pi$.

Thus, if each uniform interval on the angles $\theta_{j\neq 1}$ has width at least $0.9\pi$, any choice of product probe state and any choice of unentangled measurement basis will require exponentially many shots for the discrimination task. On the other hand, since there exists one entangled state with polynomial scaling, we have exponential sensing advantage from entanglement. It is possible that for some tasks the elements of the feature matrix shrink exponentially with system size, only to taper off for a sufficiently large number of qubits as in Fig.~\ref{Fig:Main5} c). This would offer a practically large, confstant advantage not captured in the asymptotic guarantees of Proposition 1. So far, we have results on sufficiently concentrated distributions of feature matrix entries that guarantee that an unentangled probe state will require exponentially many shots for the distribution discrimination. A natural question to ask is whether there are conditions under which the non-vanishing entries of the feature matrix are spread out enough to guarantee that a product probe state will require only polynomially many shots. The following proposition addresses this, for any choice of $\usense$. \\

\textit{\textbf{Proposition 2 (Theorem D.4 in Appendix \ref{s: theorems and proofs})-- Sufficient condition for easy tasks for product probe states and arbitrary measurement bases:} If there are at least $\sim 2^N/N^{k}$ choices of disjoint pairs $a,b$ with polynomially small $|F_{a,b}|$, then there exists some unentangled probe state $\rho_{\text{probe}}$ and some measurement basis in which the discrimination task can be accomplished in a polynomial number of shots.} \\

We prove this in the appendix by constructing a measurement basis such that the probe state $\ket{+^N}$ can be used to discriminate between two distributions for which the feature matrix has the above property. This is possible because there are enough non-vanishing elements such that a uniform weighing of all of them by a probe product state leads to a non-vanishing separation in measurement statistics. As an illustrative example, consider a multivariate Gaussian with all zero means and a covariance matrix $\Sigma = \sum_{j=1}^N \sigma_j \bm{v_j}\bm{v_j}^T$ with $\sigma_1$ in the interval $(0, \frac{1}{N}]$ and all other $\sigma_j \in [1/2, 1]$. Choose $\bm{v_1}$ to have all equal, positive entries and all other eigenvectors of the covariance matrix to lie in some orthogonal space to $\bm{v_j}$. The characteristic function of this probability distribution is given by $\chi(\bm{k}) =\exp\{-\left(\sum_j \sigma_j (\bm{v_j}^T\bm{k})^2\right)\}$. Now suppose we want to differentiate between two distributions, one where $\sigma_1= 1/N$ and one where $\sigma_1 = 2/N$. The difference in characteristic functions is then $(1-\exp\{-(\bm{v_1}^T\bm{k})^2/N\})\exp\{-\left(\sum_j \sigma_j (\bm{v_j}^T\bm{k})^2\right)\}$. Note that the choice of vector $\bm{k}$ with all 1's or all $-1$'s (corresponding to feature matrix entries $F_{0^N,1^N}$, $F_{1^N,0^N}$ ) produces a non-vanishing quantity independent of $N$, whereas terms orthogonal to this vector vanish or, if not orthogonal, introduce exponentially large weighing factors through overlap with other $\bm{v_j}$. Consequently, it would seem that this difference of characteristic functions, and the resulting feature matrix, is concentrated enough that no product state could do well. However, this is not the case - in particular, we can simply choose a $\bm{k}'$ that is zero at all but one index where it is set to $1$. This corresponds to states that have identical bits at all but one index and so there are $2^{N-1}$ such disjoint choices of pairs. Note $(\bm{v_j}^T\bm{k}')^2$ add up to 1 (this is summing over the squares of the columns of the orthogonal matrix that has $\bm{v_j}$ as its rows), and so the exponent of Gaussian term is no less than -1 (i.e. averaging terms between -1/2 and -1). Since this term only shrinks as $\sim 1/N^2$, by Proposition 2 an unentangled probe state can distinguish the two distributions in polynomially many shots. Even better, we only need to measure the qubit corresponding to the non-zero entry of $\bm{k'}$. This illustrates another important point - despite having an exponentially concentrated difference of characteristic functions, the local $\usense$ evaluates it at non-vanishing points so many times that both entangled and unentangled probes can perform the discrimination in polynomially many shots. This, again, illustrates that the emergence of exponential sensing advantage is due to both the choice of distribution and $\usense$. This proposition has implications for quantum machine learning, which we discuss in Appendix~\ref{s: theorems and proofs}. 

While we have used a linear constraint to illustrate the emergence of exponential sensing advantage, the framework and all of our theorems apply to nonlinear constraints as well. However, in this setting, achieving a practical exponential advantage becomes significantly more difficult. In transitioning to this setting, three problems emerge, the first two theoretical and the third more practical. We will first address the two theoretical issues. Firstly, since the manifold we sample from is no longer locally flat, there is no longer a single vector $\bm{a}$ that enjoys an invariant-under-$\bm{\theta}$ (i.e. a noiseless subspace under parameter sampling) value $e^{-i\bm{a}\cdot \bm{\theta}}$. This is true for a hyperplane, where all points sampled from the plane have a fixed overlap with a vector normal to said plane, but not true for a general manifold, where overlap with a vector normal to the surface at one point varies over any small perturbation away from that point. 

Secondly, it is more difficult to achieve a noise distribution that wipes out information at all but a small number of $\bm{a}$. For a hyperplane, noise may be distributed arbitrarily without affecting the normal vector - for non-flat surfaces, noise distribution affects all vectors in some way, resulting in either many vectors being sensitive directions of change or none. That is, the feature matrix becomes more spread out. For instance, we cannot get an exponential advantage from entanglement for sensing changes to the radius of an $(n-1)-$sphere with a uniform distribution over said sphere. This is because all axes $\bm{a}$ are equally noisy, so there is no advantage from picking any particular ones out (more generally, see Corollary D.4 in Appendix~\ref{s: general framework proofs}).

The third, practical issue that arises with sampling from non-linear constraints is that the most important set of vectors is no longer determined purely by the constraint. Now, it depends heavily on the specific distributions over said manifold. For a hyperplane, we know the vector normal to it will always capture information about its displacement, regardless of what the distribution of points over the hyperplane is like. For curved manifolds, however, which vector (e.g. an $(n-1)-$sphere with the radius as the conserved quantity) has the most sensitivity to changes in the conserved quantity depends heavily on the specific distribution over the manifold, which might not be known ahead of time. 

It should be noted that all three of these issues can, in principle, be solved for some cases if we allow $\usense$ to introduce information about $\bm{\theta}$ into our network using  non-commuting $\hat{G}_j$. In fact, we construct precisely such example in Appendix~\ref{s:nonlinear constraint noncommuting example} to construct a noiseless feature that is a nonlinear function of the sensed parameters $\theta_j$. Since $\usense$ with non-commuting $\hat{G}_j$ lie outside of our framework, this presents an avenue for future research.

In conclusion, in this section we introduced the formalism of the feature matrix, an object capturing the difference between the characteristic functions of the discriminant distribution, specifically at points determined by the sensing unitary $\usense$. We then discussed how the properties of this feature matrix may be used to find other cases of exponential sensing advantage, with particular constraints on probe states and observables.  We presented sufficient conditions under which discriminating between two distributions becomes exponentially difficult for product probe states, and conditions when it is polynomially easy. We then proved exponential advantage holds for a linear constraint with sufficiently noisy distributions, and discuss limitations of extending these advantages to distributions arising from nonlinear constraints.

\section{Discussion and Outlook} 
\label{sec:conc}

\subsection{Summary of the work}

In this work, we considered quantum sensing tasks where the parameters to be sensed are \textit{stochastic}: each execution of the quantum-sensing protocol experiences a different sensing parameter sample from an underlying probability distribution. The goal of the protocol is to estimate properties of the probability distribution, or distinguish multiple probability distributions. We illustrated how entanglement can improve sensitivity for such tasks. First, we showed how a Bell state quantum-sensing protocol can outperform the optimal unentangled quantum-sensing protocol in a binary classification task of phase sensing. The quantum sensing advantage arises from the Bell state's ability to be resilient to noisy features of the stochastic parameters, which cannot be achieved by an unentangled quantum state. We then generalized to tasks involving highly correlated $N$-parameter probability distributions. We provided two examples of tasks where an $N$-qubit entangled quantum-sensing protocol can either perform estimation with low error or perform classification with high accuracy using only a constant-in-$N$ number of samples, whereas an unentangled quantum-sensing protocol would require an exponential number of samples ($\sim 2^N$) to perform the same. This arises from the ability of the entangled-quantum protocol to be robust to all irrelevant stochastic fluctuations of the parameters. 

Based on this intuition, we considered tasks that can potentially arise in physical systems. One such task is that of sensing a conserved global quantity among locally fluctuating observables, for example, the total spin component in the Z direction of a classical interacting XXZ spin-chain system. We described how an entangled protocol can discriminate among two ensembles of spins with different values for this conserved quantity, with a sample complexity independent of the system size and thermodynamical temperature of the spin-chain ensemble. In contrast, the optimal unentangled quantum-sensing protocol performs poorly as a function of the system size and the thermodynamical temperature of the XXZ spin-chain. The only way for the unentangled sensor to be insensitive to this stochasticity is to operate on a timescale faster than the timescale of fluctuations in the system. For many applications, this might be experimentally unfeasible. However, an appropriately entangled quantum sensor can safely operate on a much slower timescale. In the above example, the sensor has to operate within the timescale of the fluctuations of the conserved total magnetization of the spins, which can be much longer than the timescale of the fluctuations of the individual magnetization of each spin. In Ref.~\cite{bate2025experimental}, the authors experimentally demonstrate a similar concept with trapped ions, where entanglement is harnessed to render the sensor insensitive to correlated spatial fluctuations of electromagnetic signals. Finally, we introduced a theoretical framework for assessing the possibility of exponential stochastic sensing advantage based on the characteristic functions of the underlying probability distributions and the choice of $\usense$, allowing us to consider sensing tasks beyond just phase sensing with qubits.

\subsection{Exponential entanglement advantage in sample complexity}

Quantum computers can provide an exponential advantage over classical computers in certain tasks~\cite{arute_quantum_2019, aaronson2011computational}. In these cases, the exponential advantage is in the computational time needed to solve the task. In this work, we explored the possibility of a different kind of exponential advantage, that in sample complexity. In this scenario, an entangled quantum system requires exponentially fewer samples than a classical system (which, by definition, is unentangled) to solve a task. If the sensing time is assumed to be fixed per sample of the protocol, then this result can be understood as an exponential advantage in the time taken to achieve the task.

Such an advantage has been explored in previous works, such as tasks involving learning properties of quantum states with quantum systems~\cite{huang_experiments_2022, aharonov_quantum_2022, oh2024entanglement, chen_quantum_2022}. For example, in Ref.~\cite{huang_experiments_2022} the authors showed how an exponential advantage can arise in predicting the expectation of certain observables when entanglement is allowed in processing multiple copies of a state, compared with when entanglement is not allowed. In Ref.~\cite{chen_quantum_2022}, the authors demonstrate an exponential quantum advantage in simultaneously learning all the eigenvalues of a $N$-qubit Pauli channel within an additive error. In this case, entanglement involving $N$-ancilla qubits can perform this task with polynomial sample complexity that would instead take any protocol without ancillas an exponential number of samples. This was extended to learning $N$-mode displacement channels of bosonic modes in Ref.~\cite{oh2024entanglement}. In Ref.~\cite{aharonov_quantum_2022}, the authors introduce the framework of quantum algorithmic measurements (QUALMS), and explain how entanglement can enable an exponential quantum advantage in learning properties of quantum operators.

In contrast, for the conventional task of sensing the phase rotation applied to a qubit, an $N$-qubit entangled GHZ-state can achieve the Heisenberg Limit (HL), with an error of $1/N$ in estimating the phase. This is in contrast to the Standard Quantum Limit (SQL), with an error of $1/\sqrt{N}$, achieved by considering the measurements of $N$-unentangled qubits. In this example, entanglement enables a quadratic quantum advantage in sample complexity. A natural question is whether an exponential quantum advantage in sample complexity can be achieved in the framework of quantum sensing. We answered this positively by providing examples of tasks where the parameters to be sensed are correlated. In this scenario, the values of the parameters vary between shots of the quantum-sensing protocol. It is interesting to consider whether there exist other quantum sensing paradigms, potentially inspired by similar concepts such as learning properties of quantum states, or quantum discord~\cite{ollivier_quantum_2001}, where quantum resources can enable an exponential advantage in sample complexity.

\subsection{Future directions: generalizations beyond binary classification and scalar estimation}

While in this work we considered binary-classification tasks and estimation tasks of scalar values, a natural generalization to multi-class classification and estimation of more complex quantities can be considered in the context of stochastic sensing. For example, nitrogen-vacancy centers in diamond are suitable for use in sensing magnetic and electric fields in biological samples, with applications such as studying the metabolism of cells and recording the electromagnetic signals generated by the brain~\cite{aslam_quantum_2023}. There, the sensed parameters can vary in space and time, and can fluctuate on the timescale of measurements. Creating the optimal entangled states can help sensors harness correlations between these fluctuations to improve their sensitivity. Unlike the examples we have considered in this work, analytical optimal protocols might not exist in such instances. Instead, variational methods with parameterized entangling unitaries can help approximate the optimal quantum-sensing protocol~\cite{kaubruegger_quantum_2021, kaubruegger_optimal_2023}. Improving sensitivity in these applications can not only reduce the total sensing time to achieve a target accuracy, but also enable detecting novel biological processes that operate on a timescale much faster than the sensing integration timescale. The large improvement in sensitivity we observed in the numerical simulations of sensing the conserved total magnetization of the classical XXZ model (see Sec.~\ref{sec:xxz}) illustrates the potential for similar large quantum sensing advantages in these applications.

Another potential application of our work is to joint-detection receivers. These are quantum receivers that are designed to sense communicated codewords~\cite{delaney_demonstration_2022} in the low signal-to-noise regime (which can arise in long-distance communication where the dominant noise channel is the loss of photons). Using quantum resources such an entanglement across multiple receivers can enhance the accuracy of decoding the transmitted codeword compared to an unentangled receiver. It is interesting to consider how the quantum sensing advantage can change when we consider correlated loss channels. This can arise, for example, when the transmitted signal passes through some medium in space, where the effect on the signal is similar (or correlated) among the modes. Our work, especially a generalization of Sec.~\ref{sec:theory} to multiple classes, might help not only in estimating the maximal quantum sensing advantage achievable, but also in designing codewords that are robust to correlated noise channels.

More broadly, we hope that generalizations and extensions of our work lead to the discovery of more use cases for which entangled quantum sensors can deliver practically relevant, large advantages in reduced sensing times.

\vspace{0.6cm}
\emph{Note added.} During the preparation of this manuscript we became aware of Ref.~\cite{wang2024exponential}, which also identifies a task for which it is possible to design a protocol with entanglement that achieves an exponential advantage in sample complexity over protocols without entanglement, and Ref.~\cite{brady2024correlatednoiseestimationquantum}, which also considers sensing of correlated parameters. Similar to Ref.~\cite{wang2024exponential}, we also consider $N$-qubit sensors and compare the sample complexity of an entangled and unentangled protocol. However, while Ref.~\cite{wang2024exponential, brady2024correlatednoiseestimationquantum} consider the task of estimating parameters associated with decoherence (non-unitary) dynamics, we consider phase-sensing tasks of stochastic parameters.

\section*{Data and code availability}

All data generated and code used for this work is available at \url{https://doi.org/10.5281/zenodo.15298423}.

\section*{Author contributions}

S.P., S.A.K., R.Y. and P.L.M. conceived the project. S.P. and S.A.K. performed all the numerical simulations. V.K., with guidance from R.Y. and S.A.K., came up with the theoretical framework for assessing whether a quantum sensing advantage exists for a binary-classification task. All authors contributed to the writing of the manuscript. P.L.M. supervised the project.

\section*{Acknowledgements}

We gratefully acknowledge financial support from the Air Force Office of Scientific Research under award number FA9550-22-1-0203, and thank NTT Research for their financial and technical support. P.L.M. gratefully acknowledges financial support from a David and Lucile Packard Foundation Fellowship. We acknowledge helpful discussions with Shi-Yuan Ma, Erich Mueller, Tatsuhiro Onodera, Shyam Shankar, Mandar Sohoni, and Logan Wright.


\bibliographystyle{mcmahonlab}
\bibliography{references}

\appendix

\newpage
\section{Two-qubit example: discriminating bivariate distributions}
\label{s: corrgauss}

\subsection{Statistical properties of stochastic parameters}

In this section, we describe in detail the task of discrimination two distributions with two-qubit quantum sensors, introduced in Sec.~\ref{sec:corrgauss}. We consider the sensing of properties of stochastic random parameters $\mathbf{q} = (\theta_1,\theta_2)^T$ that are governed by a multivariate Gaussian distribution:
\begin{align}
    P(\bm{\theta}) = \frac{1}{\sqrt{(2\pi)^2 \det \mathbf{V}}} \exp \Big\{ -\frac{1}{2} (\mathbf{q}-\mathbf{\bar{q}})^T \mathbf{V}^{-1}(\mathbf{q}-\mathbf{\bar{q}}) \Big\}
\end{align}
which is characterized by the mean vector $\mathbf{\bar{q}} \equiv (\bar{\theta}_1,\bar{\theta}_2)$, and the covariance matrix $\mathbf{V}$. For the latter we consider the specific form:
\begin{align}
    \mathbf{V} = 
    \begin{pmatrix}
        \sigma^2 & \sigma_{\rm corr}^2 \\
        \sigma_{\rm corr}^2 & \sigma^2
    \end{pmatrix}
\end{align}
For later use, it will also prove convenient to define the marginal probability distributions of $\theta_1$ and $\theta_2$ respectively:
\begin{subequations}
\begin{align}
    P_1(\theta_1) \equiv \int d\theta_2 P(\bm{\theta}) &=  \frac{1}{\sqrt{2\pi\sigma^2}}\exp\Big\{-\frac{1}{2\sigma^2}(\theta_1-\bar{\theta}_1)^2 \Big\} \\
    P_2(\theta_2) \equiv \int d\theta_1 P(\bm{\theta}) &=  \frac{1}{\sqrt{2\pi\sigma^2}}\exp\Big\{-\frac{1}{2\sigma^2}(\theta_2-\bar{\theta}_2)^2 \Big\}
\end{align}  
\end{subequations}
Therefore, $\sigma^2$ defines the local (marginal) variances for the random parameters, and is assumed equal for both. Then, the off-diagonal component $\sigma_{\rm corr}^2$ describes the correlation of the random parameters. This can be further clarified by exploring the properties of the covariance matrix. The eigenvalues $\sigma_{\pm}^2$ of the covariance matrix satisfy the characteristic polynomial:
\begin{align}
    (\sigma^2 - \sigma_{\pm}^2)^2 - \sigma_{\rm corr}^4 = 0
\end{align}
and can be read off as 
\begin{align}
    \sigma_{\pm}^2 = \sigma^2 \pm \sigma_{\rm corr}^2
\end{align}
The eigenvectors corresponding to these eigenvalues respectively are:
\begin{align}
    v_{+} =
    \frac{1}{\sqrt{2}}
    \begin{pmatrix}
        1 \\ 
        1
    \end{pmatrix},~
    v_{-} =
    \frac{1}{\sqrt{2}}
    \begin{pmatrix}
        1 \\ 
        -1
    \end{pmatrix}.
\end{align}
Then, the determinant of the covariance matrix is simply the product of eigenvalues:
\begin{align}
    \det \mathbf{V} = \sigma_{+}^2\sigma_{-}^2 = (\sigma^2+\sigma_{\rm corr}^2)(\sigma^2-\sigma_{\rm corr}^2)
\end{align}

The covariance matrix, being a real symmetric matrix, can therefore be written in diagonal form by the unitary transformation:
\begin{align}
    \mathbf{V} = 
    \mathbf{P}\mathbf{D}\mathbf{P}^{T} \implies \mathbf{V}^{-1} = \mathbf{P} \mathbf{D}^{-1}\mathbf{P}^T,
\end{align}
where:
\begin{align}
    \mathbf{P} = 
    \frac{1}{\sqrt{2}}
    \begin{pmatrix}
        1 & 1 \\
        1 & -1
    \end{pmatrix},~
    \mathbf{D} = 
    \begin{pmatrix}
        \sigma_+^2 & 0 \\
        0 & \sigma_-^2
    \end{pmatrix}.
\end{align}
Using this form, we can rewrite the probability distribution in a simpler form,
\begin{align}
    P(\bm{\theta}) = \frac{1}{\sqrt{(2\pi)^2 \det \mathbf{V}}} \exp \Big\{ -\frac{1}{2} [\mathbf{P}^T(\mathbf{q}-\mathbf{\bar{q}})]^T \mathbf{D}^{-1}[\mathbf{P}^T(\mathbf{q}-\mathbf{\bar{q}})] \Big\}
\end{align}
It will prove useful to make a change of variables:
\begin{align}
    \begin{pmatrix}
        \theta_+ \\
        \theta_-
    \end{pmatrix}
    =
    \mathbf{P}^T\mathbf{q} = \mathbf{P}^T
    \begin{pmatrix}
        \theta_1 \\
        \theta_2
    \end{pmatrix} 
\end{align}
Expressed in the variables $\theta_{\pm}$ that lead to a diagonal covariance matrix and therefore define uncorrelated variables, the probability distribution simplifies to:
\begin{align}
    P(\bm{\theta}) = \frac{1}{\sqrt{2\pi\sigma_+^2}}\frac{1}{\sqrt{2\pi\sigma_-^2}}\exp\Big\{-\frac{1}{2\sigma_+^2}(\theta_+-\bar{\theta}_+)^2 \Big\}\exp\Big\{-\frac{1}{2\sigma_-^2}(\theta_--\bar{\theta}_-)^2 \Big\} = P_+(\theta_+)P_-(\theta_-)
\end{align}
which is now simply the product of two single-variable Gaussian distributions.

Finally, in all our analysis, we have assumed equal covariance matrix for the two distributions. We use this simplification for analytical convenience. The advantage generally remains for unequal covariance, since the information distinguishing the two classes in the mean value of $\theta_1 - \theta_2$ can be sensed more efficiently than the difference of variance. However, we emphasize the amount of advantage will not always be the same, and will depend on the covariances. Since this example is to build intuition for the potential of quantum sensing advantage in the stochastic sensing, the precise nature of the advantage for this two-qubit setting is beyond the scope of this work.

\subsection{Quantum measurement outcomes}
\label{app:qmeas}

We now briefly discuss the quantum measurements of the quantum sensor that are used to estimate properties of the probability distributions. We define $X_k^{(s)}$ as the measurement outcome after a single projective measurement $s$ of the quantum sensor state. The statistical properties of $X_k$ are determined by the specific POVMs $\{M_k\}$. In particular considering computational basis measurements $M_k = \ket{\bm{b}_k}\bra{\bm{b}_k}$ where $\bm{b}_k, k \in 1,\ldots, 2^N$ are bitstrings in the computational basis for $N$ qubits, we find: 
\begin{align}
    \mathbb{E}[X_k^{(s)}] &= {\rm Tr}\{ M_k \rhou \} = p_k \\
    \mathbb{E}[X_k^{(s)}X_{k'}^{(s')}] &= {\rm Tr}\{ M_k M_{k'} \rhou \} = p_k\delta_{kk'}\delta_{ss'} \\
    \mathbb{E}[X_k^{(s)}X_{k'}^{(s')}] &= p_kp_{k'}(1-\delta_{ss'})
\end{align}
where we have used the fact that $M_k M_{k'} = M_k \delta_{kk'}$. We can combine the final two terms in the form:
\begin{align}
    \mathbb{E}[X_k^{(s)}X_{k'}^{(s')}] = p_kp_{k'}(1-\delta_{ss'}) + p_k\delta_{kk'}\delta_{ss'}
\end{align}

For $\NS$ shots, we estimate the normalized frequency defined as:
\begin{align}
    x_k = \frac{1}{\NS} \sum_s X_k^{(s)}
\end{align}
whose statistical properties are easily determined using the statistics of $X_k$:
\begin{align}
    \mathbb{E}[x_k] = \frac{1}{\NS}\sum_s \mathbb{E}[X_k] =  p_k
\end{align}
and
\begin{align}
    {\rm Cov}[x_k,x_{k'}] &= \frac{1}{\NS^2}\sum_{s s'}\mathbb{E}[X_k^{(s)}X_{k'}^{(s')}] - \left( \frac{1}{\NS}\sum_s \mathbb{E}[X_k] \right) \left( \frac{1}{\NS}\sum_s \mathbb{E}[X_{k'}] \right) \nonumber \\
    &= p_kp_{k'} - \frac{\NS }{\NS^2} p_kp_{k'} + \frac{\NS}{\NS^2} p_k\delta_{kk'} - \left( p_k \right) \left( p_{k'} \right) \nonumber \\
    &= \frac{1}{\NS} \left( \delta_{kk'} p_k - p_kp_k' \right) 
\end{align}
We summarize the above results as:
\begin{align}
    \mathbb{E}[x_k] &= p_k, \\
    {\rm Cov}[x_k,x_{k'}] &= \frac{1}{\NS} \left( \delta_{kk'} p_k - p_kp_k' \right),
\end{align}
which describe standard statistics for the normalized linear combination of independent multinomial random variables. Another useful quantity will be the statistical properties of a combination of measurement outcomes, $y = x_k + x_{k'}$. We have:
\begin{align}
    \mathbb{E}[y] = p_k + p_{k'}
\end{align}
and
\begin{align}
    {\rm Var}[y] &= \mathbb{E}[y^2] - (\mathbb{E}[y])^2 \nonumber \\
    &= \mathbb{E}[x_k^2 + x_{k'}^2 + 2x_kx_{k'}] - (\mathbb{E}[x_k])^2 - (\mathbb{E}[x_{k'}])^2 - 2\mathbb{E}[x_k]\mathbb{E}[x_{k'}] \nonumber \\
    &= {\rm Cov}[x_k,x_{k}] + {\rm Cov}[x_{k'},x_{k'}] + 2{\rm Cov}[x_k,x_{k'}] \nonumber \\
    \implies {\rm Var}[y] &= \frac{1}{\NS} \left( p_k(1-p_k) + p_{k'}(1-p_{k'}) - 2p_kp_{k'} \right) 
\end{align}
A similar analysis can be performed for variables $y$ that consist of linear combinations of more variables $\{x_k\}$.

\subsection{Unentangled quantum-sensing protocol}

The unentangled quantum sensing protocol proceeds as follows:

\begin{enumerate}
    \item Starting from $\ket{00}$, the ground state for the two-qubit quantum sensor, apply the single-qubit Hadamard gate on both qubits
    \begin{align}
        \upr\ket{\psi_0} = \bigotimes_{j=1,2} {\rm H}_j \ket{0}_j = \bigotimes_{j=1,2} \frac{1}{\sqrt{2}}\left(\ket{0}_j + \ket{1}_j \right)
    \end{align}
    where ${\rm H}_j$ is the Hadamard gate on qubit $j$, and $\ket{\psi}_j$ defines the state of qubit $j$ alone.
    \item Perform $\usense$, following which the quantum sensor state becomes 
    \begin{align}
        \usense\upr\ket{\psi_0} = \bigotimes_{j=1,2} \frac{1}{\sqrt{2}}\left(e^{-\frac{i}{2}(\theta_j+\nu_j)}\ket{0}_j + e^{+\frac{i}{2}(\theta_j+\nu_j)}\ket{1}_j \right)
    \end{align}
    \item Perform $\udec = \upr^{\dagger}$ to arrive at the final quantum sensor state
    \begin{align}
         \ket{\psi_f} = \udec\usense\upr\ket{\psi_0} = \bigotimes_{j=1,2} \cos\left(\frac{\theta_j+\nu_j}{2}\right)\ket{1}_j + i\sin \left(\frac{\theta_j+\nu_j}{2}\right)\ket{0}_j
    \end{align}
\end{enumerate}
We can therefore simply read off the bitstring probabilities corresponding to the readout of states $\ket{11},\ket{10},\ket{01},\ket{00}$ respectively:
\begin{align}
    p_{11}(\bm{\theta}) &= \cos^2 \left( \frac{\theta_1 + \nu_1}{2} \right)\cos^2 \left( \frac{\theta_2 + \nu_2}{2} \right), \nonumber \\
    p_{10}(\bm{\theta}) &= \cos^2 \left( \frac{\theta_1 + \nu_1}{2} \right)\sin^2 \left( \frac{\theta_2 + \nu_2}{2} \right), \nonumber \\
    p_{01}(\bm{\theta}) &= \sin^2 \left( \frac{\theta_1 + \nu_1}{2} \right)\cos^2 \left( \frac{\theta_2 + \nu_2}{2} \right), \nonumber \\
    p_{00}(\bm{\theta}) &= \sin^2 \left( \frac{\theta_1 + \nu_1}{2} \right)\sin^2 \left( \frac{\theta_2 + \nu_2}{2} \right),
\end{align}
which sum to unit probability, as they must. To determine the input-averaged probabilities, we must calculate the averages of $p_k(\bm{\theta})$ over the input probability distribution, to obtain $p_k \equiv \int d\bm{\theta}P(\bm{\theta})p_k(\bm{\theta})$. To this end, we will also require the standard complex Gaussian integral:
\begin{align}
    \frac{1}{\sqrt{2\pi\sigma^2}}\int d\theta~\exp\Big\{-\frac{1}{2\sigma^2}(\theta-\bar{\theta})^2 \Big\} e^{ia\theta} = e^{ia\bar{\theta}}e^{-\frac{1}{2}\sigma^2a^2}
    \label{appeq:complexg}
\end{align}

It will prove sufficient to compute $p_{11}$; the remaining probabilities will be shown to be computable using the results for $p_{11}$. The input-averaged probability $p_1$ is thus given by:
\begin{align}
    p_{11} &= \int d\bm{\theta}~P(\bm{\theta})p_{11}(\bm{\theta}) = \int d\theta~P(\bm{\theta})\cos^2 \left( \frac{\theta_1 + \nu_1}{2} \right)\cos^2 \left( \frac{\theta_2 + \nu_2}{2} \right)
\end{align}
We first use trigonometric identities to write $p_{11}$ as:
\begin{align}
p_{11}
&= \frac{1}{4}\int d\bm{\theta}\,P(\bm{\theta})
   \big[ 1 + \cos (\theta_1+\nu_1) \big]
   \big[ 1 + \cos (\theta_2+\nu_2) \big] \nonumber \\
&= \frac{1}{4}\int d\bm{\theta}\,P(\bm{\theta})\Big[
   1 + \cos (\theta_1+\nu_1) \\
&\qquad
   + \cos (\theta_2+\nu_2)
   + \tfrac{1}{2}\cos (\theta_1+\theta_2+\nu_1+\nu_2)
   + \tfrac{1}{2}\cos (\theta_1-\theta_2+\nu_1-\nu_2)
   \Big].
\end{align}

The above requires the calculation of five integrals. The first is simply the integral over $P(\bm{\theta})$ which reduces to unity due to normalization. To compute the remaining integrals, it will again prove useful to make the unitary change of variables:
\begin{align}
    \begin{pmatrix}
        \theta_1 \\
        \theta_2
    \end{pmatrix}
    =
    \mathbf{P} 
    \begin{pmatrix}
        \theta_+ \\
        \theta_-
    \end{pmatrix}
    =
    \frac{1}{\sqrt{2}}
    \begin{pmatrix}
        1 & 1 \\
        1 & -1
    \end{pmatrix}
    \begin{pmatrix}
        \theta_+ \\
        \theta_-
    \end{pmatrix}
    = 
    \frac{1}{\sqrt{2}}
    \begin{pmatrix}
        \theta_+ + \theta_- \\
        \theta_+ - \theta_-
    \end{pmatrix}
\end{align}
In this basis, we can address the required integrals one-by-one. We first consider the first of the final two integrals:
\begin{align}
    \int d\bm{\theta}~P(\bm{\theta})\cos (\theta_1+\theta_2+\nu_1+\nu_2) &= \int d\theta_+d\theta_- P_+P_-\cos (\sqrt{2}\theta_+ + \nu_1+\nu_2) \nonumber \\
    &= \int d\theta_+~P_+\cos (\sqrt{2}\theta_+ + \nu_1+\nu_2) \nonumber \\
    &= \frac{1}{2}\int d\theta_+~P_+ \left(e^{i\sqrt{2}\theta_+}e^{i(\nu_1+\nu_2)} + e^{-i\sqrt{2}\theta_+}e^{-i(\nu_1+\nu_2)}  \right)
\end{align}
where we are able to perform the integral over $\theta_-$ in going from the first to the second line as the integrand is independent of $\theta_-$. Using Eq.~(\ref{appeq:complexg}), the above can be immediately evaluated to yield:
\begin{align}
    \int d\theta~P(\bm{\theta})\cos (\theta_1+\theta_2+\nu_1+\nu_2) &= \frac{1}{2} \left( e^{i\sqrt{2}\bar{\theta}_+}e^{-\sigma_+^2}e^{i(\nu_1+\nu_2)} + e^{-i\sqrt{2}\bar{\theta}_+}e^{-\sigma_+^2}e^{-i(\nu_1+\nu_2)} \right) \nonumber \\
    &= e^{-\sigma_+^2}\cos (\sqrt{2}\bar{\theta}_+ +\nu_1+\nu_2)
\end{align}
By analogy, we can write down:
\begin{align}
    \int d\theta~P(\bm{\theta})\cos (\theta_1-\theta_2+\nu_1-\nu_2) = e^{-\sigma_-^2}\cos (\sqrt{2}\bar{\theta}_- +\nu_1-\nu_2)
\end{align}
The remaining two integrals are best computed in the original basis, as they depend only on $\theta_1$ or $\theta_2$. In particular, we have:
\begin{align}
    \int d\bm{\theta}~P(\bm{\theta})\cos (\theta_1+\nu_1) &= \int d\theta_1~P(\theta_1)\cos (\theta_1+\nu_1) \nonumber \\
    &= \frac{1}{2}\int d\theta_1~P(\theta_1)\left( e^{i\theta_1}e^{i\nu_1} + e^{-i\theta_1}e^{-i\nu_1} \right)
\end{align}
Once again making using of Eq.~(\ref{appeq:complexg}), the above simplifies to:
\begin{align}
    \int d\bm{\theta}~P(\bm{\theta})\cos (\theta_1+\nu_1) = e^{-\frac{1}{2}\sigma^2}\cos(\bar{\theta}_1 + \nu_1)
\end{align}
Again by analogy, we have for the final integral:
\begin{align}
    \int d\bm{\theta}~P(\bm{\theta})\cos (\theta_2+\nu_2) = e^{-\frac{1}{2}\sigma^2}\cos(\bar{\theta}_2 + \nu_2)
\end{align}
We can finally write for the probability $p_{11}$:
\begin{align}
    p_{11} &= \frac{1}{4} + \frac{1}{4}e^{-\frac{1}{2}\sigma^2}\left( \cos(\bar{\theta}_1 + \nu_1)+ \cos(\bar{\theta}_2 + \nu_2) \right) \nonumber \\
    &+ \frac{1}{8}e^{-\sigma_+^2}\cos (\sqrt{2}\bar{\theta}_+ +\nu_1+\nu_2) + \frac{1}{8}e^{-\sigma_-^2}\cos (\sqrt{2}\bar{\theta}_- +\nu_1-\nu_2) .
    \label{appeq:2qp11}
\end{align}
We note that $p_{10} = p_{11}(\nu_2 \to \nu_2 + \pi)$, so that:
\begin{align}
    p_{10} &= \frac{1}{4} + \frac{1}{4}e^{-\frac{1}{2}\sigma^2}\left( \cos(\bar{\theta}_1 + \nu_1)- \cos(\bar{\theta}_2 + \nu_2) \right) \nonumber \\ &- \frac{1}{8}e^{-\sigma_+^2}\cos (\sqrt{2}\bar{\theta}_+ +\nu_1+\nu_2) - \frac{1}{8}e^{-\sigma_-^2}\cos (\sqrt{2}\bar{\theta}_- +\nu_1-\nu_2)  
    \label{appeq:2qp10}
\end{align}
Similarly, $p_{01} = p_{11}(\nu_1 \to \nu_1 + \pi)$ so that:
\begin{align}
    p_{01} &= \frac{1}{4} + \frac{1}{4}e^{-\frac{1}{2}\sigma^2}\left( -\cos(\bar{\theta}_1 + \nu_1)+ \cos(\bar{\theta}_2 + \nu_2) \right) \nonumber \\ &- \frac{1}{8}e^{-\sigma_+^2}\cos (\sqrt{2}\bar{\theta}_+ +\nu_1+\nu_2) - \frac{1}{8}e^{-\sigma_-^2}\cos (\sqrt{2}\bar{\theta}_- +\nu_1-\nu_2) ,
    \label{appeq:2qp01}
\end{align}
 and $p_{00} = p_{10}(\nu_1 \to \nu_1 + \pi)$, so we finally have:
\begin{align}
    p_{00} &= \frac{1}{4} + \frac{1}{4}e^{-\frac{1}{2}\sigma^2}\left( -\cos(\bar{\theta}_1 + \nu_1)- \cos(\bar{\theta}_2 + \nu_2) \right) \nonumber \\ &+ \frac{1}{8}e^{-\sigma_+^2}\cos (\sqrt{2}\bar{\theta}_+ +\nu_1+\nu_2) + \frac{1}{8}e^{-\sigma_-^2}\cos (\sqrt{2}\bar{\theta}_- +\nu_1-\nu_2) .
    \label{appeq:2qp00}
\end{align}

\subsubsection{Local qubit measurements only}

An interesting limit to consider is in the situation of only local qubit measurements and ignoring cross-correlations between qubits (which in this case exist entirely due to correlations in the input variables). In this case, the excitation probability for qubit~1 is given by averaging over the probability distribution for qubit~2:
\begin{align}
    p_1 = p_{11} + p_{10} = \frac{1}{2} + \frac{1}{2}e^{-\frac{1}{2}\sigma^2} \cos (\bar{\theta}_1+\nu_1)
    \label{appeq:locp1}
\end{align}
while the excitation probability for qubit~2 is analogously given by:
\begin{align}
    p_2 = p_{11} + p_{01} = \frac{1}{2} + \frac{1}{2}e^{-\frac{1}{2}\sigma^2} \cos (\bar{\theta}_2+\nu_2)
    \label{appeq:locp2}
\end{align}
Clearly, any classification based on $p_1,p_2$ alone is entirely insensitive to the correlation properties of $\theta_1,\theta_2$, and will be limited by the marginal variance $\sigma^2 \gg \sigma_-^2$ for strongly correlated inputs.

\subsection{Entangled quantum-sensing protocol}

\begin{figure}[H]
    \centering
    \includegraphics[width=0.75\linewidth]{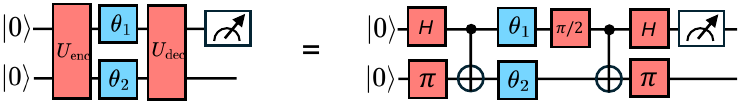}
    \caption{\textbf{Explicit circuit diagram for the two qubit entangled quantum sensing protocol} The protocol involves preparing a Bell state and measuring in this basis. Only one of the qubit needs to be measured, and the other is always in the ground state at the end of the protocol.}
    \label{Fig:Appendix:TwoQubitCircuitDiagram}
\end{figure}

The entangled protocol proceeds as follows (as illustrated in Fig~\ref{Fig:Appendix:TwoQubitCircuitDiagram}):

\begin{enumerate}
    \item Starting from $\ket{00}$, the ground state for the 2-qubit quantum sensor, prepare the entangled state
    \begin{align}
        \ket{\Psi^+} = \upr\ket{\psi_0} = \frac{1}{\sqrt{2}}\left(\ket{01} + \ket{10} \right)
    \end{align}.
    \item Perform $\usense$, following which the quantum sensor state becomes
    \begin{align}
        \usense\upr\ket{\psi_0} = \frac{1}{\sqrt{2}} \left( e^{-\frac{i}{2}(\theta_1-\theta_2+\nu_1-\nu_2)}\ket{01} + e^{+\frac{i}{2}(\theta_1-\theta_2+\nu_1-\nu_2)} \ket{10} \right)
    \end{align}
    \item Perform $\udec = \upr^{\dagger}$ to arrive at the final quantum sensor state
    \begin{align}
        \ket{\psi_f} = \udec\usense\upr\ket{\psi_0} &= \cos\left(\frac{\theta_1-\theta_2+\nu_1-\nu_2}{2}\right)\ket{10} \nonumber \\ &+ i\sin \left(\frac{\theta_1-\theta_2+\nu_1-\nu_2}{2}\right)\ket{00}
    \end{align}
\end{enumerate}
Qubit-2 is therefore always in state $\ket{0}$, indicating that the measurement of a single qubit is sufficient to extract all useful information in this entangled protocol. The excitation probability of this qubit is therefore simply given by:
\begin{align}
    p_1(\bm{\theta}) = \cos^2\left( \frac{\sqrt{2}\theta_- +\nu_1-\nu_2}{2} \right) = \frac{1}{2} \left(1 + \cos (\sqrt{2}\theta_- +\nu_1-\nu_2)  \right) 
\end{align}
By suitable choice of an entangling $\upr$, the excitation probability is therefore only sensitive to the low-noise combination of random variables $\bm{\theta}$, more precisely given by $\bm{v}_-^T \bm{\theta}$. As before, we must compute the input-averaged probability:
\begin{align}
    p_1 = \int d\bm{\theta}~P(\bm{\theta})p_1(\bm{\theta}) = \frac{1}{2} + \frac{1}{2} \int d{\theta_-}~P_-\cos (\sqrt{2}\theta_- +\nu_1-\nu_2)
\end{align}
The required integral was calculated earlier; we therefore have:
\begin{align}
    p_1 = \frac{1}{2} + \frac{1}{2} e^{-\sigma_-^2} \cos (\sqrt{2}\bar{\theta}_- +\nu_1-\nu_2)
    \label{appeq:ep1}
\end{align}

\subsection{Post-processing}

\subsubsection{Maximum likelihood estimator}

Measurement outcomes $x_k$ are estimates of bitstring probabilities in the computation basis. For a two-qubit quantum sensor, the measurement outcomes are $\{x_{11},x_{10},x_{01},x_{00}\}$. Defining a vector of quantum sensor outputs:
\begin{align}
    \mathbf{y} = 
    \begin{pmatrix}
        x_{11} \\
        x_{10} \\
        x_{01} \\
        x_{00}
    \end{pmatrix}
\end{align}
we can define the post-processing step to provide an output class label as
\begin{align}
    L_{\rm pred} = \mathcal{F}[\mathbf{y}]
\end{align}
where $L_{\rm pred}$ is the predicted class label, and $\mathcal{F}[\cdot]$ is a generally nonlinear function that maps the estimates of bitstring probabilities constructed using $\NS$ shots to a scalar class label, $\mathcal{F} : \mathbb{R}^{2^N} \to \mathbb{R}$; here $N=2$. For a maximum-likelihood based classifier, we define the simple post-processing layer:
\begin{align}
    L_{\rm pred} = \mathcal{F}[\mathbf{y}] = \stackbin[L \in A,B]{}{{\rm argmin}} \{ ||\mathbb{E}[\mathbf{y}^{(L)}]-\mathbf{y}||^2 \}
\end{align}
where $\mathbf{y}^{(L)}$ are estimates of the bistring probabilities $\{x_k\}$ for the known class $L \in A,B$, obtained using a training dataset. The above approach is used to obtain the classification accuracy plots in Fig.~\ref{Fig:Main2} of the main text. 

\subsubsection{Linear scalar output layer: entangled protocol}

For a more intuitive analysis of the performance of the various stochastic quantum sensing protocols, it proves useful to define a simpler post-processing step that obtains a linear, scalar estimator from quantum sensor measurement outcomes. This has the advantage of providing analytic results for Fisher's discriminant, which provides insight into the advantages of entanglement in stochastic quantum sensing. We start with the simplest case, which from the prior analysis is clearly the entangled protocol. Here, the scalar output variable is simply given by
\begin{align}
    y_{e} = x_1 
\end{align}
Using Eq.~(\ref{appeq:ep1}), we have:
\begin{align}
    \mathbb{E}[y_e] = \mathbb{E}[x_1] = p_1 = \frac{1}{2} + \frac{1}{2} e^{-\sigma_-^2} \cos (\sqrt{2}\bar{\theta}_- +\nu_1-\nu_2)
\end{align}
This again yields the optimal operating point $\nu_1 - \nu_2 = \pm\frac{\pi}{2}$; setting $\nu_1 = 0,\nu_2 = \frac{\pi}{2}$, we immediately find in the small angle limit:
\begin{align}
    \mathbb{E}[y_e] \simeq \frac{1}{2} + \frac{1}{2} e^{-\sigma_-^2}\left( \bar{\theta}_1-\bar{\theta}_2 \right)
\end{align}
which is as described in the main text.

To compute Fisher's discriminant we also require the variance ${\rm Var}[y_e]$, 
\begin{align}
    {\rm Var}[y_e] &= \frac{1}{\NS}p_1(1-p_1) \nonumber \\ &= \frac{1}{\NS}\left( \frac{1}{2} + \frac{1}{2} e^{-\sigma_-^2} \cos (\sqrt{2}\bar{\theta}_- +\nu_1-\nu_2) \right)\left( \frac{1}{2} - \frac{1}{2} e^{-\sigma_-^2} \cos (\sqrt{2}\bar{\theta}_- +\nu_1-\nu_2) \right) \nonumber \\ &\approx \frac{1}{4\NS}
\end{align}
where the final result is obtained to lowest order in $|\bar{\theta}_1-\bar{\theta}_2| \ll 1$ and using the aforementioned values of $\nu_j$. Under this approximation, which we verify numerically as shown in Fig.~\ref{Fig:Main2}, the variance is independent of class label $L$ to lowest order. The above results for the expected value and variance allow us to compute Fisher's discriminant, as discussed in the main text. 

\subsubsection{Linear scalar output layer: unentangled protocol with two qubit measurements}

For the unentangled protocol but now accounting for qubit correlations, we define the output variable
\begin{align}
    y_{u,\rm 2q} = x_{11} + x_{00} 
\end{align}
Using Eqs.~(\ref{appeq:2qp11}),~(\ref{appeq:2qp00}), we find:
\begin{align}
    \mathbb{E}[y_{u,\rm 2q}]= \mathbb{E}[x_{11} + x_{00}]  &= p_{11} + p_{00}  \nonumber \\
    &=\frac{1}{2} + \frac{1}{4}e^{-\sigma_+^2}\cos (\sqrt{2}\bar{\theta}_+ + \nu_1+\nu_2) + \frac{1}{4}e^{-\sigma_-^2}\cos (\sqrt{2}\bar{\theta}_- + \nu_1-\nu_2)
\end{align}
The above expression can be simplified if we consider the large variance limit where $\sigma^2 \gg \sigma_-^2$. Then the term depending on $\bar{\theta}_+$ can be neglected, and we have:
\begin{align}
    \mathbb{E}[y_{u,\rm 2q}] &\simeq \frac{1}{2} + \frac{1}{4}e^{-\sigma_-^2}\cos (\sqrt{2}\bar{\theta}_- + \nu_1-\nu_2)
\end{align}
Recalling that $\bar{\theta}_- = \frac{\bar{\theta}_1-\bar{\theta}_2}{\sqrt{2}}$, for lowest-order sensitivity to the phase difference we now require $\nu_1 - \nu_2 = \pm\frac{\pi}{2}$. This leads to slightly different optimal sensor operating points when compared to the case of local qubit measurements only; we now set $\nu_1 = 0,\nu_2 = \frac{\pi}{2}$. Once again assuming the small angle limit, we finally arrive at:
\begin{align}
    \mathbb{E}[y_{u,\rm 2q}] &\simeq \frac{1}{2} + \frac{1}{4}e^{-\sigma_-^2}\left( \bar{\theta}_1-\bar{\theta}_2 \right)
\end{align}
as described in the main text. To compute Fisher's discriminant, we also require the variance ${\rm Var}[y_{u,2q}]$, which from Appendix~\ref{app:qmeas} is given by:
\begin{align}
    {\rm Var}[y_{u,2q}] &= \frac{1}{\NS} \left( p_{11}(1-p_{11})  + p_{00}(1-p_{00}) - 2 p_{11}p_{00} \right) \nonumber \\
    &= \frac{1}{\NS} \left( \frac{1}{4}\cdot\frac{3}{4} + \frac{1}{4}\cdot\frac{3}{4} - 2\cdot \frac{1}{4}\cdot\frac{1}{4}  \right) \approx \frac{1}{4\NS}
\end{align}
where in the second line we have made use of the small angle limit again.  

\subsubsection{Linear scalar output layer: unentangled protocol with local qubit measurements only}

For the case of local qubit measurements only, we can define a simpler output variable:
\begin{align}
    y_{u,\rm 1q} = x_1 - x_2 = (x_{11}+x_{10})-(x_{11}+x_{01})
\end{align}
Using Eqs.~(\ref{appeq:locp1}),~(\ref{appeq:locp2}), we find:
\begin{align}
    \mathbb{E}[y_{u,\rm 1q}] &= \mathbb{E}[(x_{11}+x_{10})-(x_{11}+x_{01})]  = \mathbb{E}[x_{10}-x_{01}] \nonumber \\
    &= \frac{1}{2}e^{-\frac{1}{2}\sigma^2} \cos(\bar{\theta}_1+\nu_1) - \frac{1}{2}e^{-\frac{1}{2}\sigma^2} \cos(\bar{\theta}_2+\nu_2) \nonumber \\
    &= -e^{-\frac{1}{2}\sigma^2} \sin \left( \frac{\bar{\theta}_1-\bar{\theta}_2 + \nu_1-\nu_2}{2} \right)  \sin \left( \frac{\bar{\theta}_1+\bar{\theta}_2 + \nu_1+\nu_2}{2} \right) \nonumber \\
    &= -e^{-\frac{1}{2}\sigma^2} \sin \left( \frac{\bar{\theta}_1-\bar{\theta}_2 + \nu_1-\nu_2}{2} \right)  \sin \left( \frac{\nu_1+\nu_2}{2} \right)
\end{align}
where we have used $\bar{\theta}_1+\bar{\theta}_2 = 0$. Demanding the output variable to be sensitive to the phase difference $\bar{\theta}_1-\bar{\theta}_2$ to lowest order requires setting $\nu_1 = \nu_2$. Secondly, maximizing the remaining multiplicative factor to enhance sensitivity further demands setting $\frac{\nu_1 + \nu_2}{2} = \pm\frac{\pi}{2}$. For convenience we choose $\nu_1=\nu_2 = -\frac{\pi}{2}$ to satisfy both conditions. We therefore have:
\begin{align}
    \mathbb{E}[y_{u,\rm 1q}] = e^{-\frac{1}{2}\sigma^2} \sin \left( \frac{\bar{\theta}_1-\bar{\theta}_2}{2} \right)  
\end{align}
We can now proceed with the small angle approximation, following which we arrive at:
\begin{align}
     \mathbb{E}[y_{u,\rm 1q}] \simeq \frac{1}{2}e^{-\frac{1}{2}\sigma^2} \left( \bar{\theta}_1-\bar{\theta}_2 \right)
\end{align}
as described in the main text. To compute Fisher's discriminant, we also require the variance ${\rm Var}[y_{u,\rm 1q}]$, which from Sec.~\ref{app:qmeas} is given by:
\begin{align}
    {\rm Var}[y_{u,\rm 1q}] &= \frac{1}{\NS} \left( p_{10}(1-p_{10})  + p_{01}(1-p_{01}) - 2 p_{01}p_{10} \right) \nonumber \\
    &= \frac{1}{\NS} \left(  2\cdot\frac{1}{4}\cdot\frac{3}{4} - 2\cdot\frac{1}{4}\cdot\frac{1}{4} \right) \approx \frac{1}{4\NS}
\end{align}
where in the second line we have made use of the small angle limit again. 

\subsection{Performance beyond $|C|\ll 1$}

In the main text and in the prior appendices, we have shown the performance difference between an unentangled sensor and our entangled sensor approach for stochastic sensing of two correlated variables in the case of small $|C|\ll 1$. 
Here we show that this advantage is not restricted to this regime (which was used for analytic convenience), but holds for larger $|C|$ values as well.

\begin{figure}[H]
    \centering
    \includegraphics[scale=0.75]{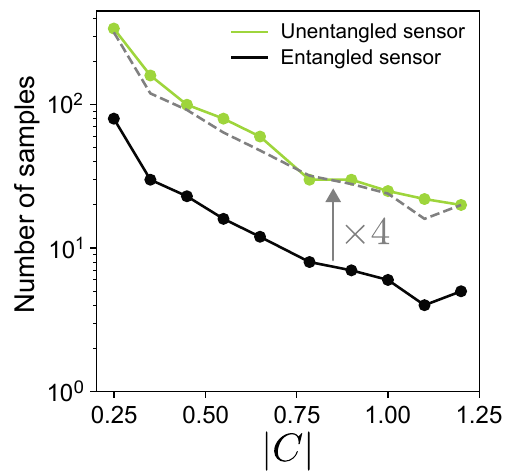}
    \caption{\textbf{Entangled vs. unentangled sensor performance at larger $|C|$.} We show the number of samples required by both protocols to reach a 95\% classification accuracy. The entangled sensor protocol exhibits a constant advantage of requiring 4 times fewer samples, as expected from analytic results. Gray dashed line multiplies the entangled sensor performance by this factor of 4 to illustrate the difference between the two curves. Here $\sigma=5, \sigma_{\rm corr}^2 = 0.99\sigma^2$.}
    \label{Fig:Appendix:scaleC}
\end{figure}

Using simulations, we evaluate the performance of the unentangled and entangled sensing protocols for the binary classification task considered in the main text, starting with the value $|C|=0.25$ considered there, but increasing up to values on the order of $\pi/2$ that defines the dynamic range of typical sensors. We have once again chosen $\sigma_{\rm corr}^2 = 0.99\sigma^2$, here with $\sigma = 5$. We determine the number of samples needed by both protocols to reach a classification accuracy of 95\%, which is plotted in Fig.~\ref{Fig:Appendix:scaleC}. As $|C|$ increases, the difference between the two distributions that need to be classified increases; the task becomes easier, and therefore the number of samples required decreases. 

However, we note that the entangled sensor requires a constant factor fewer samples, indicated by the fixed difference between the performance curves in logscale. In particular, if the entangled sensor performance curve is multiplied by the expected advantage of the factor of 4, we obtain the gray dashed line, which lies very close to the unentangled sensor performance curve. This indicates that the factor of 4 advantage of the entangled sensor scheme persists at large $|C|$.

\section{Exponential stochastic sensing advantage} 
\label{s: exponential advantage}

\subsection{Simulations}

\begin{figure}[H]
    \centering
    \includegraphics[width=0.75\linewidth]{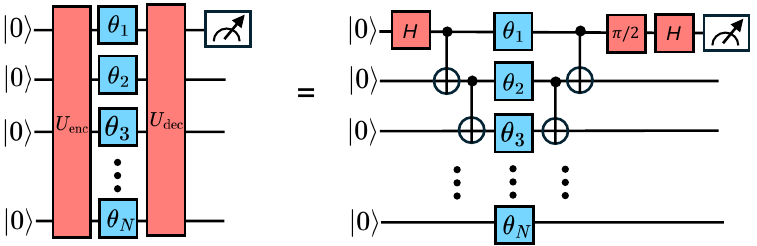}
    \caption{\textbf{Explicit circuit diagram for the N qubit entangled quantum sensing protocol which provides an exponential sensing advantage} The protocol involves preparing a GHZ state and measuring in this basis. Only one of the qubit needs to be measured, and the other is always in the ground state at the end of the protocol.}
    \label{Fig:Appendix:NQubitCircuitDiagram}
\end{figure}

In this section, we discuss the simulations performed for the results presented in Sec.~\ref{sec:exp}. We discuss in detail the simulations for the binary classification task. The simulations are very similar for the estimation task (since they involve the same procedure to generate the dataset and the outcomes of the quantum sensing protocols)

We perform simulations to support the proofs about the exponential scaling of the sample requirement for an unentangled sensor. For this we perform simulations on the random phase sensing task, where each individual phase is sampled uniformly from the interval $[0, 2\pi)$ range. The phase of the last qubit is set to satisfy the constraint for the two datasets. In this work, we choose the task of binary discriminating two datasets, defined to have the sum of phases equal to either $\pm C$. Here we choose $C = 0.3$. Operationally, we generate a batch of data and pass it through either the quantum or classical sensor. Averaging the probability distribution over the dataset of phases provides an estimate of the probability distribution over the possible outcomes of the sensor. For the classical sensor, we keep track of all the $2^N$ possibilities of qubit measurement outcomes. We then repeat this process with a new sampled batch of phases until there is minimal change in the probability distribution. Since the probabilities themselves reduce exponentially, we define convergence when the ratio of the smallest probability over the largest change in probability is above a certain value, which we choose to be $5000$ (there is a minimal change to the results increasing this value). Therefore, we have a list of probabilities for each case of the number of qubits, for either classes. To obtain the accuracy as a function of samples, we use NumPy's multinomial sampler~\cite{harris2020array}. We then compute the total variation distance of this estimate with the two classes and output as the prediction whichever is lower. From this curve, we can estimate the number of samples required to achieve a certain accuracy as a function of the number of qubits. As seen in~\ref{Fig:Main3}, we see that the classical sensor requires an exponential number of shots as a function of the number of qubits while the quantum sensor is constant.

There is degree of freedom of the optimal sensing protocol, which we describe here. This doesn't affect the scaling with respect to $N$, but is a function of $C$. For the case of the quantum sensor, which uses the GHZ state, this is straightforward. In this case, a single qubit measurement is sufficient. The probability of the qubit being excited is $P = \frac{1 + \sin{\sum_i \phi_i}}{2}$, where $\phi_i$ is the phase rotation on the $i$ qubit around the Pauli-Z axis. Since this sum is small and averages to zero over both classes, operating at this point makes this the most sensitive protocol. The whole entangled quantum sensing protocol is illustrated in Fig.~\ref{Fig:Appendix:NQubitCircuitDiagram}.

The classical protocol is less trivial. Since each qubit is only sensitive to the corresponding phase, there is no information about the class label in the expectation on the single qubit expectation values. The only observable which is dependent on the total sum involves product of expectations over all the Pauli operators. Since this will be exponentially small, this is what makes this task hard for an unentangled sensor. To understand the maximal sensitivity operating point, let's first consider the two and three qubit case. For the two qubit case, we would want the two qubits to sense at an axis $\pi/2$ off from each other. This would result in the following joint probability for both qubits being in the up state (for example):

\begin{align}
    p_{11} &= \frac{1}{2\pi} \int_0^{2\pi} d\theta \frac{1 + \sin{(\theta)}}{2} \frac{1 + \cos{(C - \theta)}}{2}
    \nonumber \\
    &= \frac{1}{4} + \frac{1}{8\pi} \int_0^{2\pi} d\theta \sin{(\theta)}\cos{(C - \theta)}
    \nonumber \\
    &= \frac{1}{8} + \frac{\sin{(C)}}{8},
\end{align}
which is first order sensitive to $C$. This also similarly works out for all other qubit measurement outcomes. For the case of three qubits, we do not require any phase offset between the qubits. For example, the expression for the expected probability for all qubits being in the excited state is:

\begin{align}
    p_{111} &= \frac{1}{(2\pi)^2} \int_0^{2\pi} d\theta_1 \int_0^{2\pi} d\theta_2 \frac{1 + \sin{(\theta_1)}}{2} \frac{1 + \sin{(\theta_2)}}{2} \frac{1 + \sin{(C - \theta_1 - \theta_2)}}{2}.
\end{align}
Like before, the only term which survives the integration and depends on $C$ is \\ $\sin{(\theta_1)}\sin{(\theta_2)\sin{(C - \theta_1 - \theta_2)}}$, which can be simplified to $\sin{C}/4$ when averaged over the domain, to first order in $C$. It can then be shown, recursively, that for even number of qubits, one of the qubits needs to measured at a $\pi/2$ phase shift from the other qubits. In the simulations, this is set to the last qubit (without loss of generality).

The simulations for the estimation task follow in the same procedure. We use a linear layer to map the outcome of the sensing protocol to estimate the phase. The optimal linear layer is found by fitting the dataset using least squares regression. To quantify the accuracy of the estimate, we compute the mean square error. To obtain this, we run the simulation many times (for a given number of qubits and number of samples of the quantum sensing protocol), until there is a negligible change in the results.

\subsection{Mapping on to a single qubit}

\begin{figure}[H]
    \centering
    \includegraphics[width=0.75\linewidth]{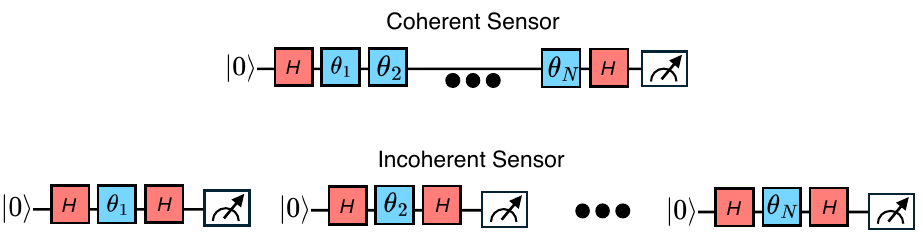}
    \caption{\textbf{Exponential sensing advantage with a single qubit.} The coherent sensor is defined as one which can coherently sense all the $N$ phases. In the task considered, simply sensing all the phase successively suffices. The incoherent sensor is restricted to perform a measurement after sensing each phase. This therefore produces $N$ samples.}
    \label{Fig:Appendix:SingleQubit}
\end{figure}

In this section, we motivate how the exponential sensing advantage with $N$ qubits (each experiencing a single phase), can be mapped onto a single qubit experiencing all $N$ phases in time. In this case, the relevant quantum resource is the coherence of the qubit. In this context, we define the quantum protocol as having quantum memory long enough to measure all the phases coherently. On the other hand, the classical sensor is defined to have limited coherence, which is confined to measure each phase one by one. Both protocols involve preparing and measuring the qubits along the equator of the Bloch sphere. For the case of the quantum protocol, we can just integrate all the $N$ phases together. This imparts a total phase $\phi_{\rm sum} = \sum_i \phi_i$, where $\phi_i$ is the phase of the $i$th pulse. $\phi_{\rm sum}$ is the phase to be sensed, and therefore this protocol performs the same as the entangled protocol described in Sec.~\ref{sec:exp}. On the other hand, the classical sensor resorts to $N$ measurements, one for each phase $\phi_i$ (see Fig.~\ref{Fig:Appendix:SingleQubit}). This therefore produces the same distributions of qubit outcomes as the $N$ qubit unentangled protocol. Hence, this protocol requires an exponentially more number of samples to achieve the same accuracy as a function of $N$. While in this case, we consider a task where the sensing protocol is very simple, it is interesting to consider tasks with more complex control. This example illustrates the interesting sensing advantages which can be achieved in time-varying and correlated signals.

\section{Sensing conserved quantities}
\label{s: xxz}

\subsection{Conserved quantities for the XXZ Hamiltonian}

In this section, we detail the simulation results presented in Sec.~\ref{sec:xxz}. For convenience we reproduce the XXZ Hamiltonian considered in the main text (we set $\hbar = 1$ and $\boltzmann = 1$ for convenience):
\begin{align}
    H_{\rm XXZ} = - J \sum_j (S_j^xS_{j+1}^x + S_j^y S_{j+1}^y +\Delta  S_j^z S_{j+1}^z) 
    \label{appeq:hxxz}
\end{align}
We first show that dynamics governed by $H_{\rm XXZ}$ ensure that the value of $S_j^z$ locally for spin $j$ is not a constant of the motion (i.e. is not conserved). While formally a classical Hamiltonian, for purposes of readability we instead use the language of commutators more familiar from a quantum description of the Hamiltonian to analyze conserved quantities~\cite{pires_classical_1996}. Then taking the classical limit $[f,g]/i\hbar \to \{f,g\}$~\cite{yang_generalizations_1980} allows transcribing results to the classical Hamiltonian. As this transformation does not change the value of a commutator from zero to non-zero (or vice versa), the conserved quantities are unchanged in moving between the two limits.

We therefore simply evaluate the commutator using the canonical commutation relations $[S_j^{\alpha},S_{k}^{\beta}] = i \epsilon_{\alpha\beta\gamma} S_j^{\gamma} \delta_{jk}$ where $\epsilon$ is the usual Levi-Cevita symbol:
\begin{align}
    [H_{\rm XXZ},S_k^z] &= -\Big[ J\sum_j (S_j^xS_{j+1}^x + S_j^y S_{j+1}^y +\Delta  S_j^z S_{j+1}^z), S_k^z \Big] \nonumber \\
    &= -\Big[ J\sum_j (S_j^xS_{j+1}^x + S_j^y S_{j+1}^y), S_k^z \Big] \nonumber \\
    &= J\sum_j \left(  -S_j^x\Big[S_{j+1}^x,S_k^z \Big]\delta_{(j+1)k} - S_{j+1}^x\Big[S_{j}^x,S_k^z \Big]\delta_{jk} -S_j^y\Big[S_{j+1}^y,S_k^z \Big]\delta_{(j+1)k} - S_{j+1}^y\Big[S_{j}^y,S_k^z \Big]\delta_{jk} \right) \nonumber \\
    &=  J\left( + S_{k-1}^xS_k^y + S_{k+1}^xS_k^y - S_{k-1}^yS_k^x - S_{k+1}^yS_k^x \right) \neq 0
\end{align}
However, the total $z$-component of the spin $S^z = \sum_j S_j^z$ instead does commute with $H_{\rm XXZ}$ and is conserved:
\begin{align}
    [H_{\rm XXZ},\sum_k S_k^z] &= J\sum_k \left( + S_{k-1}^xS_k^y + S_{k+1}^xS_k^y - S_{k-1}^yS_k^x - S_{k+1}^yS_k^x \right) \nonumber \\
    &= J\sum_k \left( + S_{k-1}^xS_k^y - S_{k-1}^yS_k^x  \right) + \sum_k \left( + S_{k+1}^xS_k^y  - S_{k+1}^yS_k^x \right) \nonumber \\
    &= J\sum_k \left( + S_{k}^xS_{k+1}^y - S_{k}^yS_{k+1}^x  \right) + \sum_k \left( + S_{k+1}^xS_k^y  - S_{k+1}^yS_k^x \right) \nonumber \\
    &= J\sum_k \left( + S_{k}^xS_{k+1}^y - S_{k+1}^yS_k^x - S_{k}^yS_{k+1}^x  + S_{k+1}^xS_k^y   \right) = 0,
\end{align}
where going from the second to the third line, we have made use of the translation invariance of the classical spin system enforced by periodic boundary conditions. This allows translating the summand indices from $k-1 \to k$, following which the cancellation of terms is clear.

It will prove useful to introduce a set of angular variables $\phi_j$ and spin variables $s_j$ such that:
\begin{align}
    S_j^x &= f(s_j)\cos \phi_j \\
    S_j^y &= f(s_j)\sin \phi_j \\
    S_j^z &= s_j
\end{align}
where $f(s_j) = \sqrt{1-s_j^2}$ and $s_j \in [-1,1]$. Under this change of variables, the XXZ Hamiltonian takes the simple form:
\begin{align}
    H_{\rm XXZ} = - J\sum_j \left( f(s_j)f(s_{j+1})\cos(\phi_{j+1}-\phi_j) + \Delta s_j s_{j+1} \right) 
\end{align}
This equivalent form of the XXZ Hamiltonian is used for generating spin configurations in the following subsection.

\subsection{Generating a Gibbs ensemble of spins interacting via the XXZ Hamiltonian}

For a system of $N$ classical spins in a periodic ring interacting via $H_{\rm XXZ}$ as defined by Eq.~(\ref{appeq:hxxz}), and in thermal equilibrium characterized by the normalized inverse temperature $\beta$, the final spin configurations will follow the Boltzmann distribution $\mathcal{P} \propto e^{-\beta H_{\rm XXZ}}$. We would like to sample spin configurations from this Boltzmann distribution at inverse temperature $\beta$ to obtain an ensemble of states of the $N$ classical spins, to then be sensed by the quantum sensors we consider in the main text. Unfortunately, sampling from the Boltzmann distribution using standard rejection samplers is highly inefficient, due to the high-dimensionality of the distribution for large $N$, and the need to determine the normalization constant, namely the partition function.

A much more computationally-efficient approach that is routinely used instead to sample different spin configurations from the Boltzmann distribution is the Metropolis(-Hastings) algorithm~\cite{metropolis_equation_1953, hastings_monte_1970}. The Metropolis algorithm requires only the transition probability between two configurations, and not the absolute probability which would necessitate knowledge of normalization. The specific Metropolis algorithm we employ is detailed in Algorithm~\ref{alg:metropolis}, which differs from standard implementations via the use of a constrained state transition rule~\cite{prosen_macroscopic_2013} that ensures the global magnetization $\sum_j S_j^z$ is conserved, as demanded by the XXZ Hamiltonian. We use the constrained Metropolis algorithm to obtain $\NS$ spin configurations for $N$ spins interacting via the XXZ Hamiltonian with parameters $J,\Delta$, and at inverse temperature $\beta$, all of which are defined in Fig.~\ref{Fig:Main4} of the main text. The Metropolis algorithm also requires a `thermalization' time $\tau_{\rm therm}$ for information of the initial configuration to be lost, and a sampling period $\tau_{\rm sweep}$ between samples to minimize correlation between different sampled spin configurations. We set $\tau_{\rm therm} = 10000$ and $\tau_{\rm sweep} = 500$, in units of Metropolis algorithm steps. The total number of Metropolis algorithm steps is therefore $\tau_{\rm therm} + \tau_{\rm sweep} \times \NS$.

\RestyleAlgo{ruled}
\SetKwComment{Comment}{/* }{ */}
\SetKwInOut{input}{Model Input}
\SetKwInOut{alginput}{Algorithm Input}
\SetKwInOut{output}{Output}
\SetKwFor{For}{for}{}{endfor}

\begin{algorithm}[H]
    \caption{Constrained Metropolis algorithm}
    \label{alg:metropolis}
    \input{$N, \beta, J, \Delta$}
    \alginput{$\tau_{\rm therm}, \tau_{\rm sweep}, \delta s, \delta \phi, S$}
    \output{Gibbs ensemble of $S$ random spin configurations $\{\bm{S}^z\}$}
    Randomly initialize spin configuration $\bm{\phi},\bm{s}$ with $\phi^{(0)}_j \sim \mathcal{U}(0,2\pi)~\forall~j$, $s^{(0)}_j = \frac{M}{N}~\forall~j$ s.t. $\sum_j s^{(0)}_j = M$ \\
    \For{$i \gets 1$ to $(\tau_{\rm therm} +  \tau_{\rm sweep}\times\NS$)}{
        Evaluate energy of current configuration, $E_0 = H_{\rm XXZ}(\bm{s}^{(i-1)},\bm{\phi}^{(i-1)})$\; 
        Randomly choose any two distinct spins $j,k$\;
        Perturb each phase independently, $\phi^{(t)}_{j} \gets \phi^{(i-1)}_{j} + d\phi_j$ for $d\phi_j \sim \mathcal{U}(-\delta\phi,+\delta\phi)$\;
        Perturb spins together $s^{(t)}_j \gets s^{(i-1)}_j + ds, s^{(t)}_k \gets s^{(i-1)}_k - ds$, $ds \sim \mathcal{U}(-\delta s,+\delta s)$ so $s_j + s_k$ is unchanged\;
        Obtain resulting test configuration $\bm{s}^{(t)},\bm{\phi}^{(t)}$\;
        Evaluate energy of test configuration, $E_t = H_{\rm XXZ}(\bm{s}^{(t)},\bm{\phi}^{(t)})$\; 
        Calculate change in energy due to flip, $\delta E = E_t - E_0$\; 
        \eIf{$\delta E < 0$}{
        $\bm{s}^{(i)},\bm{\phi}^{(i)} \gets \bm{s}^{(t)},\bm{\phi}^{(t)}$
        }
        {
        Calculate transition probability $P_{\rm tr} = e^{-\beta \delta E}$\;
        Sample random variable $r \sim \mathcal{U}(0,1)$\;
        \eIf{$r \leq P_{\rm tr}$}{
        $\bm{s}^{(i)},\bm{\phi}^{(i)} \gets \bm{s}^{(t)},\bm{\phi}^{(t)}$
        }{$\bm{s}^{(i)},\bm{\phi}^{(i)} \gets \bm{s}^{(i-1)},\bm{\phi}^{(i-1)}$}
        }
        \If{$i > \tau_{\rm therm}$ and $\lfloor\frac{i-\tau_{\rm therm} } {\tau_{\rm sweep}}\rfloor=0$ {\rm (}where $\lfloor\cdot\rfloor$ is the ${\rm floor}$ function{\rm )}}{
        Add current configuration to ensemble $\{\bm{S}^z\} \gets \bm{s}^{(i)}$
        }
    }
\end{algorithm}

\subsection{Simulations}

We first create a dataset of $S^z_i$ ensemble values using the Constrained Metropolis algorithm. We choose a dataset size large enough such that increasing the dataset has a negligible effect of the classification accuracy curves. From this we convert to a phase imparted on the qubit. For the simulations presented in Sec.~\ref{sec:xxz}, we set this scaling factor to be $\pi$. Since $S^z_i$ can take any value between $-1$ and $1$, this results in the phase taking on any value between $-\pi$ and $\pi$. We choose this range to study the effect of stochasticity in the system. We then pass these phases through the quantum sensing protocol, either entangled or unentangled. The protocol described is essentially that of Appendix~\ref{s: exponential advantage}. The only difference is that for the unentangled product state protocol, there is no apriori known optimal values for the phase offset to each qubit which sets the bound on sensitivity. Therefore, we sweep all possibilities of these values, allowing this to be $0$ and $\pi/2$ (the two sweet spots) for each qubit. We then choose the combination which produces the best result for each choice of system size $N$ and temperature $T$. The results are illustrated in Fig.~\ref{Fig:Main4}.

\section{A general framework for exponential separation on stochastic parameters with eigenbasis-invariant unitary encodings}
\label{s: general framework proofs}

\subsection{Defining our framework of quantum stochastic sensing advantage}
\label{s: appendix framework}
\begin{figure}[h]
    \centering
    \includegraphics[width=\linewidth]{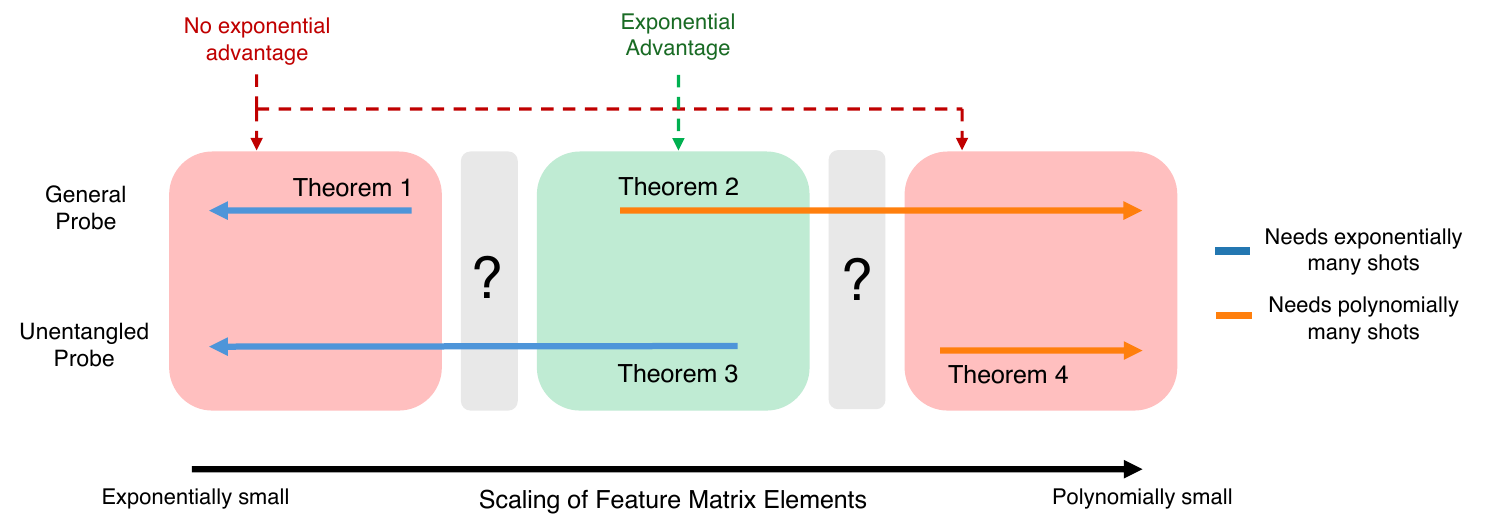}
    \caption{A schematic presenting the main idea behind each theorem and the number of shots required to discriminate between two distributions $\mathcal{P}_{\rm A},\mathcal{P}_{\rm B}$ with some fixed level of confidence. Theorem 1 identifies a scaling of feature matrix elements where even an entangled probe state needs exponentially many shots. Theorem 2 identifies a scaling beyond which it an entangled probe state requires polynomially many shots. Theorems 3 determines distributions of feature matrix elements where, with a general and product state measurement basis respectively, an unentangled probe state requires exponentially many shots. Finally, Theorem 4 identifies a scaling where an unentangled probe state requires only polynomially many shots. Overlaps in these regions determine the presence or absence of exponential advantage. There are gaps in the scaling uncovered by our theorems, where we only know the difficulty for one class of probe states - these are left for future work.}
    \label{fig:theorem overview}
\end{figure}

In this Appendix, we describe in further details the discussions of Sec.~\ref{sec:theory}. We assume that we prepare a probe state, $\rhou$, on $N$ qubits which is then acted on by a unitary, $\usensegen(\bm{\theta})$, with $N$ angles that are sampled from a distribution $\mathcal{P}_\Phi(\bm{\theta})$ - this sample changes every shot. Further, it is assumed that the eigenvectors of $\usensegen$ are invariant under parameters $\theta$. We then measure in a projective basis given by projector $\hat{O} = \udec^\dag M_0 \udec$, where $M_0$ is some projective measurement in the $\sigma_z$ (computational) basis. We specifically choose $\hat{O}$ so that $\hat{O}$'s eigenvalues are all equal to zero or one. For discrimination between two cases, this is equivalent to choosing an optimal POVM basis. We allow the measurements to be performed in a potentially entangled basis.The expected value of such a measurement is given by

\begin{align}
    p_0^\Phi = {\rm Tr}\Big\{ (\udec^{\dagger}{M}_0 \udec) \int d\bm{\theta}~\mathcal{P}_{\Phi}(\bm{\theta})\usense^{\text{general}}(\bm{\theta})\rhou_{\rm probe}\usense^{\text{general}\ \dagger}(\bm{\theta}) \Big\} = \langle \hat{O}\rangle_\Phi.
    \label{eq:p0 appendix}
\end{align}
\begin{align}
    \usense^{\text{general}} &:= \exp\left\{-i\sum_{j=1}^N \theta_j \hat{G}_j \right\} = \exp\{-i\hat{G}(\bm{\theta})\},\nonumber \\ &[\hat{G}_j,\hat{G}_k]=0 \forall j,k \text{ and } \hat{G}(\bm{\theta})\text{ has a product state eigenbasis}
\label{eq: usense general appendix}
\end{align}

Note that, as defined above, $\usensegen$ is guaranteed to have a product state eigenbasis that is independent of  $\bm{\theta}$. Let $\lambda_j^{\ket{a}}$ be the eigenvalue of $\ket{a}$ for $\hat{G}_j$, and let $\bm{\lambda}(\bm{a})$ be the vector of eigenvalues for each $\hat{G}_j$ for $\ket{a}$. Then $\usensegen(\bm{\theta}) \ket{a} = \exp\{-i \bm{\lambda(\bm{a}) \cdot \bm{\theta}} \}\ket{a}$ where $\bm{\theta}$ is the vector of angle parameters $\theta_i$. Rewriting Eq.~(\ref{eq:p0 appendix}) in the eigenbasis of $\usensegen$, and taking advantage of the linearity of the trace, gives us

\begin{align}
    p_0^\Phi = \sum_{\ket{a},\ket{b}}{\rm Tr}\Big\{ \hat{O} \bm{\rho}_{\bm{a},\bm{b}} \int d\bm{\theta}~\mathcal{P}_{\Phi}(\bm{\theta}) e^{-i(\bm{\lambda}(\bm{a})-\bm{\lambda}(\bm{b}))\cdot \bm{\theta}} \ket{a}\bra{b}   \Big\} 
    \label{eq:p0 appendix rewritten 1}
\end{align}

\begin{align}
    p_0^\Phi = \sum_{\ket{a},\ket{b}} \bm{O}_{\bm{a},\bm{b}}\bm{\rho}_{\bm{a},\bm{b}}\chi^\Phi(\bm{\lambda}(\bm{a})-\bm{\lambda}(\bm{b}))
    \label{eq:p0 appendix rewritten 2}
\end{align}

\begin{align}
    \bm{O}_{\bm{a},\bm{b}} := \bra{b}\hat{O}\ket{a},\ \ \bm{\rho}_{\bm{a},\bm{b}}:= \bra{a}\rhou_{\text{probe}}\ket{b},\ \ \chi^\Phi(\bm{k}):= \int d\bm{\theta} ~\mathcal{P}_\Phi(\bm{\theta})e^{-i\bm{k}\cdot \bm{\theta}}
    \label{eq:def o rho chi}
\end{align}
where, in Eq~\ref{eq:def o rho chi}, $\chi^\Phi$ is the non-unitary Fourier transform of the distribution $\mathcal{P}_\Phi$, also known as its ``characteristic function". To differentiate two distinct distributions, $\Phi=A,B$, we must examine the difference between the two expected measured values, i.e. the ``separation value":
\begin{align}
    \Delta_{A,B} := p_0^A - p_0^B &= \langle \hat{O}\rangle_A - \langle \hat{O}\rangle_B = \sum_{\ket{a},\ket{b}} \bm{O}_{\bm{a},\bm{b}}\bm{\rho}_{\bm{a},\bm{b}}\left( \chi^A(\bm{\lambda}(\bm{a})-\bm{\lambda}(\bm{b})) - \chi^B(\bm{\lambda}(\bm{a})-\bm{\lambda}(\bm{b}))\right) \nonumber \\ &= \sum_{\ket{a},\ket{b}} \left[\bm{\rho} \circ F \circ \bm{O} \right]
\label{eq: separation value appendix}
\end{align}
where $\bm{\rho},\bm{O}$ are matrix representations of $\rhou_{\text{probe}},\hat{O}$ in the eigenbasis of $\usensegen$ and ``$\circ$" denotes the Schur product. Finally, $F$ is the characteristic feature matrix, defined as
\begin{equation}
    F_{a,b} = \chi^A(\bm{\lambda}(\bm{a})-\bm{\lambda}(\bm{b})) - \chi^B(\bm{\lambda}(\bm{a})-\bm{\lambda}(\bm{b}))
\label{eq: feature matrix appendix}
\end{equation}

Note that for a given choice of $\rhou_{\text{probe}}$ and $\hat{O}$, $\Delta_{A,B}$ corresponds to the change in a probability between two sets of outcomes due to the change in the distributions. Consequently, the number of shots required to distinguish these two probability values scales as $\Omega(1/\Delta_{A,B})$. Even worse, for cases where $p_0^A,p_0^B$ are both near $1/2$, the number of shots scales as $ \Theta(1/\Delta_{A,B}^2)$.  Importantly, if there existed some choice of measurement distribution where the total variational distance between the two parameter distributions was polynomially small, it would be possible to construct a projector onto all the outcomes that are more likely for one class than the other - this would have a polynomially small separation value.

For small $\Delta_{A,B}$, and $p_0^{A},p_0^{B}$ near $\frac{1}{2}$, we can use a normal approximation $S^{90\%}\geq z_*^2/(4\Delta_{A,B}^2)$ where $S^{90\%}$ is the number of shots required to discriminate between the distributions with $90\%$ confidence and $z_*\approx 1.28$. Consequently, a small upper bound on $\Delta_{A,B}$ leads to a high lower bound on the number of shots needed for discrimination with some fixed confidence level. \\

\subsection{Feature matrix construction: two examples}
\label{s: feature matrix examples}
An equally important point to illustrate, through example, the construction of the feature matrix for the \text{same} choice of distributions and two different choices of $\usensegen$. Consider a simple, two-dimensional distribution of Gaussians on $\theta_1, \theta_2$ with two different cross-correlations with means zero and covariance matrices 
\begin{equation}
    \Sigma_A = \begin{bmatrix}
        1 & 0.5\\
        0.5 & 1
    \end{bmatrix},\ \ 
    \Sigma_B =  \begin{bmatrix}
        1 & 0.6\\
        0.6 & 1
    \end{bmatrix}
\end{equation}
The characteristic functions of these two Gaussian distributions is:
\begin{equation}
    \chi_{A,B} (\bm{k})= e^{-i\bm{k}^T\bm{\mu}_{A,B}-\frac{1}{2}\bm{k}^T\Sigma_{A,B}\bm{k}} = e^{-\frac{1}{2}\bm{k}^T\Sigma_{A,B}\bm{k}} 
\end{equation}
The difference in characteristic functions functions is therefore:
\begin{equation}
    \chi_A(\bm{k}) - \chi_B(\bm{k}) = \left(1-e^{-0.1k_1k_2} \right)e^{-\frac{1}{2}\left(k_1^2+k_1k_2+k_2^2\right)}
\end{equation}

Suppose that our choice of $\usense$ is given by:
\begin{equation}
    U_{\text{sense}} = \exp\left\{-i\theta_1 Z_1Z_2 -i\theta_2Z_2 \right\}
\end{equation}
where $Z_j = \frac{1}{2}(\sigma^z_j+\mathbb{I})$, so $\usense$ applied to state $\ket{a_1, a_2}$ would get phase $\exp\{-i(a_1a_2 \theta_1 + a_2\theta_2)\}$. Consequently, $\bm{\lambda}(\bm{a}) = \langle a_1a_2, a_2\rangle$. The entry of the feature matrix $F_{\ket{a},\ket{b}}$ is therefore given by:
\begin{equation}
    F_{a,b} = \chi_A(\langle a_1a_2-b_1b_2, a_2-b_2\rangle) - \chi_B(\langle a_1a_2-b_1b_2, a_2-b_2\rangle)  
\end{equation}
\begin{equation}
    = \left(1-e^{-0.1(a_1a_2-b_1b_2)(a_2-b_2)} \right)\exp\left\{-\frac{1}{2} \left((a_1a_2-b_1b_2)^2 + (a_1a_2-b_1b_2)(a_2-b_2) + (a_2-b_2)^2\right)\right\}
\end{equation}

The resulting feature matrix, with elements going in order as $\ket{00},\ket{01},\ket{10},\ket{11}$ becomes:

\begin{equation}
    \hat{F}  = 
    \begin{bmatrix}
        0 & 0 & 0 & (1-e^{-0.1})e^{-\frac{3}{2}}\\
        0 & 0 & 0 & 0 \\
        0 & 0 & 0 & 0 \\
        (1-e^{-0.1})(e^{-\frac{3}{2}}) & 0 & 0 & 0
    \end{bmatrix}
\label{e: feature matrix example 1}
\end{equation}
Note that, in the above case, the optimal probe state and measurement bases are ones that concentrate their off-diagonal terms  at the non-zero entries i.e. $\frac{1}{\sqrt{2}}(\ket{00}+\ket{11})$. Since the choice of $\usense$ is entangling, it is perhaps not surprising that we obtain a feature matrix with a clear advantage for an entangling probe state. For contrast, consider an unentangled $\usense$ defined by: 
\begin{equation}
    \usense = \exp\left\{-i\theta_1 Z_1 -i\theta_2 X_2 \right\}
\end{equation}
where $Z_1 = \frac{1}{2}(\sigma_1^z + \mathbb{I})$ and $X_2 = \frac{1}{2}(\sigma^x_2 + \mathbb{I})$. Defining our basis as $\ket{00} = \ket{0^z,+},\ket{01} = \ket{0^z,-},\ket{1,0} = \ket{1^z,+}, \ket{11} = \ket{1^z,-}$, we get that $\usense \ket{a_1,a_2} = e^{-ia_1\theta_1-ia_2\theta_2}$. Consequently, $\bm{\lambda}(\bm{a}) = \langle a_1, a_2\rangle$. From this, we can get that the entry of feature matrix $F_{a,b}$ is given by:
\begin{equation}
    F_{a,b} = \left(1-e^{-0.1(a_1-b_1)(a_2-b_2)} \right)\exp\left\{-\frac{1}{2} \left((a_1-b_1)^2 + (a_1-b_1)(a_2-b_2) + (a_2-b_2)^2\right)\right\}
\end{equation}
This gives us the feature matrix:

\begin{equation}
    \hat{F}  = 
    \begin{bmatrix}
        0 & 0 & 0 & (1-e^{-0.1})e^{-\frac{3}{2}}\\
        0 & 0 & (1-e^{0.1})e^{-\frac{1}{2}} & 0 \\
        0 & (1-e^{0.1})e^{-\frac{1}{2}} & 0 & 0 \\
        (1-e^{-0.1})(e^{-\frac{3}{2}}) & 0 & 0 & 0
    \end{bmatrix}
\label{e: feature matrix example 2}
\end{equation}
From Eq.~(\ref{e: feature matrix example 2}), (and in reference to Eq.~(\ref{eq: separation value appendix})) the best probe state and measurement basis are ones that pick out the smaller off-diagonal elements (i.e. $\frac{1}{\sqrt{2}} ( \ket{01}+\ket{10})$ in our defined basis). 

As a more practical illustration, we now go on to use the characteristic feature matrix to identify the optimal measurement basis and probe state for the example problem of section~\ref{s: exponential advantage}.

\subsection{Choice of probe state and measurement basis for exponential advantage example}
\label{s: optimal measure for exponential advantage}

The characteristic function of the example distribution for $\sum_j \theta_j = \pm C$ is given by
\begin{equation}
    \chi_{\pm C}(\bm{k}) = e^{ik_NC}\prod_{j=1}^{N-1}\text{sinc}(\pi(k_j-k_N))
\label{e: example characteristic function}
\end{equation}

The difference in characteristic functions is therefore:
\begin{equation}
    \chi_C(\bm{k})-\chi_{-C}(\bm{k}) = 2i\sin(C)\prod_{j=1}^{N-1}\text{sinc}(\pi(k_j-k_N))
\label{e: example characteristic function p2}
\end{equation}

Note that Eq.~\ref{e: example characteristic function p2} goes to zero unless either $k_j=1\forall j$ or $k_j=-1\forall j$. This means that the only non-zero entries of the feature matrix are
$$F_{\ket{0^N},\ket{1^N}} = 2i\sin(C), F_{\ket{1^N},\ket{0^N}}=-2i\sin(C) $$

Consequently, given a probe state $\rhou$ and projector $\hat{O}$, the separation value becomes equal to 
\begin{equation}
    \Delta_{A,B}=2\sin(C)(i\bra{0^N}\hat{\rho}\ket{1^N}\bra{0^N}\hat{O}\ket{1^N} + h.c.) = 4\sin(C)\text{Re}\left\{i\bra{0^N}\hat{\rho}\ket{1^N}\bra{0^N}\hat{O}\ket{1^N}\right\}
\label{e:separation value exponential advantage example}
\end{equation}
Note from Eq.~\ref{e:separation value exponential advantage example} that the triangle inequality implies the separation value magnitude is at most $4\bigg|\sin(C)\bra{0^N}\hat{\rho}\ket{1^N}$ $\bra{0^N}\hat{O}\ket{1^N}\bigg|$. The separation is at its maximum for a pure state, and so  $\left|\bra{0^N}\hat{\rho}\ket{1^N}\right| $ is at most $1/2$. Further, since $\hat{O}$ is a projector, the sum of the squared norms of each row is equal to the diagonal of that row - from this we get that any off-diagonal element is at most $1/2$. In other words, the separation value magnitude is at most $|\sin(C)|$. We can construct $\hat{\rho},\hat{O}$ that saturate this bound. First, we have $\left|\bra{0^N}\hat{\rho}\ket{1^N}\right|=1/2$ for any pure probe state given by $\ket{\psi_{\text{probe}}}=(1/\sqrt{2})(\ket{0^N}+e^{i\phi}\ket{1^N})$. Second, we have $\left|\bra{0^N}\hat{O}\ket{1^N}\right|=1/2$ for $\hat{O}=\ket{\psi_{\text{meas}}}\bra{\psi_{\text{meas}}}$ where $\ket{\psi_\text{meas}}=(1/\sqrt{2})(\ket{0^N} + e^{i\xi}\ket{1^N})$. Third, and finally, we must choose the values of $\phi,\xi$ so that the real part of the product in Eq.~\ref{e:separation value exponential advantage example} is maximized. This is given if $\phi = \xi \pm \pi/2$. In other words, the probe state and measurement basis that minimize the lower ($\Omega(1/\Delta_{A,B})$) and upper $O(1/\Delta_{A,B}^2)$ bounds on the number of required shots are given by:

\begin{equation}
    \hat{\rho} = \frac{1}{2}\left(\ket{0^N} + e^{i\phi}\ket{1^N}\right)\left(\bra{0^N}+e^{-i\phi}\bra{1^N}\right),\ \ \ \hat{O}= \frac{1}{2}\left(\ket{0^N} \pm ie^{i\phi}\ket{1^N}\right)\left(\bra{0^N} 
\mp ie^{-i\phi}\bra{1^N}\right)
\end{equation}

Suppose that instead $\hat{\rho}$ is an unentangled state, and $\hat{O}$ is a measurement in an unentangled basis. Consulting Eq.~\ref{e:separation value exponential advantage example}, once more we note that the separation magnitude is at most 4$\left|\sin(C)\bra{0^N}\hat{\rho}\ket{1^N}\bra{0^N}\hat{O}\ket{1^N} \right|$. $|\bra{0^N}\hat{\rho}\ket{1^N}|$, in the case of an unentangled state, is at most $1/2^N$. $|\bra{0^N}\hat{O}\ket{1^N} |$ is at most $1/2$ and this bound is saturated in a measurement basis of single-qubit Hadamard states, with $\hat{O}$ being a sum of projectors over product states with an even number of $\ket{-}\bra{-}$ terms. Finally, to make the imaginary part of the product in Eq.~~\ref{e:separation value exponential advantage example} go to zero (and therefore the real part accounts for all of the magnitude) we choose $\hat{\rho}$ to be a pure product state with one qubit initialized in the state $(1/\sqrt{2})(\ket{0}+i\ket{1})$ and the other qubits initialized in the Hadamard state $\ket{+}$. This gives us the largest possible separation value for an unentangled probe state and unentangled measurement basis possible of $\sin(C)/2^{N-1}$, and requiring at least $\Omega(2^{N}/\sin(C))$ shots for discrimination in scaling. 

With a clearer understanding of how an underlying characteristic function and choice of $\usense$ interplay to produce a feature matrix, we now go on to our general proofs of conditions of exponential stochastic sensing advantage based on properties of the feature matrix.

\subsection{Theorems and proofs}
\label{s: theorems and proofs}

Within this setting, we are able to construct a more general framework of when to expect an exponential separation, and to what degree, for different kinds of distributions. \\

\textit{\textbf{Theorem D.1:} If $\left|F_{a,b}\right| = O\left(r^{-N}\right)$ for fixed constant $r>\sqrt{2}$ and for all $\ket{a}, \ket{b}$, then for any choice of probe state $\rho_0$ and projector $\hat{O}$ the number of shots required to distinguish $C$ from $C'$ scales exponentially in $N$.} \\

\textit{Proof:} Let $\rho_0$ be the input probe state and define the output density matrices for constraint values $C=C_1,C_2$ as
\begin{equation}
    \rho_{1,2} := \int_{f(\vec{\theta})=C_{1,2}}p_{C_{1,2}}(\vec{\theta})U_{\text{sense}}(\vec{\theta})\rho_0U_{\text{sense}}^\dagger(\vec{\theta}) d\vec{\theta}
\end{equation}
By the Helstrom bound, the probability of successful classification of $\rho_1$ given initial priors $\pi_1,\pi_2$ (where $\pi_1=1-\pi_2$) is
\begin{equation}
    p_{\text{success}} = \frac{1}{2} + \frac{1}{2}||\pi_1\rho_1 - \pi_2\rho_2||_1 = \frac{1}{2} + \frac{1}{2}||\pi_1\rho_1 - \pi_2\rho_1 + \pi_2(\rho_1-\rho_2)||_1 
\end{equation}

\begin{equation}
    \leq  \pi_1 + \frac{1-\pi_1}{2}||\rho_1-\rho_2||_1  
\end{equation}
where $||*||_1$ denotes the trace norm. Note that \begin{equation}
    ||\rho_1-\rho_2||_1 \leq \sqrt{2^N}\sqrt{||\rho_1-\rho_2||_2} = \sqrt{2^N}\sqrt{\text{Tr}\left[ (\rho_1-\rho_2)(\rho_1-\rho_2) \right]} = \sqrt{2^N}\sqrt{\text{Tr}\left[ (F\circ \rho_0)^2 \right]}\leq \sqrt{2^N}r^{-N}
\end{equation}

Consequently, if $r>\sqrt{2}$, the separation between the output density matrices is exponentially small and therefore requires exponentially many shots. Note that this cannot be overcome with adaptive protocols. Since the probability of success at any one shot only offers an exponentially small distribution over the priors, for \textit{any} choice of priors, any individual shot, at every stage, can offer only an exponentially small improvement to the confidence in classification - otherwise the Helstrom bound would be violated at some point. Consequently, even adaptive schemes would require exponentially many shots. An important consequence of this theorem, for later proofs, is that elements of the feature matrix that scale as $O(\sqrt{2}^{-N})$ may be ignored. \\

\textit{\textbf{Theorem D.2:} If for all $N$ there exists at least one pair of states $\ket{a},\ket{b}$ such that $\left|F_{a,b}\right| = \Omega(1/N^k)$ for fixed constant $k$, then there exists an input probe state and measurement observable $\hat{O}$ such that the number of shots required to distinguish $C$ from $C'$ scales polynomially in $N$}. \\ 

\textit{Proof:} Choose $\rho_0 = \ket{\psi}\bra{\psi}$ where $\ket{\psi}= \frac{1}{\sqrt{2}}(\ket{a}+e^{i\phi}\ket{b})$ for some choice of $\phi$, and choose $\hat{O} = (\ket{a}-\ket{b})(\bra{a}  - \bra{b})$. The resulting choice of operators picks out the outer product of states with a polynomially small coefficient. This enables a polynomially small separation measure and thus a polynomially large number of shots required to distinguish the constraints. \\

We now introduce the notion of an ``unentangled" probe state which we present, without loss of generality, as 
\begin{equation}
    \rho_{\text{unentangled}} = \otimes_{j=1}^N\rho^j,
\label{e: rho unentangled}
\end{equation}

where each $\rho^j$ is a density matrix on a single qubit. Note that it is, in principle, possible to construct a more general unentangled probe through the statistical mixtures of probes of form Eq.~\ref{e: rho unentangled} - however, the separation value would simply be the average of individual separation values, and therefore at most as large as the best separation for an unentangled probe of form Eq.~\ref{e: rho unentangled}. To study this setting more carefully, we need to introduce some definitions. Let $d$ be the Hamming distance between states $\ket{a},\ket{b}$ (i.e. the number of bits that they differ by), and $D$ be the set of indices specifying the bits where they differ and $\delta(i)$ specifying the bit value of $a_i$ at the point where they differ. Finally, let $\sigma(i)$ return the bit value at an index $i$ where $\ket{a}$ and $\ket{b}$ have the same bit value.

For example, if we have $d=2$, $D=\{1, 3\}$, $\delta_1=0, \delta_3=1$ and finally $\sigma_2=0, \sigma_4=1$, this corresponds to the outer product of states $\ket{0,0,1,1}\bra{1,0,0,1}$. That is, they have Hamming distance 2 ($d=2$), differing at the first and third bit ($D=\{1,3\}$) with the first state having values 0 and 1 at those bit indices ($\delta(1)=0, \delta(3)=1$) and the bit indices where they are the same, at the second and fourth bits, have values of 0 and 1 respectively ($\sigma(2)=0,\sigma(4)=1$). Finally, let $\mathcal{D}^d$ denote the set of all possible tuples $(D,\delta)$ for states with Hamming distance $d$.

Finally, we also emphasize that the following theorems hold for, specifically, a \textit{local} $\usense$- that is, one of form
\begin{equation}
    \usense = \exp\left\{-i\sum_{j=1}^{N} \theta_j\hat{G}_j\right\}
\end{equation}
where each generator $\hat{G}_j$ acts on exactly one qubit. \\

\textit{\textbf{Lemma D.3:} For any unentangled probe state $\rho_{\text{unentangled}}$ and local $\usense$, the contribution to the absolute value of $\Delta^{A,B}$ from terms with Hamming distance $d$ is at most}\\

\begin{equation}
    \sum_{D,\delta\in\mathcal{D}^d} \frac{|F_{a_\delta, b_\delta}| \text{max}_\sigma |\bra{a_{\delta,\sigma}}\hat{O}\ket{b_{\delta,\sigma}}|}{2^d},
\end{equation}
where the summation is over all possible $D,\delta$ and $\max_{\sigma}$ denotes the maximum cross-diagonal term of $\hat{O}$ for a fixed choice of $D,\delta$ (i.e. choosing the identical bits to maximize this off-diagonal).

\textit{Proof:} Let $d$ denote the number of bits two states differ by,  $D$ denoting the set of indices where $\ket{a},\ket{b}$ differ with $\delta_j$ specifying the $j$-th bit of $a$, and $\mathcal{D}^d$ is the set of all such index differences/directions. Finally, let $\sigma_j$ denote the value of the identical bit between $\ket{a},\ket{b}$ at the $j$-th index. The magnitude of the contribution to the separation value for terms differing by $d$ bits is given by
\begin{equation}
    \left|\sum_{D,\delta\in\mathcal{D}^d} \prod_{j\in D}\rho^j_{\delta_j,1-\delta_j} \left(\sum_{\sigma_k=0,1}\prod_{k\notin D}\rho_{\sigma_k,\sigma_k}^k F_{a_{\delta,\sigma},b_{\delta,\sigma}}\bra{a_{\delta,\sigma}}\hat{O}\ket{b_{\delta,\sigma}}\right)  \right|
\label{e:fixed d separation}
\end{equation}
\begin{equation}
    = \left|\sum_{D,\delta\in\mathcal{D}^d} \prod_{j\in D}\rho^j_{\delta_j,1-\delta_j} F_{a_{\delta},b_{\delta}}\left(\sum_{\sigma_j=0,1}\prod_{j\notin D}\rho_{\sigma_j,\sigma_j}^j \bra{a_{\delta,\sigma}}\hat{O}\ket{b_{\delta,\sigma}}\right)  \right|.
\label{e:fixed d separation local sensing assumption}
\end{equation}
We move $F_{a,b}$ inward in the summation since, due to our assumed, purely local $U_{\text{sense}}$, this value depends only on the differences between $a,b$. Next, we bound this quantity using the triangle inequality and exploit the fact that $|\rho^j_{\delta_j,1-\delta_j}|\leq 1/2$ to get: 
\begin{equation}
    \leq \sum_{D,\delta\in\mathcal{D}^d} \frac{|F_{a_\delta, b_\delta}|}{2^d}\left( \sum_{\sigma_j=0,1}\prod_{j\notin D}\rho_{\sigma_j,\sigma_j}^j |\bra{a_{\delta,\sigma}}\hat{O}\ket{b_{\delta,\sigma}}| \right)
\end{equation}

\begin{equation}
    \leq \sum_{D,\delta\in\mathcal{D}^d} \frac{|F_{a_\delta, b_\delta}| \text{max}_\sigma |\bra{a_{\delta,\sigma}}\hat{O}\ket{b_{\delta,\sigma}}|}{2^d}\left( \sum_{\sigma_j=0,1}\prod_{j\notin D}\rho_{\sigma_j,\sigma_j}^j  \right) = \sum_{D,\delta\in\mathcal{D}^d} \frac{|F_{a_\delta, b_\delta}| \text{max}_\sigma |\bra{a_{\delta,\sigma}}\hat{O}\ket{b_{\delta,\sigma}}|}{2^d},
\label{e: remove sum of rhoss}
\end{equation}
where the final equality in Eq.~\ref{e: remove sum of rhoss} comes from reorganizing the sum of products as product of sums $\rho^j_{0,0}+\rho^j_{1,1}=1$. \\

\textbf{Difficult distributions for product probe states and arbitrary measurements - conditional on mean values of feature matrix elements}

\textit{\textbf{Theorem D.3.1:} Let $F_{a_\delta,b_\delta}$ be the feature matrix entry indexed by states $\ket{a}$ and $\ket{b}$ with Hamming distance $d$ with $\delta(j)$ specifying the bits of $\ket{a}$ at indices where the bits differ. Let $\left| \bar{F}_{a,b} \right|$ be the average magnitude of all feature matrix elements where $a,b$ have this Hamming distance. Suppose} $\left| \bar{F}_{a,b} \right|$ goes as $O(r^{-N})$ for a local $\usense$. Defining $\epsilon:= d/N$, if $r>\left(\frac{1}{\epsilon}-1\right)^\epsilon/(1-\epsilon)$ for every value of $d$, then for any choice of projector $\hat{O}$ or unentangled probe state $\rho_{\text{unentangled}}$, the number of shots required to distinguish distributions $\mathcal{P}_{\rm A}$,$\mathcal{P}_{\rm B}$ scales exponentially in $N$.\\

\textit{Proof:} Since we consider a local $\usense$ and $\rho_{\text{unentangled}}$, we utilize Eq.~\ref{e: remove sum of rhoss} to bound the contribution to the separation value. Note that the maximum value for any element of the projector matrix is 1, so this gives us:

\begin{equation}
    \leq \sum_{D,d\in \mathcal{D}^d} \frac{|F_{a_\delta,b_\delta}|}{2^d} = \frac{|\bar{F}_{a,b}|}{2^d}2^d {N \choose d} = {N \choose d}\left|\bar{F}_{a,b}\right|,
\label{e:final inequality theorem a131}
\end{equation}
where $\left|\bar{F}_{a,b}\right|$ is the average magnitude of the feature matrix across elements $a,b$ differing by $d$ bits. Choosing $d=\epsilon N$, we can use the Stirling approximation (accurate within a factor near 1~\cite{robbins_sterling}) to get that 
\begin{equation}
    {N \choose \epsilon N} \approx \frac{\sqrt{2\pi N}}{2\pi N \sqrt{\epsilon (1-\epsilon)}} \frac{(\frac{N}{e})^N}{(\frac{\epsilon N}{e})^{\epsilon N} (\frac{(1-\epsilon) N}{e})^{(1-\epsilon) N}} = \frac{1}{\sqrt{2\pi N \epsilon (1-\epsilon)}}\left(\frac{(\frac{1}{\epsilon}-1)^\epsilon}{1-\epsilon}\right)^N.
\label{e: stirling approximation combinations}
\end{equation}
Together with Eq.~\ref{e:final inequality theorem a131}, Eq.~\ref{e: stirling approximation combinations} proves the theorem.\\

\textbf{Difficult distributions for product probe states and arbitrary measurements - conditional on distribution of feature matrix elements}

\textit{\textbf{Theorem D.3.2:} Suppose that $\left|F_{a,b}\right| = \Omega(1/N^k)$ for fixed constant $k$ only for at most $O\left(2^{N(1-\xi)} \right)$ choices of $\ket{a},\ket{b}$ and some constant $0< \xi\leq 1$ and a local $\usense$. Further, suppose there exist an ordered vector $\bm{\epsilon}$ of length $O(\text{poly}(N))$ such that $0<\bm{\epsilon}_{j-1} <\bm{\epsilon}_j < \bm{\epsilon}_{j+1}\leq \frac{1}{2}$ such that the number of feature matrix entries with magnitudes $\Omega(2^{-\epsilon_{j+1}N}) = \left|F_{a,b}\right| = O(2^{-\epsilon_j N})$ goes as at most $O(2^{(1+\xi_j)N})$ elements, for $0<\xi_j<\epsilon_j<1$. Then for any unentangled probe state and any projector $\hat{O}$, exponentially many shots in $N$ are required to distinguish distributions $\mathcal{P}_{\rm A},\mathcal{P}_{\rm B}$.} \\

\textit{Proof:} As before, we start from Eq.~\ref{e: remove sum of rhoss} from Lemma D.3. Recalling that the maximum value of an off-diagonal projector is 1, this gives us 
\begin{equation}
    \leq \sum_{D,d\in \mathcal{D}^d}\frac{|F_{a_\delta, b_\delta}|}{2^d}
\label{e: theorem 132 eq bound}.
\end{equation}
First, let's consider the number of polynomially scaling elements. Note that for every choice of off-diagonal differences, there are $2^{N-d}$ elements that have the same magnitude of feature matrix entry. Since there are only $O(2^{N-\xi N})$ polynomially scaling elements of the feature matrix, there must be at most $O(2^{d-\xi N})$ such elements. For exponentially small scaling terms, there must be at most $O(2^{d+\xi_j N})$ such scaling terms. For the polynomially scaling terms, we plug in the scaling value into Eq.~\ref{e: theorem 132 eq bound} and the possible number of states to get:
\begin{equation}
    \sum_{D,d\in \mathcal{D}^d}\frac{|F_{a_\delta, b_\delta}|}{2^d} \leq \sum_{D,d\in \mathcal{D}^d}\frac{1}{2^d} \leq C2^{d-\xi N}\frac{1}{2^d} = C 2^{-\xi N},
\label{e: bound for polynomially scaling terms}
\end{equation}
where the first bound arises from the fact that any entry of the feature matrix is at most 1 in magnitude, and the $C$ from the second bound being a fixed constant such that there are at most $C2^{(1-\xi)N}$ many polynomially-small entries. Extending that same reasoning to the exponentially scaling terms, for each $\xi_j$, we have inequality:

\begin{equation}
    \sum_{D,d\in \mathcal{D}^d}\frac{|F_{a_\delta, b_\delta}|}{2^d} \leq \sum_{D,d\in \mathcal{D}^d}\frac{C_1 2^{-\epsilon_j N}}{2^d} \leq C_1C_2 2^{d+\xi_j N}\frac{2^{-\epsilon_j N}}{2^d} = C_1 C_2 2^{-(\epsilon_j-\xi_j) N},
\label{e: bound for exponentially scaling terms}
\end{equation}
Here, $C_1,C_2$ are the coefficients that bound the entries of $|F_{a,b}|\leq C_1 2^{-\epsilon_j N}$ and the number of such elements as $\leq C_2 2^{(1+\xi_j)N}$. Since $\bm{\epsilon}$ is only polynomially long in the number of qubits $N$, it follows that even if we add together all the separate bounds of Eq.~\ref{e: bound for exponentially scaling terms} for each $j$ together, we will still get an exponentially vanishing bound. This concludes the proof.\\

\textbf{Difficult distributions for product probe states and unentangled measurements - conditional on root mean square of feature matrix elements}

\textit{\textbf{Theorem D.3.3 (Proposition 1 in the text): } Suppose $\rho_{\text{probe}}$ is a product state and $\hat{O}$ is a projector to an unentangled basis. Let $F_{\text{RMS}}(d)$ be the root mean square magnitude of $|F_{a,b}|$ for states differing by $d$ bits and a local $\usense$. Defining $\epsilon = d/N$, if  $F(d) = O(r^{-N})$ with $r>\sqrt{\frac{(1/\epsilon -1)^\epsilon}{1-\epsilon}}$  for every $d$, then exponentially many shots are required for the discrimination.}\\

\textit{Proof:} As with the proof for theorem D.3.1, we start with Eq.~\ref{e: remove sum of rhoss} as our starting point. We consider a local measurement basis, which must be chosen to lie on the equator of each qubit's Bloch sphere to extract the maximum amount of information about the $z$-sensors (more generally, must lie in equal superposition of the two eigenstates of the single qubit encoder in $U_{\text{sense}}$). In other words, the projector must maximally non-commute with the sensing unitary. From this, we have that $\bra{a_{\delta, \sigma}}\hat{O}\ket{b_{\delta,\sigma}}$ does not depend on what bits $\ket{a},\ket{b}$ have in common, but only on the bits that differ. Consequently, we have
\begin{equation}
    \sum_{D,\delta\in\mathcal{D}^d} \frac{|F_{a_\delta, b_\delta}| \text{max}_\sigma |\bra{a_{\delta,\sigma}}\hat{O}\ket{b_{\delta,\sigma}}|}{2^d} = \sum_{D,\delta\in\mathcal{D}^d} \frac{|F_{a_\delta, b_\delta}| |\bra{a_{\delta}}\hat{O}\ket{b_{\delta}}|}{2^d},
\label{e: separable O assumption}
\end{equation}
where, in Eq.~\ref{e: separable O assumption}, since the off-diagonal entries of $\hat{O}$ depend only on the differing bits, the off-diagonals have the same value for all $\sigma_j$.  

Now, note that for any projector $\hat{O}$, $\hat{O}^2=\hat{O}$, so for the sum $\sum_{a,b}|\bra{a}\hat{O}\ket{b}|^2 = \text{Tr}(\hat{O}^2) = \text{Tr}(\hat{O}) \leq 2^N$. Now, note that since $\hat{O}_{a,b}$ is identical over all $2^{N-d}$ choices of identical bits  between $a,b$, we have
\begin{equation}
    \sum_{D,\delta\in \mathcal{D}^d} 2^{N-d}\left|\bra{a_\delta} \hat{O} \ket{b_\delta} \right|^2 \leq 2^{N}
\end{equation}
\begin{equation}
    \sum_{D,\delta\in \mathcal{D}^d}\left|\bra{a_\delta} \hat{O} \ket{b_\delta} \right|^2 \leq 2^{d}.
\label{e:bound on square product observable}
\end{equation}
Returning to Eq.~\ref{e: separable O assumption}, we can upper bound this expression by via the Cauchy-Schwarz inequality, giving us a final inequality:
\begin{equation}
    \sum_{D,\delta\in\mathcal{D}^d} \frac{|F_{a_\delta, b_\delta}| |\bra{a_{\delta}}\hat{O}\ket{b_{\delta}}|}{2^d} \leq \frac{1}{2^d}\sqrt{\sum_{D,\delta\in \mathcal{D}^d}|F_{a,b}|^2}\sqrt{\sum_{D,\delta\in \mathcal{D}^d} \left|\bra{a_\delta} \hat{O} \ket{b_\delta} \right|^2} 
\end{equation}
\begin{equation}
    \leq \frac{1}{2^{d/2}} \sqrt{\sum_{D,\delta\in \mathcal{D}^d}|F_{a,b}|^2} = \frac{F_{\text{RMS}}(d)}{2^{d/2}}\sqrt{{N \choose d}2^d} = F_{\text{RMS}}(d)\sqrt{{N \choose d}} ,
\end{equation}
where $F_{\text{RMS}}$ is the root mean square magnitude of terms $|F_{a,b}|$ where $a,b$ have Hamming distance $d$. Now, let $d=\epsilon N$. Then, using Eq.~\ref{e: stirling approximation combinations}, an approximation accurate to within a factor of two for positive integers, we get
\begin{equation}
    \leq  \sqrt{2 \frac{1}{\sqrt{2\pi N \epsilon (1-\epsilon)}}\left(\frac{(1/\epsilon -1)^\epsilon}{1-\epsilon}\right)^N}F_{\text{RMS}}(d).
\label{e: inequality unentangled measure}
\end{equation}
Supposing that $F_{\text{RMS}}(d) = O(r^{-N})$ and dropping terms that do not matter for exponential scaling, we get:
\begin{equation}
    r> \sqrt{\frac{(1/\epsilon-1)^\epsilon}{1-\epsilon}}
\label{e: final inequality unentangled measure}
\end{equation}

This completes the proof. \\

\textbf{Difficult distributions for product probe states and unentangled measurements - conditional on distribution of feature matrix elements}

\textit{\textbf{Theorem D.3.4:} Suppose that $\left|F_{a,b}\right| = \Omega(1/N^k)$ for fixed constant $k$ only for at most $O\left(2^{N(1-\xi)} \right)$ choices of $\ket{a},\ket{b}$ and some constant $0< \xi\leq 1$. Further, suppose there exist an ordered vector $\bm{\epsilon}$ of length $O(\text{poly}(N))$ such that $0<\bm{\epsilon}_{j-1} <\bm{\epsilon}_j < \bm{\epsilon}_{j+1}\leq \frac{1}{2}$ such that the number of feature matrix entries with magnitudes $\Omega(2^{-\epsilon_{j+1}N}) = \left|F_{a,b}\right| = O(2^{-\epsilon_j N})$ goes as at most $O(2^{(1+\xi_j)N})$ elements, for $0<\xi_j<2\epsilon_j<1$. Then for any unentangled probe state and any projector $\hat{O}$ to an unentangled basis, exponentially many shots in $N$ are required to distinguish distributions $\mathcal{P}_{\rm A},\mathcal{P}_{\rm B}$.} \\

\textit{Proof:} The proof for the polynomial case is identical as in the proof for Theorem D.3.2. For the exponential case, the proof is the same up to Eq.~\ref{e: bound for exponentially scaling terms} (i.e. arguments for how many distinct terms are allowed) past which we have, using the inequality from Eq.~\ref{e:bound on square product observable},
\begin{align}
    \sum_{D,d\in\mathcal{D}^d} \frac{\left|F_{a_\delta,b_\delta} \right| \left| 
\bra{a_\delta}O\ket{b_\delta} \right|}{2^d} \leq \frac{1}{2^d}\sqrt{\sum_{D,d\in\mathcal{D}^d}\left|F_{a_\delta,b_\delta} \right|^2}\sqrt{\sum_{D,d\in\mathcal{D}^d} \left| 
\bra{a_\delta}O\ket{b_\delta} \right|^2} \nonumber \\ \leq \frac{C_1C_2}{2^{d/2}}2^{\frac{d+\xi_jN}{2}}2^{-\epsilon_jN} = C_1C_22^{-\left(\epsilon_j -\frac{\xi_j}{2}\right)N}
\label{e:final bounds on distribution based separation}
\end{align}
Here, $C_1,C_2$ are the coefficients that bound the entries of $|F_{a,b}|\leq C_1 2^{-\epsilon_j N}$ and the number of such elements as $\leq C_2 2^{(1+\xi_j)N}$. Since $\bm{\epsilon}$ is only polynomially long in the number of qubits $N$, it follows that even if we add together all the separate bounds of Eq.~\ref{e:final bounds on distribution based separation} for each $j$ together, we will still get an exponentially vanishing bound. This concludes the proof.\\

\textit{\textbf{Theorem D.4 (Proposition 2 in the text): } Consider an arbitrary $\usense$ as defined in eq.~\ref{eq: usense general appendix}. If there are $\Omega(2^N/N^{k_1})$ disjoint pairs $a,b$ such that $\left|F_{a,b}\right| = \Omega(1/N^{k_2}) $ for some fixed constants $k_1, k_2$, then there exists an unentangled probe state $\rho_{\text{probe}}^{\text{unentangled}}$ and a measurement observable $\hat{O}$ such that the number of shots required to distinguish $C$ from $C'$ scales polynomially with $N$ for any fixed confidence level}. \\

\textit{Proof:} Choose the unentangled probe state to be the Hadamard state for each qubit, i.e. $\rho_{\text{probe}}^{\text{unentangled}} = \ket{+^N}\bra{+^N}$, where $\ket{+}$ is an equal superposition over $\usense$ eigenstate bits at each qubit. For each pair $\ket{a},\ket{b}$, define the projective term 
\begin{equation}
    P_{a,b} := \frac{1}{2}(\ket{a} + e^{i\phi}\ket{b})(\bra{a} + e^{-i\phi}\bra{b}),
\label{e: pair projector}
\end{equation}
where $\phi = \arg\{F_{a,b}\}$. Note that this term is Hermitian and $P_{a,b}^2 = P_{a,b}$, so the term is a proper projector. Finally, define our observable $\hat{O}$ as
\begin{equation}
    \hat{O} := \sum_{(a,b)\in G} P_{a,b}
\label{e:good observable}
\end{equation}
where $G$ denotes the set of disjoint pairs with polynomially scaling feature matrix elements. Note that all $P_{a,b}$ are orthogonal, so $\hat{O}^2 = \hat{O}$ and $\hat{O}$ is hermitian, implying that it is also a projector. Finally, computing the separation value gives us:
\begin{equation}
    \Delta^{A,B} = \sum_{a,b} \left(\rho_{\text{probe}}^{\text{unentangled}} \circ F \circ \hat{O} \right) = \sum_{(a,b)\in G}\frac{1}{2^N}|F_{a,b}| = \Omega\left(\frac{1}{N^{k_1+k_2}} \right).
\end{equation}
Since the separation value is only polynomially small, polynomially many shots suffice to distinguish the distributions. \\

An important consequence of Theorem D.4 is that it effectively rules out the use of machine learning to obtain exponential stochastic sensing advantage, at least without a highly targeted initialization. The reason for this is that it implies that the number of eigenstates of $U_{\text{sense}}$ that pick out useful information about the underlying distributions must scale as an exponentially small fraction of the overall Hilbert space. All other states acquire exponentially little information, and therefore a vast barren plateau emerges everywhere but some small region of Hilbert space, which one has a vanishingly small chance of simply guessing without additional prior knowledge. Conversely, if one already knows what this set of useful states is, one can simply select the appropriate superposition of them in some entangled state to obtain the advantage, without the aid of quantum machine learning (though certainly polynomial improvements to shots past this point are conceivably possible).

\textit{\textbf{Corollary D.4:} Consider a local $\usense$ and suppose distributions $\mathcal{P}_{\rm A}$, $\mathcal{P}_{\rm B}$ are even in each argument $\theta_j$, up to identical constant shifts in each $\theta_j$ for both distributions. If at least one element of the feature matrix shrinks as $\Omega(1/N^k)$ for some fixed constant $k$, there can be no exponential advantage in shots from using an entangled probe state to discriminate between the distributions. More strongly, there isn't even an exponential advantage from using an entangled measurement basis.}

\textit{Proof:} For a local $\usense$, the points we evaluate are given by the difference between the expected values of $\exp\{-i(\bm{a}-\bm{b})\cdot\bm{\theta}\}$ of the two distributions, where $\bm{a},\bm{b}$ correspond to the bit strings of states, $\ket{a}\ket{b}$. If the two distributions are even in all their arguments, then for a particular choice of  $\ket{a}\ket{b}$, any other choice of states differing at the same bit indices will produce an identical difference of expectations (difference in bits determines the sign of a particular $\theta_j$ in the exponent which, for even probability distribution functions, is unimportant). Consequently, if there exists a polynomially scaling element of the feature matrix for terms $\ket{a},\ket{b}$ with Hamming distance $d$ then there exist $2^d$ (for all the ways bits at those indices can differ) times $2^{N-d}$ (all the ways the similar bits at the indices can be the same) disjoint pairs $\ket{a},\ket{b}$ for which the corresponding feature matrix is polynomially scaling. Therefore, by Theorem D.4, there exists an unentangled probe state and some measurement basis where it is possible to discriminate between the two distributions in polynomially many shots. Importantly, this Corollary also applies to pairs of distributions that are identical constant shifts of $\bm{\theta}$ away from being even in each argument. This is because the shifts could be canceled as fixed, single qubit rotations applied to the probe state or observable. Even more strongly, since the complex phase $\phi = \arg\{F_{a,b}\}$ is the same for all pairs $a,b$, we can choose our projector to have an eigenbasis in product states over the qubits whose bits differ (in particular, the Hadamard basis), so there isn't even an exponential advantage from entanglement in the measurement basis.

\subsection{Removal of exponential advantage under measurement-time optimization in the continuous process setting}
\label{s: general framework proofs: continuous time}

While this work's focus is on the case of random, unentangled unitary, it is instructive to comment on what happens if we treat the source of stochasticity in the phases as a continuous process where the amount of time that the probe state is exposed to the imposed processes is allowed to vary (as is studied in detail in~\cite{wang2024exponential}). In some cases, an exponential advantage in the random-unitary-channel picture  disappears when this additional degree of freedom is introduced in the continuous-process picture. Consider the example with a fixed sum of phases in the main text (Eq.~\ref{e: linear constraint example}) where the information on the sum is hidden on one qubit. The phase experienced by this qubit goes as C+(N independent random angles). Suppose we reduce the exposure time that this qubit has to the stochastic process to a fraction $dt$ of the time that the GHZ state is exposed to it and that this gives us a phase of $dt\times$C + $dt\times$(N independent random angles). If $dt\sim 1/N$ then we have negligible noise on our qubit, and we just need to distinguish $\pm dt\times $C, which needs a factor of $N^2$ more shots than the GHZ protocol. Accounting for the fact that the full time for sensing with the GHZ state takes $N$ shots for sensing with the unentangled state (since $dt\sim1/N$), this means that the advantage gained in total measurement time is only a factor of $N$, and so we have recovered the familiar Heisenberg scaling. Whether all such cases of exponential advantage with random unitary channels reduce to a Heisenberg advantage with measurement time optimization in a continuous process setting is an open question for future research. Importantly, given that there are often practical limits on how quickly a measurement can be performed and the system reset relative to the rate of some stochastic process, there might be a strong advantage in protocol resources all the same. We provide an outlook in the Discussion Section~\ref{sec:conc}.

\section{Exponential advantage for stochastic signals with nonlinear constraints: quadratic example}
\label{s:nonlinear constraint noncommuting example}

\subsection{The task}
Suppose that we are given parameters $\theta_1,..,\theta_{N_\text{var}} \in \mathbb{R}$ which are required to satisfy the constraint 
\begin{equation}
    \left(\sum_{j=1}^{N_\text{var}/2} \theta_j\right)^2 + \left(\sum_{j=N_{\text{var}}/2 +1 }^{N_\text{var}} \theta_j\right)^2 = C.
\label{e:sum of squares constraint}
\end{equation}
The task is then to be able to differentiate a constraint of $C+\epsilon$ and $C$ with as few shots as possible. 
\subsection{Entangled sensing unitary}
One way to recreate the sum-of-squares constraint as an expected value of our entangled circuit is to encode each parameter $\theta_j$ as the coefficient of a Pauli string. In particular, given $N_{\text{var}}$ parameters, take a system of $N_{\text{qubits}}=\lceil \log(N_{\text{var}})\rceil$ qubits and assign each parameter to be the coefficient of a distinct Pauli-string of length $N_{\text{qubits}}$ consisting of either a $X$ or $Y$ Pauli operator for each qubit. In particular, $\theta_j$ is assigned to Pauli strings with an even number of $Y$ Pauli elements for $j\leq N_{\text{var}}/2$ and to Pauli strings with an odd number of $Y$ Pauli elements for $j>N_{\text{var}}/2$. The resulting sensing unitary may be given by
\begin{equation}
    \usense(\bm{\theta}) := \exp\left\{-i\left(\sum_{j=1}^{N_{\text{var}}} \theta_j \hat{P}_j\right) \right\},\ \ \hat{P}_j :=  \otimes_{n=1}^{N_\text{qubits}}P_{\sigma_j(n)}, \ \ \ P_1=X,P_2=Y , 
\label{e: sum of squares encoding unitary}
\end{equation}
where $\sigma_j(n)\in\{1,2\}$ is an indexing function. Note that if $\hat{P}_j$ and $\hat{P}_k$, have a different parity of the number of $Y$ operators, then $\left\{\hat{P}_j, \hat{P}_k\right\} = 0$. Conversely, if they have the same parity of $Y$ elements, $\left\{\hat{P}_j, \hat{P}_k\right\} = 2 \otimes_{n=1}^{N_{\text{qubits}}} Q_{|\sigma_j(n)-\sigma_k(n)|}$ where $Q_0=\mathbb{I}$ and $Q_1=Z$ . From this, we get that 
\begin{equation}
    \hat{G}(\bm{\theta})^2 = \left(\sum_{j,k=1}^{N_{\text{var}}/2} \theta_j \theta_k \otimes_{n=1}^{N_{\text{qubits}}} Q_{|\sigma_j(n)-\sigma_k(n)|}\right) + \left(\sum_{j,k=N_{\text{var}}/2+1}^{N_{\text{var}}} 
 \theta_j \theta_k \otimes_{n=1}^{N_{\text{qubits}}} Q_{|\sigma_j(n)-\sigma_k(n)|}\right)
\label{e: square of noncommuting hamiltonian}
\end{equation}
Note that the state $\ket{1^{N_{\text{qubits}}}}$ is an eigenstate of the operator in Eq~\ref{e: square of noncommuting hamiltonian}, with eigenvalue $(\sum_{j\leq N_{\text{var}}/2} \theta_j)^2 + (\sum_{j>N_{\text{var}}/2} \theta_j)^2 = C $. Consequently, it is also an eigenstate of higher powers of $\hat{G}(\bm{\theta})$, with higher powers of the eigenvalue. Finally, note that $\bra{1^{N_{\text{qubits}}}} \hat{P}_j\ket{1^{N_{\text{qubits}}}} = 0$ for any $j$. Altogether, this gives us:
\begin{equation}
    \bra{1^{N_{\text{qubits}}}} \usense(\bm{\theta}) \ket{1^{N_{\text{qubits}}}} = \cos(\sqrt{C})
\label{e: fixed expected value}
\end{equation}
Therefore, after preparing the state $\ket{1^{N_{\text{qubits}}}}$ and applying $\usense$, the probability of still remaining in that state is given by $\cos^2(\sqrt{C})$. This is a ``noise-free" quantity, in that it holds and provides a way to estimate $C$ completely independently of any choice of distributions over the constraint manifold. Further, if we can apply a bit flip gate on an ancilla qubit, conditional on every one of the main qubits still being 1, we can estimate $C$ in a number of shots completely independent of $N$, just like the case of the linear constraint and all-$Z$ $\usense$ of the main text. The difference in expected value of the probability from a perturbation to $C$ of $C+\epsilon$ is approximately $\approx \epsilon \cdot \text{sinc}(2\sqrt{C})$ to first order in $\epsilon$. Consequently, it is possible to discriminate between the perturbed and unperturbed distributions in $O(1/(\epsilon \cdot \text{sinc}(2\sqrt{C}))^2)$ shots.

\subsection{Unentangled shot cost}
Here we compute the asymptotic scaling of the minimum number of shots required for an encoding unitary given by
\begin{equation}
    \hat{U}_{\text{encoding}}^{\text{unentangled}} := \otimes_{j=1}^{N_{\text{var}}} \exp\{-i\theta_j \hat{Z}_j\},
\label{e: sum of squares unentangled encoding}
\end{equation}
with an unentangled probe state and measurement basis. Note that in the case where we require each subset of variables to sum to $\sqrt{C/2}$, the unentangled competitor would be required to solve for the fixed sum of a subset of angles, which is known to be exponential in the worst case. Thus, we have exponential separation over this task.

\section{Exponential advantage for stochastic signals with multiple copies}
\label{s: multiple copies}

\subsection{The task}

\begin{figure}[H]
    \centering
    \includegraphics[width=\linewidth]{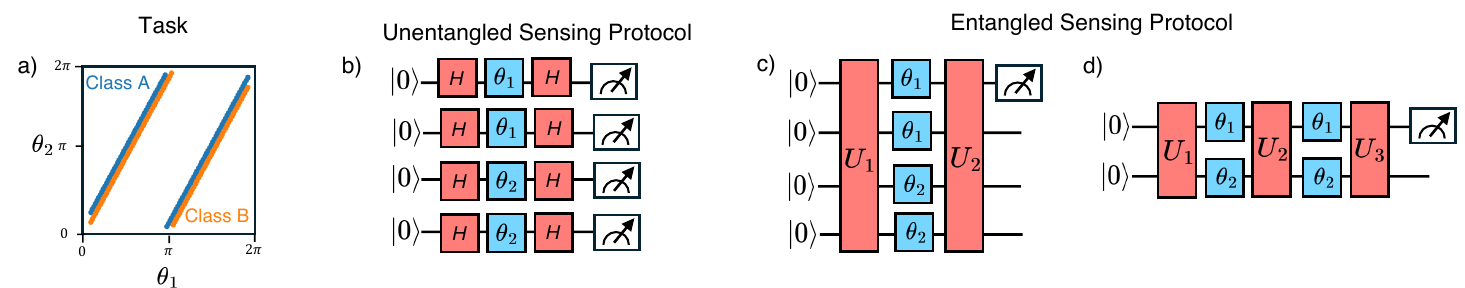}
    \caption{\textbf{Illustration of a two-qubit phase sensing task which requires either multiple applications of phases.} \textbf{a)} Dataset of the phases on each qubit for each of the class. The phase of the first qubit is roughly twice that of the second qubit. The task is to determine whether the set of phases received belong to Class A or B. At each new sample, we know the signal from the same class is applied. However, the actual values is randomly chosen from the dataset. \textbf{b)} The most general classical protocol \textbf{c)} The most general quantum protocol with two copies of the sensing unitary, which requires double the number of qubits. \textbf{d)} The most general quantum protocol with two access to the sensing unitary in time.}
    \label{Fig:Appendix:MultipleCopies}
\end{figure}

In this section, we consider the potential for a sensing advantage with multiple copies of the sensor. In this situation, the phases on the copies are the same, but still stochastic between samples. We motivate how an exponential sensing advantage (in the number of copies) can be achieved when entangled across the copies of the sensor is allowed, over a protocol which does not have entanglement. We also consider the situation (and achieve the same advantage) when we have one copy, but access to multiple sensing unitaries (all with the same phases).

Consider the following two qubit task, for which we need two copies. This is naturall connected to $N=2$ case of the task described in Sec.~\ref{sec:exp}. The only difference is that, for this task, the phase on one of the qubit is always twice that of the other qubit (this could happen, because either we know the magnetic field is twice as strong, or the qubit happens to be twice as sensitive). Concretely, if the first qubit experiences a phase $\theta_1 = 2\theta$, the second qubit experiences a phase $\theta_2 = \theta + \phi$. Between each sample, the value of $\theta$ uniformly varies between the total interval $[0, 2\pi)$ (as illustrated in Fig.~\ref{Fig:Appendix:MultipleCopies} (a)). We first ask whether this task can be solved within one copy of the system (that is, in the framework of the previous section). Curiously, the answer turns out to be no. Even in the presence of arbitrary two-qubit entanglement, there is no quantum protocol which can achieve a classification accuracy greater than random guessing. This can be seen by considering the intuition adopted in building the quantum protocol: which is to operate in a noiseless "subspace" of the Hilbert space. However, for this task, no such noiseless subspace exists. In fact, there is no protocol which can generate a state with a dependence on $\phi$. We can show this by considering the most general protocol:

\begin{enumerate}
    \item Let the initial state be any arbitrary superposition over the states of the Hilbert space: $\ket{\psi} = a\ket{00} + b\ket{01} + c\ket{10} + d\ket{11}$, where the coefficients are arbitrary up to the normalization constraint.

    \item Under the action of the sensing unitary, the state is: $\ket{\psi_{(\theta, \phi)}} = a\ket{00} + b e^{i(\theta + \phi)} \ket{01} + c e^{2i\theta}\ket{10} + d e^{i(3\theta + \phi)} \ket{11}$

    \item However, since $\theta$ is samples from a random variable, we must compute the density matrix to describe the state of the system: $\rho_{\phi} = \int_\theta \ket{\psi_{(\theta, \phi)}} \bra{\psi_{(\theta, \phi)}} d\theta = a^2 \ket{00}\bra{00} + b^2 \ket{01}\bra{01} + c^2 \ket{10}\bra{10}+ d^2 \ket{11}\bra{11}$ 

    \item The density matrix after the sensing unitary is completely diagonal. More importantly, there is no dependence on $\phi$, which means it is not possible to extract the value of $\phi$ from any general protocol.
\end{enumerate}

For this task, the smallest number of copies of the phases we require is two. This could either be in space, by having twice as many qubits, or in time, and interleaving two qubit interaction in between the two processes. For this simple task, we can easily derive the most optimal protocol for the quantum sensor. Let us first discuss the protocol which uses multiple copies in space (that is, requiring $4$ qubits in total).

\begin{enumerate}
    \item Beginning in the ground state $\ket{0000}$, we prepare the state $\ket{\psi} = \frac{1}{\sqrt{2}}(\ket{1000} + \ket{0101})$.

    \item Evolve this state under the action of the sensing unitary. For convention of the phases, see Fig~\ref{Fig:Appendix:MultipleCopies} (c). The resulting state of the system is $\ket{\psi} = \frac{1}{\sqrt{2}}(e^{2i\theta}\ket{1000} + e^{2i\theta + 2i\phi}\ket{0101})$

    \item We essentially operate in a "noiseless subspace", since the state is independent of $\theta$. We measure in the subspace of $\ket{\psi} = \frac{1}{\sqrt{2}}(\ket{1000} + i\ket{0101})$, which results in the probability of the qubit being in the excited state $P = \frac{1 + \sin{(2\phi)}}{2}$
\end{enumerate}

We can also consider the case of two qubits, but with access to two applications of the sensing unitary, as illustrated in Fig.~\ref{Fig:Appendix:MultipleCopies} (d). This scheme achieves the same sensitivity as the protocol with $4$ qubits.

\begin{enumerate}
    \item Beginning in the ground state $\ket{00}$, we prepare the state $\ket{\psi} = \frac{1}{\sqrt{2}}(\ket{10} + \ket{01})$.

    \item After sensing for the first time, the resulting state of the system is $\ket{\psi} = \frac{1}{\sqrt{2}}(e^{2i\theta}\ket{10} + e^{i(\theta + \phi)}\ket{01})$

    \item Apply the two-qubit unitary $U = \ket{00}\bra{10} + h.c.$, giving the state $\ket{\psi} = \frac{1}{\sqrt{2}}(e^{2i\theta}\ket{00} + e^{i(\theta + \phi)}\ket{01})$

    \item The second sensing unitary produces the state $\ket{\psi} = \frac{1}{\sqrt{2}}(e^{2i\theta}\ket{00} + e^{i(2\theta + 2\phi)}\ket{01})$

    \item Measure in the basis of $\frac{1}{\sqrt{2}}(\ket{00} + i\ket{01})$, which flips the qubit with a probability of $P = \frac{1 + \sin{(2\phi)}}{2}$
\end{enumerate}

In contrast, the classical sensor, illustrated in Fig.~\ref{Fig:Appendix:MultipleCopies} b), does not have access to quantum memory or entanglement. Similar to our previous discussions, the optimal scheme to be sensitive to $\phi$ involves estimating a joint correlation from many shots. In this case, the joint correlation involves three measurements, two of which are from qubits experiencing $\theta_2$. This therefore requires more samples to achieve the same classification accuracy compared to the previous protocols. We can generalize this to the $2$-qubit case, with $N$ copies by considering the situation where one of the phases is $N$ times larger than the other. Or we can consider the case where we have $N$-qubits with fewer number of copies. In both situations, it is straightforward to obtain an exponential quantum sensing advantage in sample complexity.

\subsection{Framework}
Here, rather than requiring the sensing unitary to encode a single parameter per qubit, we can define $\usense$ as having each parameter act on two qubits:
\begin{equation}
    \usense = \sum_{j=1}^{N_{\text{var}}} \frac{\theta_j}{2}(\sigma_z^j+\sigma_z^{j+N_{\text{var}}})
\label{e:multicopy usense}
\end{equation}
Considering the possible entries of the feature matrix, note that for entry corresponding to states $\ket{\bm{a_1},\bm{a_2}},\ket{\bm{b_1},\bm{b_2}}$ (where $\bm{a_j}$ is a bitstring on the $j$-th set of $N_{\text{var}}$ qubits) the $\bm{k}$ vector that determines the values we evaluate the difference of characteristic functions now has its entries as 
\begin{equation}
    \bm{k}_i = (\bm{a_1})_i-(\bm{b_1})_i + (\bm{a_2})_i-(\bm{b_2})_i
\label{e: multicopy k values}
\end{equation}

In other words, this choice of $\usense$ can evaluate the difference of characteristic functions at a larger number of different points than the single-copy version (entries corresponding to $\bm{k}_i=-2,-1,0,1,2$).  Consequently, for choices of characteristic function differences where a single-copy version evaluated it only at vanishing points, this version of $\usense$ can evaluate it at additional, non-vanishing points and achieve exponential advantage this way. Since this version of $\usense$ evaluates at a larger number of distinct points, rather than simply \textit{different} points than its single-copy counterpart, it is more powerful than simply a single-copy $\usense$ with its weights of $\sigma_z^j$ rescaled. 

For instance, consider the example problem provided in the main text under Eq.~\ref{e: linear constraint example}. As was demonstrated, it is impossible to distinguish the two distributions with only single copies using the unscaled version of $\usense$. With the version of $\usense$ in Eq.~\ref{e:multicopy usense}, choosing $\bm{a_1}= 0^{N_{\text{var}}}, \bm{b_1}=1^{N_{\text{var}}},\bm{a_2}=10..0, \bm{b_2}=0^{N_{\text{var}}}$ gives us a point in the difference of characteristic functions that we could not otherwise evaluate. In particular, given that the magnitude of the difference in characteristic functions is $\sin(k_N(C_2-C_1)/2)\text{sinc}(\pi(k_1-2k_N))\prod_{j=2}^{N-1}\text{sinc}(\pi(k_j-k_N))$, the above choice of $\ket{\bm{a_1,a_2}},\ket{\bm{b_1,b_2}}$ allows us to evaluate it at the non-vanishing points of $k_1=2, k_N=1=k_{j\neq 1}$.

\end{document}